# Cosmology for high school

Filipe Alves Pereira Bento ‡   and  José A. C. Nogales †‡

†‡*Departamento de Física (DFI)*
‡*Pós-graduação em Educação Científica e Ambiental(PPGECA)*
*Universidade Federal de Lavras (UFLA)/ICN*
*Lavras-MG,Caixa postal 3037,CEP 37000-900, Brazil.*

**Abstract**

These notes on modern cosmology were made with high school teachers and students in mind. Taking into account the lack of availability of time and classes that many teachers have nowadays in Brazil, often caused by having to comply with very extensive subject curricula, performing various extra-class tasks and the tribulations inherent in the daily lives of all As workers in this country, we seek to develop basic topics and essentials about modern cosmology for your learning and consultation. We believe that, using it, you will be very well prepared to bring this very important theme of nature to the classroom. A differential of this didactic material is that each chapter has practical activities that allow us to fix the concepts of cosmology.

Estas notas de cosmologia moderna foi feito pensando  nos professores e estudantes do ensino medio. Levando em conta a falta de disponibilidade de tempo e aulas que muitos professores têm hoje em dia no Brasil, muitas vezes causados por terem que cumprir currículos de disciplinas muito extensos, desempenhar várias tarefas extra classe e às atribulações inerentes do cotidiano da vida de todos os trabalhadores desse país, procuramos desenvolver tópicos básicos e essências sobre cosmologia moderna para o seu aprendizado e consulta. Acreditamos que, fazendo uso dele, você estará muito bem preparado para levar esse tema tão importante da natureza para a sala aula. Um diferencial deste material didático é que cada capitulo tem atividades praticas que permitem fixar os conceitos da cosmologia.

**Palavras-chave:** modern cosmology, high school

# COSMOLOGIA PARA O ENSINO MÉDIO

QUESTIONANDO E CONHECENDO UM POUCO MAIS SOBRE O UNIVERSO

Uma proposta para ambientes formais e não formais de ensino de ciências

2021

FILIPE ALVES PEREIRA BENTO
JOSÉ ALBERTO CASTO NOGALES VERA

# Introdução

Caro professor (a),

Este livro foi feito pensando em você e nos estudantes. Levando em conta a falta de disponibilidade de tempo e aulas que muitos professores têm hoje em dia, muitas vezes causados por terem que cumprir currículos de disciplinas muito extensos, desempenhar várias tarefas extra classe e às atribuições inerentes do cotidiano da vida de todos os trabalhadores desse país, procuramos desenvolver tópicos básicos e essencias sobre cosmologia moderna para o seu aprendizado e consulta. Tal preocupação, porém, não faz desse material incompleto e/ou superficial no que ele se propõe a discutir. Acreditamos que, fazendo uso dele, você estará muito bem preparado para levar esse tema tão importante da natureza para a sala aula.

Começamos o livro com uma discussão sobre ciência e método científico. O objetivo aqui é introduzir o assunto, apresentando uma visão que consideramos ser adequada e importante sobre a natureza desse tipo de conhecimento. Depois, preparamos um capítulo com alguns resumos sobre a visão da cosmologia do nosso Universo e alguns desenvolvimentos históricos do entendimento humano por esse assunto e alguns outros da ciência moderna. O terceiro capítulo é dedicado a falar sobre as cosmologias históricas e se cosntitui uma ótima oportunidade para trabalhar junto aos professores de História, Geografia e demais disciplinas relacionadas. Os capítulos 4 e 5 tratam, respectivamente, das teorias de Einstein e da Mecânica Quântica, assuntos que consideramos essenciais para o estudo de cosmologia. Enfim, o primeiro tópico específico dessa ciência começa no capítulo 5, e tratará sobre o seu nascimento assim como a descoberta da expansão do Universo. Uma atenção especial é necessária para esse capítulo, pois ele está escrito na forma de um artigo acadêmico. Dessa forma, caso vá utilizá-lo diretamente com seus alunos, você deve auxiliá-los durante o processo. Aliás, será muito interessante esse contato entre estudantes de ensino médio e a escrita acadêmica. Essa foi a intenção de escrevê-lo nesses moldes. O restante dos outros capítulos trata sobre Teoria do Big Bang, buracos negros e matéria escura.

Também desenvolvemos uma série de atividades práticas, relacionadas aos seus respectivos capítulos, as quais ser encontradas no final do livro. Além de perguntas, temos sugestões de experimentos que podem ser realizados em sala de aula com equipamentos bem simples. Inclusive, existem atividades baseadas em debates, palestras e observação do céu. Dessa forma, elas tem a intenção de enriqucer e abrir oportunidades novas para a sua aula.

<div align="center">

**Capítulo 1**

# Uma conversa sobre ciência

*Objetivos*

</div>

*Neste capítulo apresentaremos alguns conceitos essenciais à prática de ciência. Os conceitos de observação, fatos, hipóteses, experimento e teorias estão implícitos no texto. É esperado que os alunos entendam a ciência como uma forma de investigação criteriosa da natureza, baseada na observação de fatos, elaboração e argumentação de hipóteses e validação das mesmas, através de experimentos. A capacidade de argumentação e pensamento crítico também são destacadas como essenciais.*

**Introdução**

A ciência é algo fundamental para nós e a maneira como vivemos. Pode-se dizer que chega a ser instintivo do ser humano querer investigar e conhecer o ambiente ao seu redor. Observe um bebê e veja como ele se comporta com o que está à sua volta. Ele toca em tudo, quer pegar, presta atenção nos movimentos, examina com o tato, visão e, se não ficarmos atentos, até com o paladar! O momento em que ele descobre uma parte do seu próprio corpo, como, por exemplo, a mão, e passa horas examinando-a, apalpando e olhando fixamente para ela, num estado de profunda contemplação com a recente descoberta! A primeira mordida naquela "coisa" de cor verde, super chamativa e a revelação gosto igualmente azedo (limão)! Experiências, agradáveis ou não, resultantes da investigação da natureza que os cerca e que vão compondo o seu repertório de conhecimentos do mundo. Quando vão ficando mais velhos e suas faculdades mentais vão se desenvolvendo, eles expandem e aprimoram suas investigações e experimentações. Através da fala, conseguem não só investigar o mundo físico próximo a eles, mas também o mundo das ideias. "Por que o dia é claro e a noite é escura?", perguntam, "Porque tem o sol brilhando", responde a mãe. "E por que o sol brilha?", insistem, "Ah, porque sim!", retruca, já sem muita paciência, inibindo toda a curiosidade da criança, a qual deveria ser estimulada através de mais perguntas e questionamentos. Nascemos com um grande instinto de curiosidade, mas, enquanto vamos crescendo, muitas vezes somos estimulados a não praticar e a inibi-la com respostas definitivas e curtas. Se torna "um incômodo" desnecessário perguntar as coisas. E é justamente a curiosidade um dos elementos mais importantes para se fazer ciência. Questionar o mundo que nos cerca, perguntar como as coisas funcionam, por que são da maneira que são, como estão organizadas, se existem padrões e muitas outras perguntas, é uma qualidade muito importante para o cientista, pois é a partir desse tipo questionamento que somos despertados a ir em busca do desconhecido a fim de entendê-lo. E quanto mais refletimos, investigamos e conhecemos, mais saciados ficamos desse instinto que nos acompanha desde bebês, e mais lúcidos e responsáveis nos tornamos em relação ao mundo e seus elementos.

Dessa forma, dizemos que ciência é uma maneira de conhecermos a natureza e tentar explicá-la. Para fazer isso, devemos estar atentos ao método científico. Ele trata de uma série de características comuns presentes nas pesquisa e trabalhos dos cientistas. É importante salientar que a descrição adequada do método científico não o descreve tal como "uma receita de bolo" o faria. Em vez disso, iremos nos ater nessas características que são essenciais à sua prática:

- **Reconhecer e apontar alguma questão, problema, fato ou enigma que ainda não foi explicado completamente e/ou sem solução** – essa primeira etapa é muito dependente do sobre o que refletimos no primeiro parágrafo – a curiosidade. Pessoas curiosas, que fazem perguntas o tempo todo, críticas, que se indagam sobre o mundo que as cerca e não se contentam com respostas rasas, sempre estão atentas a essas questões sem resposta, e, dessa forma, anseiam por conhecer e aprender mais. Lembre-se, essa é uma das características mais belas e importantes não só dos cientistas, mas de qualquer pessoa.

- **Propor uma solução/explicação para o tal problema/questão** – em uma linguagem mais direta, seria o equivalente a propor um palpite bem fundamentado para explicar o fato que você observou. Em outra linguagem mais formal, seria a proposta de hipótese, que nada mais é que um raciocínio bem estruturado sobre a solução do seu problema ou a explicação da sua pergunta. Como será discutido nesse mesmo capítulo mais adiante, o raciocínio lógico e capacidade argumentativa ajudam muito nessa etapa.
- **Baseado nas suas hipóteses, propor possíveis consequências que elas irão gerar** – quando você elaborou a sua hipótese, também é necessário prever as consequências que ela irá causar. Essa parte é muito importante, pois definirá se ela é ou não uma hipótese científica e de que maneira será testada para verificarmos a sua veracidade. Hipóteses científicas são aquelas que são passíveis de serem testadas e avaliadas objetivamente. Quando eu afirmo "O espaço é permeado por uma substância indetectável.", não estou fazendo uma afirmação científica pois o conteúdo que apresento nela não pode ser verificável por nenhum teste. Já quando, por exemplo, Albert Einstein propôs a sua Teoria Geral da Relatividade, e afirmou que uma das suas previsões era que a matéria podia curvar a luz, ele estava realmente fazendo uma afirmação científica, pois ela era suscetível a ser testada e ser comprovada como verdadeira ou falsa (veremos essa história detalhadamente mais à frente). Portanto, não se esqueça que ao explicar as consequências das hipóteses que você propuser para uma questão científica, faça afirmações que poderão ser testadas quanto à sua veracidade.

> **"Ciência é muito mais uma maneira de pensar do que um corpo de conhecimento"**
> Carl Sagan, " Por que precisamos entender ciência? " (1990).

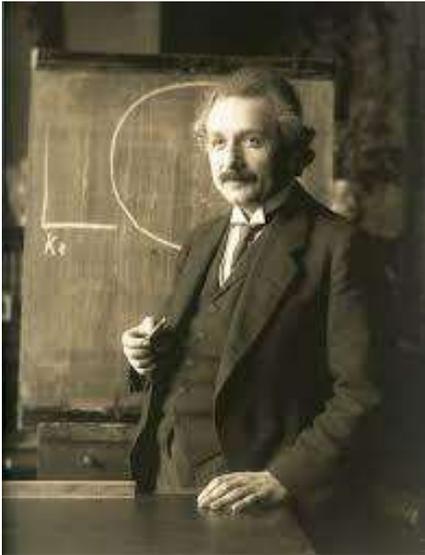

| Albert Einstein (1879 – 1955) – famoso físico alemão responsável por revolucionar a física e a maneira como enxergamos o universo. Sua Teoria Geral da Relatividade redefiniu o que se entendia na época por gravidade, espaço e energia. Ganhou um prêmio Nobel de Física em 1921 pela descoberta do efeito fotoelétrico. Dava muito valor à imaginação e esforço, motivos pelos quais afirmava serem os responsáveis pelo seu sucesso.

*"A ciência, como um todo, não é nada mais do que um refinamento do pensar diário"*
*Albert Eisntein* |

Fonte: Pixabay

- **Realização dos testes/cálculos/experimentos sobre as previsões** - nessa etapa, os testes devem ser realizados com todo cuidado e atenção possíveis. Todas as variáveis ou fatores envolvidos no processo devem ser conhecidos e identificados objetivamente. Não se pode permitir que ele seja realizado de uma forma descuidada e, muito menos, tendenciosa, ou seja, quem realiza o teste manipula o processo e resultados para encontrar um "resultado compatível" às suas hipóteses. Caso todos esses fatores forem respeitados e o resultado tenha sido em concordância com as previsões, as hipóteses foram validadas e serão tratadas como verdadeiras. Aqui, porém, existem duas ressalvas. A primeira é no caso de o teste ter demonstrado a falsidade das previsões. O caminho a ser adotado é ir reexaminá-las e propor outras observações, previsões e/ou hipóteses e repetir o processo. O outro caso diz respeito quando o teste validou as previsões. Isso não significa um resultado definitivo! Em ciência é necessário a prática constante da humildade e cautela, os quais dizem respeito à "atitude científica" que discutiremos mais adiante. Outros irão repetir e investigar seus testes, estudar com cuidado suas previsões e até mesmo propor novas maneiras de verificá-las. Se ocorreram erros durante o processo, eles certamente serão apontados e é necessária humildade para reconhecê-los independentemente do prestígio e fama do cientista que cometeu a falha. Em ciência, pouco importa a autoridade e interesses pessoais, mas sim o compromisso incondicional com a verdade e integridade. Já dizia outro físico muito importante (e igualmente bem-humorado) Richard Feynman "O teste de todo conhecimento é o experimento. O experimento é o único juiz da verdade científica"[1].

---

[1] Duas observações: estamos nos referindo ao conhecimento científico, existem outras formas de conhecimento, como o artístico e o religioso, que não podem ser interpretados dessa maneira e serão brevemente discutidos nesse capítulo; mesmo dentro das ciências existem variações em relação aos tipos de experimentos como, por exemplo, as diferenças entre os experimentos realizados em sociologia e química. Devido à natureza diversa dessas ciências e seus objetos de estudo, suas experimentações também são diferentes, com metodologias específicas a cada uma (consultar o livro "Pesquisa em ciências humanas e sociais" de Antonio Chizzoti).

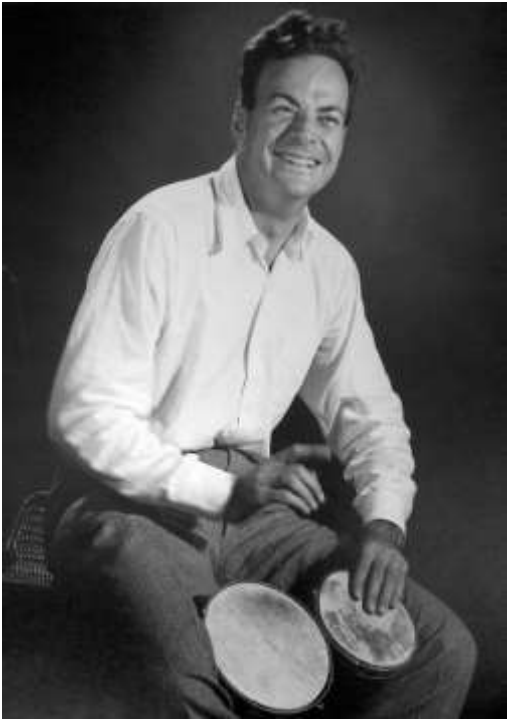

Richard P. Feynman (1918-1988) – físico norte americano que foi um dos pioneiros da eletrodinâmica quântica, a qual estuda os movimentos dos elétrons quando reagem entre si. Ganhou o prêmio Nobel de Física em 1965 pelos seus trabalhos. Era uma pessoa não convencional, não gostava de autoridade e nem burocracias. Dava muito valor a criatividade e curiosidade na formação do cientista. Tocava o bongo e falava português, pois havia passado uma temporada no Brasil.

*"A ciência é aquilo que aprendemos sobre como não deixar enganar a nós mesmos"*
*Richard Feynman*

Fonte: HypeScience

- **Formulação da lei/teoria final** - aqui serão reunidos os três elementos principais discutidos anteriormente: hipóteses, previsões e resultados experimentais. Normalmente, a elaboração da teoria final e, principalmente, a sua aceitação na comunidade acadêmica, demoram, pois, os processos descritos nos tópicos anteriores (testes de validação) podem levar anos para serem realizados e/ou reavaliados. Em ciência, é muito comum que as hipóteses sejam testadas exaustivamente, por diferentes grupos e repetidas vezes. Assim, se não negadas, elas se consolidam e ganham credibilidade perante à comunidade científica. Einstein, que publicou sua teoria em 1915, só veio a ter a sua segunda[2] constatação experimental em 1919, através de um eclipse do sol, uma história que envolveu até o Brasil (será tema de um capítulo posterior)! A terceira constatação experimental veio em 1925, através da observação do desvio gravitacional da luz provinda de algumas estrelas. Assim mesmo, esse terceiro resultado foi contestado, devido à alegação de interferências na medição, e só foi realizado satisfatoriamente para a comunidade científica em 1954 pelas medições do astrofísico Daniel Magnes Popper[3]. Já consolidada como fato desde o princípio da década de 20, a Teoria Geral da Relatividade ainda foi ter outra constatação experimental, aproximadamente, cem anos depois, com a detecção das ondas gravitacionais pelo projeto LIGO.

---

[2] A primeira foi a precessão na órbita de mercúrio, um problema já conhecido há tempos, mas que nenhuma teoria podia explicar. A Teoria Geral da Relatividade o solucionava.

[3] Treschman, Keith. (2014). Early Astronomical Tests of General Relativity: the anomalous advance in the perihelion of Mercury and gravitational redshift. Asian Journal of Physics. 23. 171-188.

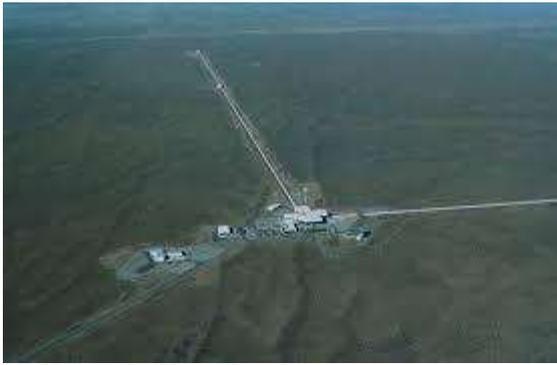
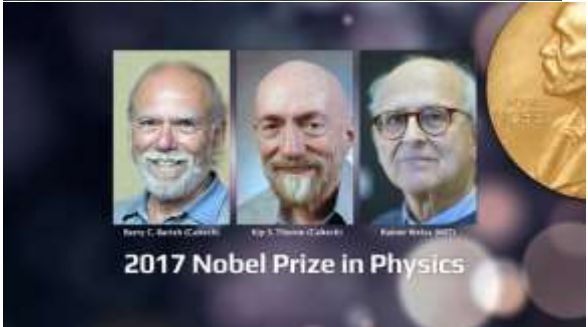

O Projeto LIGO - Laser Interferometer Gravitationalc Wave Observatory (Observatório de Ondas Gravitacionais por Interferômetro Laser) foi fundado em 1992 e começou a operar em 2002. Os seus fundadores foram os físicos Barry Barish, Kip Thorne e Reiner Weiss. As ondas gravitacionais foram previstas pela Teoria Geral da Relatividade de 1915. No dia 14 de setembro de 2015, ás 9h51, um sinal de onda gravitacional foi detectado como sendo proveniente da colisão entre 2 buracos negros com massa equivalente a 30 massas solares. Os três cientistas ganharam o Nobel de Física em 2017.

**Fonte: www.skyandtelescope.org**

Lembrem-se que no mundo científico essas características descritas não devem ser interpretadas como uma "receita de bolo". Elas são qualidades comuns e desejáveis nas atividades científicas. Foram muitas as pesquisas e avanços científicos que não possuíram alguma dessas etapas, foram realizados por tentativa e erro, experimentação na ausência de hipóteses ou descobertas acidentais e, no entanto, foram bem-sucedidas devido à atitude científica das pessoas que estavam envolvidas.

Em uma noite de setembro, em 1928, um médico e cientista, chegando ao seu laboratório, resolveu que era hora de se livrar de alguns itens sujos, contaminados e sem uso que estavam sobre ela. Vinha há tempos trabalhando nesse lugar, localizado no Hospital St. Mary, em Londres, pesquisando substâncias capazes de combater as infecções causadas por microrganismos em ferimentos. Ele notava a grande mortalidade associada a essas infecções e seus trabalhos estavam voltados para tal causa. Em 1921, já havia descoberto as propriedades antimicrobianas da lisozima, uma substância que ele primeiro observou ao analisar colônias de bactérias semeadas em placas a partir de secreções nasais de pacientes resfriados. Depois de notar que o muco destruía a bactéria ao seu redor, ele o investigou à procura da substância que possuía o efeito desejado, encontrando o que viria a ser batizado depois como lisozima, a responsável por tal ação. Fez descobertas posteriores encontrando-a também nas lágrimas humanas e de alguns animais, como defesa natural contra microorganismos (hoje em dia, ela é, inclusive, sintetizada e adicionada no leite para a alimentação de crianças em zonas de vulnerabilidade econômica e social). Avançando para o então ano de 1928, em uma noite de setembro, ao pegar uma das placas semeadas com colônias de bactérias *Staphylococcus aureus*[4] para descartar, notou que elas haviam sido contaminadas por um bolor[5]. Ele havia passado os últimos dias do feriado com a família, deixado as placas amontoadas em um canto do seu laboratório. Decidiu examiná-las com mais cuidado e viu que, ao redor do bolor, a colônia de bactéria não se desenvolvia. O cientista não ignorou esse FATO. Mostrou ao seu assistente, Merlym Price, o qual o lembrou "Foi como

---

[4] Bactérias que podem causar infecções brandas e até mesmo mais graves como pneumonias e meningite.

[5] Mofo; nome dado à infecção de alguma superfície por fungos que não formam a estrutura de semelhante a cogumelos.

o senhor descobriu a lisozima". Ele resolveu isolar aquele fungo e o identificou como do gênero *Penicillium*. Chamou inicialmente a secreção que provinha dele de "suco de mofo" e começou a testá-lo em colônias de bactérias. Constatou o caráter inibitório desse "suco" contra o desenvolvimento das colônias de bactérias. Ali estava descoberta, pelas mãos de Alexander Fleming, a penicilina – o antibiótico que revolucionou não só o mundo científico como também toda a sociedade, naquela época. De início, a comunidade científica não deu atenção à descoberta de Fleming e foi só dez anos depois, por motivos relacionados à II Guerra Mundial, que ela começou a ser desenvolvida em parceria com um laboratório e mais pesquisadores, se tornando medicamento apto para o uso em pessoas. A penicilina foi amplamente utilizada na II Guerra Mundial e salvou muitas vidas. Hoje é largamente utilizada em todo o mundo. Fleming recebeu o prêmio Nobel de Medicina em 1945, juntamente com Howard Florey e Ernst Boris Chain os quais foram os responsáveis pela produção e elaboração do medicamento final à base da penicilina.

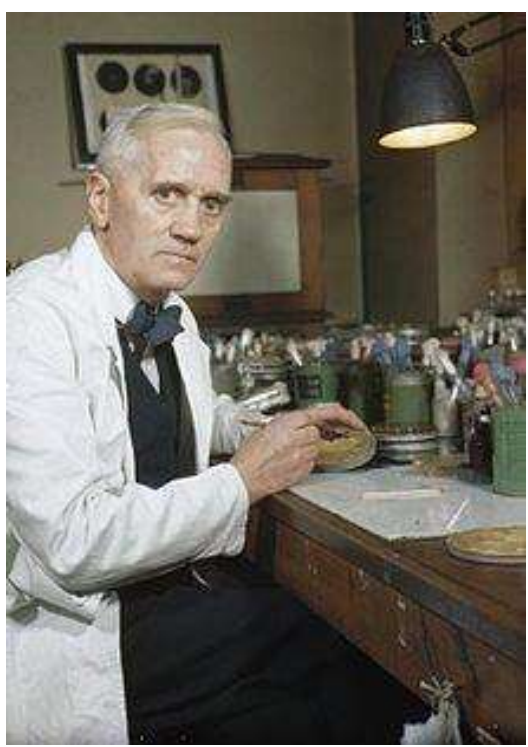

Fonte: WikiWand

Alexander Fleming (1881 - 1955) – Nasceu em Darvel, Escócia, filho de fazendeiros. Estudou medicina em Londres, na Escola de Medicina do Hospital St. Mary. Serviu o exército durante I Guerra Mundial no posto de capitão. As experiências na guerra foram decisivas para a sua dedicação nos estudos contra microrganismos que causavam infecções em ferimentos.

*"Às vezes, encontramos o que não procuramos. Quando acordei logo após o amanhecer em 28 de setembro de 1928, certamente não planejava revolucionar toda a medicina descobrindo o primeiro antibiótico do mundo, ou o assassino de bactérias. Mas suponho que foi exatamente o que fiz".*

**Como age um cientista?**

Algumas das palavras que regem a conduta de um bom cientista (ou, pelo menos, deveriam) são: senso crítico, integridade e humildade. Quando temos uma hipótese, princípio ou lei que afirmam algo, mas, em algum momento, outro pesquisador mostra evidências corretas, bem testadas e estudadas que as contrariam, é com humildade que devemos aceitá-las e abandonar as questões anteriores (lembrando, estamos assumindo que as novas evidências já foram testadas e se mostraram corretas). Não importa a posição social, renome, fama ou autoridade do cientista – se as suas hipóteses ou princípios forem invalidadas corretamente por apenas uma evidência negativa que seja, elas devem ser abandonadas ou ampliadas e reestruturadas. Não importa também quantas pessoas defendem tal ideia ou por quanto tempo ela existe - cinquenta anos, centenária ou milenar - pois uma simples prova contrária (testada e validada pelo método científico) é o suficiente para colocá-la em xeque.

Outra atitude que vai contra o bom comportamento dos cientistas é deixar que suas vontades, crenças pessoais ou interesses interfiram em seus resultados, com a finalidade de satisfazer seus desejos pessoais. Quando suas descobertas diferem daquilo que esperavam ou que gostariam que fossem verdade, o cientista precisa aceitá-las incondicionalmente. Essa integridade tão necessária à prática da ciência está em contato direto com a capacidade de refletirmos sobre as ideias, crenças, regras e, até mesmo, preconceitos que carregamos conosco. Muitas vezes, elas estão enraizadas há tanto tempo em nós, que não refletimos se realmente fazem algum sentido, se são válidas e, no caso dos preconceitos, o quanto são prejudiciais para vida em sociedade. Assim, mesmo quando se mostram sem razões suficientes para que ainda as adotemos, continuamos levando-as conosco e deixando que influenciem o nosso cotidiano. No caso de cientistas, torna-se um problema ainda maior pois isso os conduz, dentre muitos outros comportamentos ruins à ciência, a adotar uma postura de valorizar e dar atenção somente às ideias e hipóteses que estejam de acordo com as suas próprias. Em alguns casos, fingem considerar todas as hipóteses, positivas e negativas, mas, na verdade, analisam com mais entusiasmo o que lhes convém, e são propositalmente displicentes ao que os contraria, criando uma falsa sensação de imparcialidade. Portanto, os conceitos de honestidade intelectual e integridade são componentes diários na vida de um cientista e devem sempre serem muito respeitados em quaisquer de suas práticas.

*"Não importa o quão ultrajante uma mentira possa ser, ela será aceita se declarada em voz alta o suficiente e com frequência suficiente."*

Isaac Asimov – The Near East (1968), p. 31.

---

### Pensamento crítico e argumentação coerente

Quando pensamos com criticidade, estamos aptos a discernir entre a veracidade ou a falsidade de um argumento com mais chances de acertar, do que quando nos deixamos levar por pensamentos rasos e curtos. Somos o tempo todo abordados com argumentos que tentam nos persuadir a fazer algo. Igualmente, estamos o tempo todo tentando convencer e apresentar nossas ideias ao mundo e pessoas. Num debate científico, é mais que necessário fazer uso de bons argumentos, sem afirmações vagas, imprecisas e/ou morais. Quando falamos sobre moral, estamos nos referindo a um conjunto de comportamentos e posturas adotadas por um certo povo ou religião em algum momento da história. Portanto, não é correto basear um argumento em preceitos referentes a determinada moral pessoal pois ele só fará sentido para aqueles que compartilham das mesmas crenças. Um bom argumento não depende de pontos de vista ou opiniões pessoais. Eles se baseiam na discussão objetiva e clara dos fatos. "Eu acho..." "Na minha opinião..." enfraquecem o argumento em um debate, pois dão a entender que aquela ideia não é uma constatação baseada em fatos objetivos, mas algo de cunho subjetivo, referente ao autor da frase. Não podemos, no entanto, confundir isso com "não ter direito a opiniões". Não é isso. Todos nós somos livres a pensar o que bem entendemos e emitir nossas opiniões. Sobre tudo podemos e devemos discutir, conversar e emitir pontos de vista. Mas devemos sempre ter em mente "Além do meu direito, tenho razões suficientes para pensar dessa maneira?". Isso é ser crítico e coerente

Investigar sobre o mundo e seus elementos requer também a habilidade de questionar, argumentar com coerência e saber o momento certo em que velhas ideias, conceitos ou convicções devem ser abandonados. Os bons cientistas têm uma capacidade muito construtiva de mudar seus pontos de vista em relação a determinado assunto, perante evidências experimentais sólidas[6]. Para isso, é necessário saber interpretar e refletir sobre essas informações de uma forma crítica e responsável. Dessa maneira, a racionalidade é uma ferramenta presente na ciência como maneira de raciocínio e abordagem de ideias. Saber questionar racionalmente e ser sensato fazem parte do senso crítico do cientista. Muitas pessoas têm a visão de que a ciência com suas leis, teorias e modelos são imutáveis. Muitos acreditam que mudar de ideia ou convicção sobre fatos é uma fraqueza humana e sinônimo de algo negativo. Todas essas abordagens estão erradas e não agregam nada de construtivo para o processo científico. Podemos dizer, na verdade, o contrário! O fato das teorias serem algo dinâmico é uma qualidade que as torna mais fortes. O seu refinamento e aperfeiçoamento trazem confiabilidade à maneira que tentam descrever a realidade. Muito da competência do cientista se deve mais ao fato de saber aperfeiçoar progressivamente as suas ideias, quando as evidências assim o mostram, do que propriamente defendê-las.

**Arte, religião e ciência**

Embora muitas vezes se misturem, cada uma possui seu domínio próprio de atuação. A ciência busca conhecer e explicar os fenômenos naturais desde a mais simples reação química no menor ser vivo existente, até às leis que regem a estrutura em larga escala do universo. Nesse caminho ela passa pelo estudo de todos os seres vivos, suas ecologias e comportamentos, o domínio dos números e suas operações, os ambientes do nosso planeta com as diversas geografias, o bem-estar de animais e seres humanos, sua saúde física e mental e muitas outras subáreas. Para isso, como bem discutido nesse capítulo, ela lança mão da investigação científica para tentar encontrar padrões, formular teorias e leis que descrevam essas realidades da maneira mais correta e precisa possível.

O conhecimento artístico provém da interpretação pessoal e criatividade do artista perante o mundo e as emoções que nele mesmo se despertam ao fazer isso. Alguns artistas, como Picasso, chegavam a comparar sua arte como uma maneira científica de enxergar a natureza. Ele costumava dizer, ao ser questionado sobre os traços não-convencionais que utilizava, ao invés das recomendações dos seus professores, os quais o mandavam desenhar utilizando os conceitos das escolas clássicas, "Eu pinto o que eu vejo", para se referir à forma criteriosa de sua observação e investigação das figuras. Dessa forma, podemos, em muitos casos, considerar a arte como concepções estéticas científicas da natureza. Um campo riquíssimo de conhecimentos gerados e técnicas empregadas por milênios, desde que os homens primitivos desenhavam em paredes e árvores e construíam pequenos utensílios, estátuas até, por exemplo, os *raps* atuais que nos fazem refletir e conhecer mais sobre nossa realidade social, política e econômica. Alguns autores desenvolvem estudos comparativos entre a arte e a ciência, como o caso de René Vidal. Em seu artigo " The Art and Science of Problem Solving"[7] (A Arte e a Ciência da resolução de problemas) ele traça as principais diferenças e similaridades entre as duas. Algumas das principais semelhanças são que, tanto artistas como cientistas, fazem uso da criatividade, experimentação, observação e abstração (refletir sobre as características e essência da existência de um objeto, ideia ou conceito) no desenvolvimento dos seus trabalhos. A prática da arte é importante para o desenvolvimento cultural e emocional das pessoas, seja praticando-a ativamente por meio das pinturas, escrevendo poemas, músicas, dançando e muitas outras modalidades ou, simplesmente, a apreciando. A arte também é uma expressão do conhecimento humano!

---

[6] Coerentes; plausíveis; obtidas corretamente.

[7] Vidal, René. Victor Valqui. The Art and Science of Problem Solving. **Investigação Operacional, 25** (2005) 157-178.

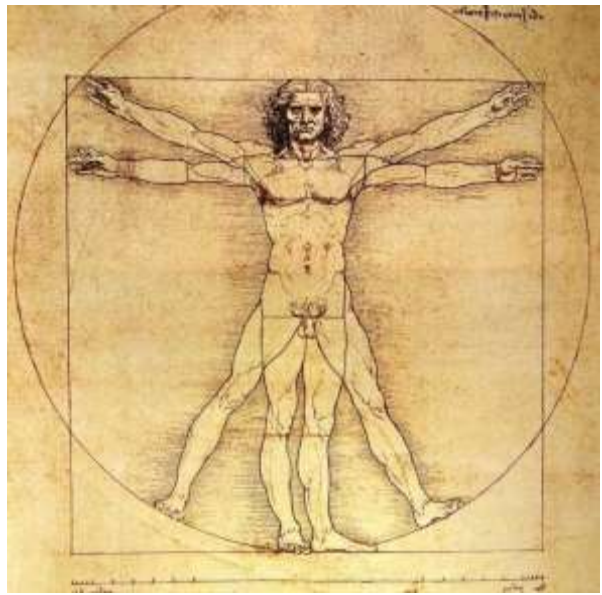

*"Homem Vitruviano"* – Leonardo da Vinci (1490) – Arte? Ciência? Ou os dois? O *Homem Vitruviano*, é o desenho feito por Leonardo da Vinci representando as proporções perfeitas do corpo humano respeitando as suas simetrias. Leonardo investigou e estudou incansavelmente a anatomia humana, dissecando corpos e membros, para que suas obras fossem as mais reais possíveis.
*"A arte diz o indizível; exprime o inexprimível, traduz o intraduzível."*
Leonardo da Vinci

A religião é outra manifestação da existência humana. Ela se ocupa em buscar, filosoficamente, propósitos, significados, sentidos e condutas à existência do homem e de todo universo. Dessa forma, em quase todas as religiões existe o conceito de adoração, fé, ser (ou seres) supremo (s) e a estabilização de uma comunidade em volta de suas crenças e costumes. Existem conteúdos fascinantes nas mais variadas religiões e crenças do mundo, de diferentes épocas e contextos. Textos com muitas informações sobre a cultura, moralidade, rituais e que ainda hoje são importantes para o dia a dia das sociedades. Alguns problemas e atritos entre ciência e religião acontecem quando religiosos e cientistas tentam intrometer um nas atividades do outro, não respeitando as suas naturezas completamente diferentes. Existem religiosos que aceitam perfeitamente as descobertas científicas assim como cientistas que praticam a sua religião sem problema algum. O grande segredo é o respeito, bom senso e saber diferenciar os propósitos de cada um.

**Para conhecer mais:**

**Livros**

O mundo Assombrado pelos Demônios – Carl Sagan.

Só pode ser brincadeira, Sr. Feynman – Richard Feynman.

**Série de TV:**

Genius 2 - Pablo Picasso , do History Channel.

<div style="text-align:center"><span style="color:blue">**Capítulo 2**</span></div>

# Cosmologia histórica – um olhar sobre as cosmovisões do Universo pelas culturas passadas

*Objetivos*

*Este capítulo tem como objetivo definir e exemplificar aos alunos o que é cosmologia e como ela surgiu nas culturas dos povos antigos ao redor do mundo, onde, na maioria das vezes, se confundia com religião, e proporcionava uma visão surpreendente do Universo. A apresentação da cosmologia cultural também tem por finalidade enriquecer o conhecimento e estimular a valorização dos costumes e culturas dos povos primitivos, contribuindo para que os mesmos sejam respeitados e tenham sua importância reconhecida em seu contexto histórico na humanidade.*

*"O Cosmos é tudo o que existiu, existe ou existirá"*

*Carl Sagan*

**O que é cosmologia?**

Quando pesquisamos no dicionário o significado da palavra "cosmologia" encontramos várias definições, as quais podem ser resumidas como sendo um ramo da astronomia que tem por objetivo investigar a estrutura, origem e evolução do universo em larga escala. Por larga escala, entendemos que os cosmólogos[8] trabalham, principalmente, com estruturas de tamanhos iguais ou maiores que galáxias[9]. Para exemplificar, algumas das questões importantes da cosmologia atual são a matéria/energia escura[10] e o Big Bang[11]. Contudo, não queremos dizer que os cosmólogos só pesquisam e estudam sobre coisas incrivelmente grandes e superenergéticas. Não! Bons cosmólogos, sejam profissionais ou não, são grandes curiosos e investigadores do universo e, dessa maneira, precisam de um vasto conhecimento das várias ciências, como, principalmente, a física e matemática. Assim, podemos dizer que a cosmologia é uma ciência que estuda o universo do ponto de vista da sua origem, estado atual e possíveis destinos, fazendo uso de outras ciências, como, por exemplo, a física, matemática, computação e química. Porém, podemos nos perguntar: "Então, a cosmologia surgiu só quando já havia uma física bem desenvolvida e computadores potentes?". De jeito nenhum! Essa é uma visão ingênua sobre o processo de desenvolvimento da cosmologia e das ciências em geral! Quando estudamos a história da humanidade e dos diversos povos, culturas e religiões que surgiram desde o aparecimento dos primeiros homens, uma das características comuns a todos eles é a busca do entendimento de onde e como surgiram o mundo e seus habitantes. Nesse ponto, chegamos a um riquíssimo e belo conjunto de histórias e mitos, dos quais muitos se tornaram religiões importantes na Antiguidade, e que tiveram a finalidade de explicar a origem e desenvolvimento do que viria a ser conhecido como Cosmos[12]. Embora utilizem padrões diferentes da atual cosmologia científica, essas cosmologias (as quais daremos o nome de cosmologias históricas) são uma fonte riquíssima de conhecimentos sobre como as civilizações viviam, como interpretavam o mundo e, propriamente, como a cosmologia nasceu e foi evoluindo até alcançar o estado em que está hoje..

---

[8] Quem estuda cosmologia

[9] Sistema formado por várias estrelas, gás, poeira e matéria escura. Galáxias podem ser desde anãs (aproximadamente alguns milhões de estrelas) até gigantes com centenas de trilhões de estrelas. Nossa galáxia, a Via Láctea, é uma galáxia intermediária com entre 100 a 400 bilhões de estrelas.

[10] Tipo de matéria que não tem interação com a matéria normal e nem com ela mesma! Veremos mais adiante do que se trata.

[11] Principal teoria científica que melhor explica a origem do universo.

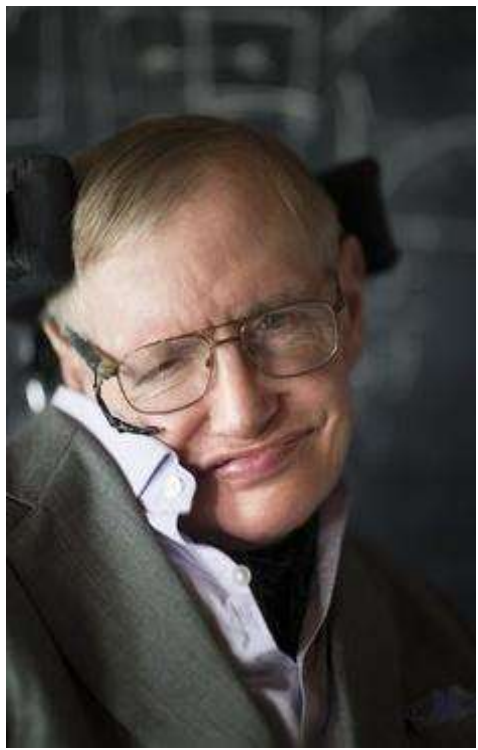

Stephen William Hawking (1942 – 2018) – Foi um cosmólogo dos nossos tempos! Também era físico e se dedicou a estudar, principalmente, buracos negros e a formulação de uma teoria unificada da física. Escreveu inúmeros livros de divulgação científica, os quais foram responsáveis por levar ciência de uma maneira clara e divertida para diversas pessoas pelo mundo. Aos 21 anos, foi diagnosticado com uma doença degenerativa chamada Doença neural motora que comprometia toda a musculatura do seu corpo. Viveu até os 76 anos de idade.

"*Não importa quanto a vida possa ser ruim, sempre existe algo que você pode fazer, e triunfar. Enquanto há vida, há esperança.*"

Fonte: www.hawking.uk.com

**Cosmologia histórica – a interpretação do Cosmos pelos povos antigos**

Como dito anteriormente, quando voltamos nossos olhares para o passado e decidimos investigar a história da humanidade, a constante busca por explicações e compreensão do mundo, de como ele foi criado, o que é o céu (muitos povos antigos nem mesmo possuíam esse conceito, como veremos adiante), Sol, Lua e estrelas e uma série de outros questionamentos dessa natureza, percebemos que elas são uma característica comum a esses povos, independente da sua cultura e localização geográfica. Cosmologia histórica pode ser entendida como o conjunto de histórias e mitos provindos das primeiras civilizações e grupos sociais da história humana que buscavam explicar o mundo. As cosmologias antigas, se não eram as próprias religiões em si, foram as bases para a sua formação. Dessa maneira, ao interpretar suas histórias, encontramos alguns elementos comuns, como deuses ou entidades poderosas, os quais normalmente são o início do mundo, ou o fazem a partir de elementos geográficos também comuns, como rios, montanhas, vulcões e outros, variando de acordo com as localizações geográficas de cada cultura. As presenças de animais nessas cosmologias também são abundantes e variam de acordo com a fauna natural de cada região. Por exemplo, para os nativos de nossas terras, o *Dasypodidae* (tatu) aparece fornecendo o acesso a um novo mundo. Já nos desenhos que retratam a história da criação contada pelos egípcios, os falcões, animais comuns naquela região, são os grandes articuladores do processo.

---

[12] Palavra de origem grega que era usada para se referir à "ordem", "regularidade" e "beleza". A palavra "cosmético" tem a mesma origem.

Por fim, estudar não só a cosmologia de civilizações passadas, mas a sua cultura em geral, é um exercício de interpretação que exige nossa imersão nos seus mundos e contextos. Por mais diferente que as histórias possam nos parecer e, até mesmo, estranhas, é necessário sempre lembrar de levar em conta todo o contexto e época em que elas foram criadas; a maneira como seus povos viviam; e quais conhecimentos detinham. Do contrário, corremos o risco de fazer análises superficiais e, muitas vezes, até preconceituosas, sem mesmo chegar à essência e importância que esses pensamentos tiveram para a história. Nosso conhecimento de hoje não é algo construído a partir do nada, mas uma continuação da busca desde que os primeiros homens tiveram curiosidade sobre o universo, através de um simples olhar para o céu, numa noite estrelada, em uma planície africana qualquer, há milhares de anos atrás...

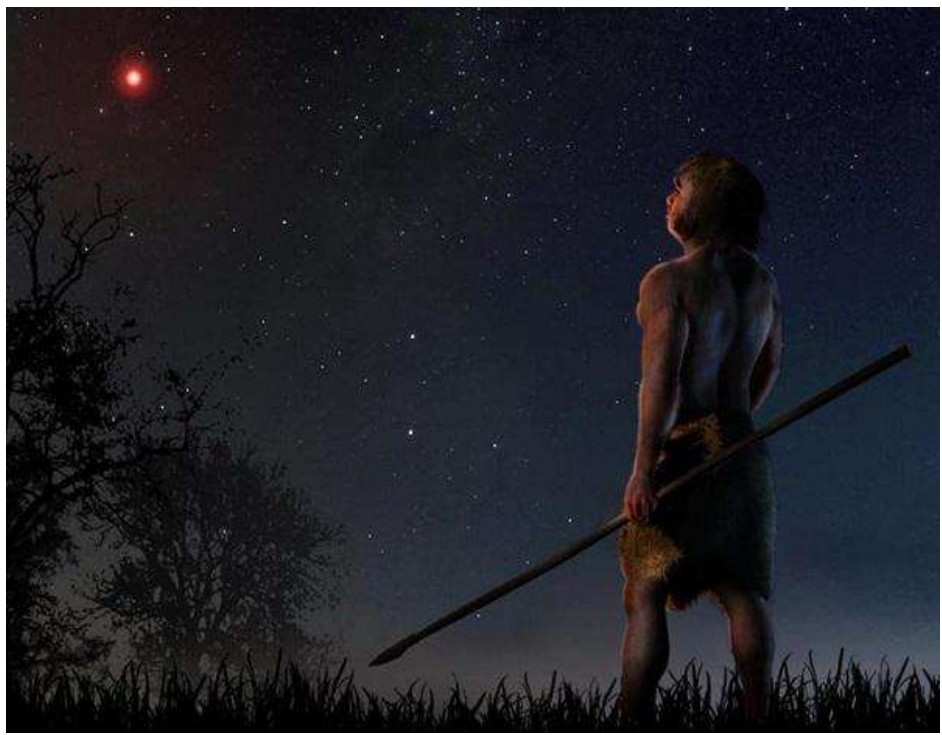

Ilustração representando um humano primitivo observando o céu e uma estrela em destaque.

Fonte: Astronomy Magazine – astronomy.com

**O novo mundo dos Caiapós através de um buraco de tatu!**

Os caiapós são um grupo étnico de indígenas brasileiros habitantes da região da Amazônia brasileira. Hoje, podem ser encontrados ao norte do estado de Mato Grosso e sul do Pará. Possuem uma cosmologia rica e cheia de elementos da natureza de sua região, a qual desperta o interesse e estudos de pesquisadores de várias áreas[13]. O conto a seguir descreve como é a sua visão do início do mundo. Para eles, o universo é um conjunto de camadas, chamadas *pykas*. Essas camadas são sobrepostas, e representam os vários mundos existentes. Para explicar como o mundo em que habitam foi *descoberto* (e não criado) eles contam essa história:

*No início, a nação indígena dos caiapós habitava uma camada superior onde não havia nem o Sol e a Lua, tampouco rios ou florestas, ou mesmo o azul do céu. Sua alimentação consistia de apenas de alguns animais e mandioca, pois não conheciam peixes, pássaros ou frutas. Certo dia, um índio caçador caiapó estava perseguindo um tatu-canastra e acabou se distanciando muito de sua aldeia. A medida que o índio se afastava da sua tribo, a figura do tatu crescia cada vez*

---

[13] Ver a revista *Scientific American - Etnoastronomia,* edição especial número 14, matéria "A cosmologia dos caiapós" de Márcio D'Olne Campos, p. 62.

*mais, se tornando um animal muito grande. Quando o índio já estava próximo de alcançá-lo, o tatu rapidamente cavou a terra e desapareceu dentro dela.*

Neste ponto do conto, podemos perceber a regionalidade presente na referência ao tatu-canastra, um animal típico da América do Sul, e também a presença de elementos mágicos, como o fato do tatu "crescer cada vez mais".

*Como a cova era imensa, o índio caiapó entrou nela e seguiu o animal. Quando já tinha percorrido grande parte de sua extensão, ficou muito surpreso ao perceber que na extremidade brilhava uma faixa de luz. Então ele e o tatu caíram pelo buraco direto dentro de um vazio, e um forte vento os fez retornar de volta para a borda do buraco. De lá ele observou o que tinha embaixo e, maravilhado, viu que existia um outro mundo, repleto de Buritis,[14] com um céu muito azul, e o sol a iluminar e aquecer as criaturas; na agua, muitos peixes coloridos e tartarugas. Nos lindos campos floridos, destacavam-se as frágeis borboletas; florestas exuberantes abrigavam belíssimos animais e insetos exóticos, contendo ainda diversas árvores carregadas de frutos. Os pássaros embelezavam o espaço com suas lindas plumagens.*

A descrição do "novo mundo" segue as referências aos elementos da sua região.

*Deslumbrado, o índio ficou admirando aquele paraíso, até o cair da noite. Entristecido em ver o pôr do Sol, pensou em retornar, mas já estava escuro. Novamente surge à sua frente outro cenário maravilhoso: uma enorme Lua nasce detrás dos picos, clareando com sua luz de prata toda a natureza. E assim permaneceu, até que a Lua se foi surgindo novamente o Sol. Muito emocionado, o índio voltou à tribo e relatou as maravilhas que viu. O grande ancião caiapó, diante do entusiasmo de seu povo, consentiu que todos descessem um a um, pela cova, através de uma imensa corda de algodão, até aquela nova pyka. Lá seria o magnífico Mundo Novo, onde todos viveriam felizes. Porém, nem todos tiveram coragem de descer e, os pequenos pontos brilhantes que eram vistos no céu noturno desse novo mundo, eram as fogueiras daqueles que ficaram habitando a camada superior e que, por medo, não desceram".*

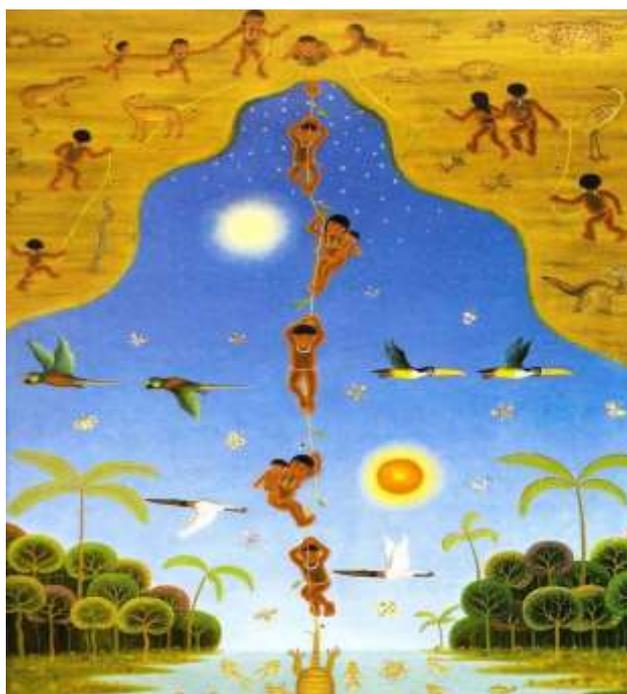

Os caiapós deixando seu mundo antigo e descendo para o maravilhoso e fértil novo mundo.

---

[14] Palmeira comum no Brasil, Venezuela e América Central.

## ... E o mundo surge das águas turbulentas – A cosmologia egípcia

Para os egípcios antigos (aproximadamente entre 2.000 e 3000 a.C) o mundo era dividido em três partes. A Terra, a qual era plana, estava situada entre os outros dois mundos. Também era cortada pelo rio Nilo e cercada por um imenso oceano. Acima da Terra estava o mundo do céu, que era sustentado por 4 pilares, os quais, muitas vezes, foram representados por montanhas. O terceiro e mais escuro mundo era o que eles se referiam como *Duat*, um mundo que existia nas profundezas da Terra. Eles acreditavam que tudo o que não era visível no mundo superior, estaria escondido no Duat. Assim, as estrelas que sumiam ao amanhecer, o Sol poente no final da tarde e, até mesmo, os falecidos habitavam lá. Acreditava-se que o Sol, ao se pôr, percorria um longo caminho pelo *Duat* durante a noite, até retornar no outro dia.

Embora os egípcios acreditassem em universo que não alterava seu estado, eles também pensavam que ele nem sempre existiu da maneira como era. Dessa forma, existiam três versões que contavam como este universo teria sido criado. Um ponto comum a todas essas histórias é que as águas eram a origem de tudo, e deveriam existir eternamente nos três mundos, sempre os cercando e percorrendo seus ambientes. Também para os egípcios o universo e seus componentes eram criaturas vivas, normalmente representadas por deuses ou personagens.

Em uma das principais versões da criação, é contado que das águas primordiais, representadas pelo deus *Nun*, nasceu *Atum*, o deus criador de tudo. Ele surgiu como uma montanha que brotou das águas (*Nun*) e, a partir dele mesmo, criou o deus *Shu*, deus do ar, e *Tefenet*, a deusa da chuva e do orvalho. Logo após, *Atum* criou as divindades de *Geb* e *Nut*. *Geb* representava a Terra e *Nut* o céu. No entanto, eles foram criados em unidade, até que *Shu* levantou o corpo de *Nut* para bem alto, criando assim o céu separado da Terra. Essa foi a criação inicial de tudo e todos os outros deuses e entidades do mundo viriam posteriormente a partir dela[15].

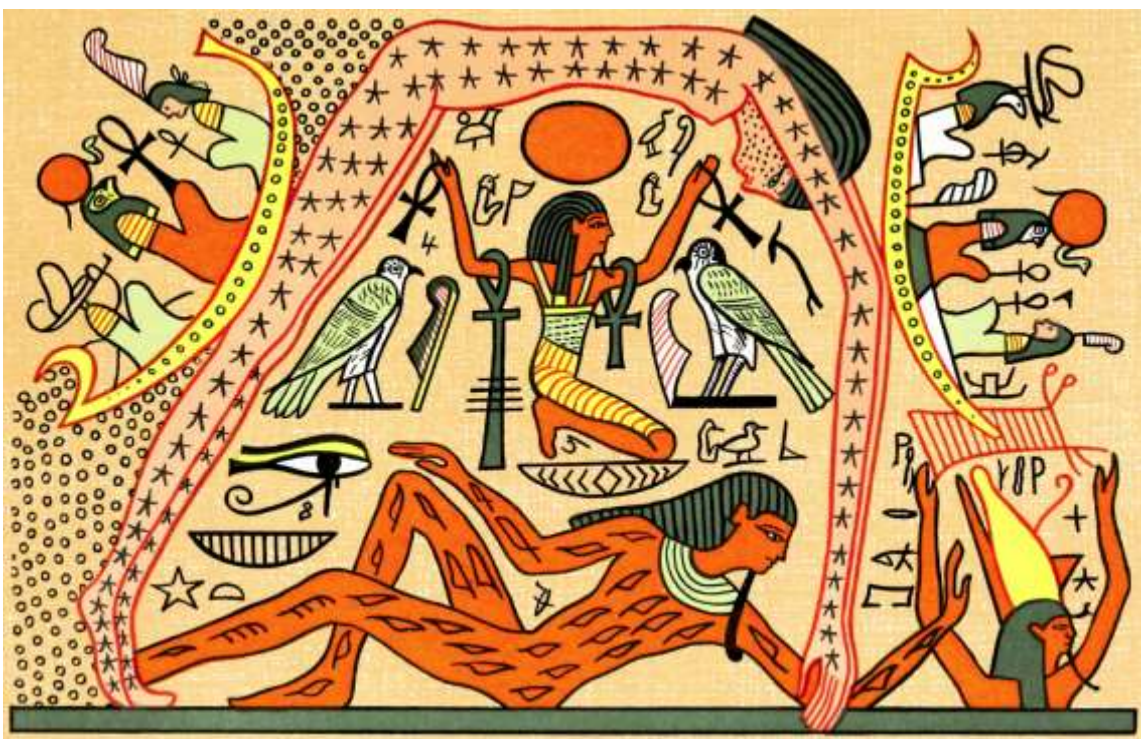

O deus Shu, ao centro, separando a Terra (Geb) do céu (Nut) no momento da criação.

---

[15] Para outra versão similar da criação contada pelos egípcios ver o livro *"Conceptions of Cosmos"* de Helge Kragh, páginas 7 e 8.

**Cosmologia dos povos mesopotâmicos[16]**

Os povos que viveram entre os rios Tigres e Eufrates, há, aproximadamente, 5000 a.C também possuíam histórias nas quais expressavam suas ideias sobre como o universo teve a sua origem. A versão principal conta que três deuses governavam o universo: *Anu*, o deus do Céu, *Ea*, o deus da Terra e das águas e *Enlil* era o governante do ar, o qual ficava entre os domínios dos outros dois deuses[17] (notamos aqui uma similaridade com a cosmologia egípcia). Embora *Anu* seja considerado o pai dos deuses, ele compartilha o governo do universo juntamente com *Ea* e *Enlil*. Similarmente aos deuses egípcios, os deuses mesopotâmicos eram descendentes das águas. *Tiamat*, a deusa das águas salgadas, e *Apsu*, o deus das águas doces, deram origem à mistura primordial de onde os outros deuses surgiram. *Anu* e *Ea* também estavam unidos no início, e foi *Enlil* quem soprou entre eles e os separou.

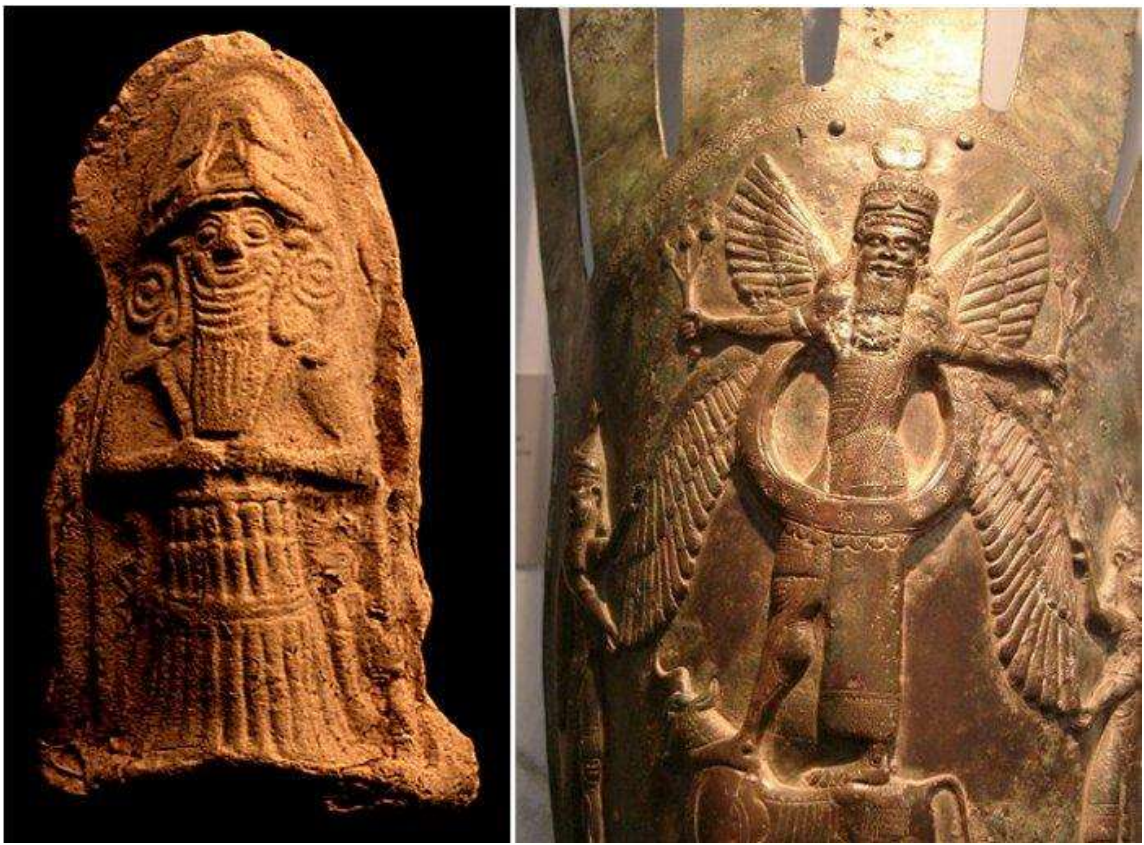

Representações de Anu – o deus do Céu – da Mesopotâmia antiga.

Fonte: AncientPages.com

Embora os babilônios fossem desenvolver uma astronomia e matemática muito avançadas para sua época, e fossem exímios observadores e conhecedores do céu e das estrelas, isso não influenciou significativamente suas concepções sobre a origem do universo, permanecendo essas em seu caráter mitológico. No entanto, existem alguns relatos isolados onde as observações astronômicas se misturam com a sua visão da ordem do universo. No mito da criação babilônico, *Enuma Elish,* escrito em tábuas de argila, totalizando aproximadamente mil linhas sobre a história da criação do universo, é possível encontrar referências às observações da lua e suas fases. A lua

---

[16] Refere-se mais especificamente aos sumérios, amoritas (babilônios), assírios e caldeus.

[17] Os nomes dos três deuses apresentados aqui provêm da nomenclatura dada pelos amoritas (babilônios). Existem variações de acordo com o povo em questão.

é descrita como um marcador do tempo, um deus que, usando uma coroa em posições diferentes (fases da lua), definiria o período de um mês. O deus da cidade da Babilônia, Marduque, foi quem organizou o calendário. Ele ordenou que a Lua surgisse e a tornou uma "criatura das trevas", ficando ela a responsável pela noite e por medir o tempo. Conta a lenda que todo mês, sem falhar nem uma única vez, ele a presenteava com uma coroa, a qual ela sempre usava. O seguinte trecho é referente ao texto do *Enuma Elish*[18]:

"*No início do mês, ao subir sobre a terra, teus chifres brilhantes seis dias medirão; no sétimo dia, apareça metade da tua coroa. Na Lua cheia, enfrentarás o Sol ... [Mas] quando o Sol começar a ganhar em ti nas profundezas do céu, diminua o seu esplendor, inverta o seu crescimento.*"

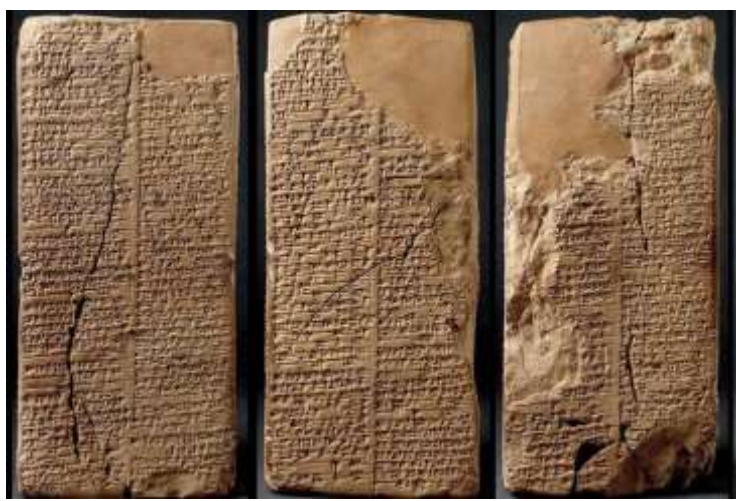

Tabuletas de barro contendo o texto *Enuma Elish* – o mito da criação babilônica.

Fonte: www.mitografias.com.br

**Para aprender mais – Sugestão de leitura**

Na história da cosmologia dos povos antigos um vasto e rico material pode ser encontrado. O historiador da ciência e autor dinamarquês Helge Kragh, já citado nesse texto, escreveu o livro "*Conceptions of Cosmos – From Myths to Accelerating Universe, a history os Cosmology*" (Oxford University Press, ainda sem tradução para o português) no qual ele descreve, detalhadamente nos primeiros capítulos, muito da história das cosmologias antigas.

O que procuramos apresentar nesse capítulo foi uma breve introdução ao assunto das cosmologias históricas e sua importância para o desenvolvimento da cosmologia. Embora tenhamos abordado apenas três exemplos, existem inúmeras outras cosmologias antigas, advindas de vários povos e diferentes tempos da nossa existência, as quais também podem ser trabalhadas em sala de aula.

**Sugestão de atividade**

1) Deixe sua imaginação fluir e se pense que você vive em uma civilização do Mundo Antigo qualquer. Você pode até inventar uma se quiser! Como é a *cosmologia* do seu povo? Explique a visão de vocês sobre como o universo foi criado e como ele é. Quanto mais imaginação e criatividade, melhor! Faça uso também de desenhos, poesias e o que mais desejar.

---

[18] Livro "*Conceptions of Cosmos*," Helge Kragh, página 9.

# Capítulo 3

## Afinal, como é o universo? Como sabemos o que sabemos?

*Objetivos desse capítulo*

*Os alunos são convidados a conhecer uma visão científica geral do universo. Através da pergunta "Como podemos afirmar que sabemos o que sabemos?", surge a reflexão de que a visão científica do universo não vem por acaso, tampouco é fruto de somente simples imaginações e subjetividades, mas o trabalho de anos de pesquisas e investigações científicas rigorosas baseadas no método científico. Algumas reflexões sobre o método científico ressurgem nesse capítulo para reforçar essa ideia. Também é apresentado uma pequena história de como essas investigações foram evoluindo e se tornando mais precisas. Ao final, seria significativo que, além desses conceitos, os alunos soubessem que a atual visão científica do universo o mostra como um universo em expansão (não estático), evoluindo de algum momento no passado, como mostra a teoria do Big Bang, onde a sua densidade e temperatura eram muito elevadas. Também seria significativo que saibam conceituar bem resumidamente e diferenciar a Teoria da Relatividade Geral e a Mecânica Quântica, se possível, fazendo uso do conceito de "recorte da realidade" apresentado aqui. Importante: não se espera que nesse momento adquiram conceitos um pouco mais técnicos referentes a quaisquer uma dessas teorias apresentadas no capítulo.*

Comecemos com uma história contada por Stephen Hawking no primeiro capítulo do seu livro "Uma breve história do tempo". Uma certa vez, um renomado cientista chamado Bertrand Russel estava dando uma palestra sobre astronomia. Ele falava sobre como os planetas orbitavam a nossa estrela o Sol e, como esse, por sua vez, também orbita outras estrelas – um conjunto que conhecemos por nossa galáxia, a Via Láctea. Conta a história que, do fundo da plateia, levantou-se uma senhora que disse: " O que o senhor acabou de falar é uma imensa bobagem. Na verdade, o mundo é um prato achatado apoiado no dorso de uma tartaruga gigante". Sorrindo, Bertrand perguntou: "No que a tartaruga está apoiada? " "O senhor é muito esperto, rapaz, muito esperto. Mas existem tartarugas até lá embaixo! ", respondeu a mulher.

Qual das duas perspectivas faz mais sentido para nós - a de Bertrand ou da instigante senhorinha? Para a grande maioria de nós, o universo representado como uma torre de tartarugas soaria um tanto quanto impossível. Mas por que, afinal, afirmamos saber o que sabemos? Provavelmente, aquela senhorinha era uma grande admiradora de tartarugas e achava belo relacionar a estrutura do "mundo" (universo) com seu animal favorito. As colocações de Bertrand, na verdade, não eram simples colocações e nem muito menos de sua autoria! Foram teorias construídas através de séculos de observações[19] e estudos, por diferentes homens e em diferentes partes do mundo, seguindo o tão imprescindível "método científico" (o qual discutimos no primeiro capítulo). E, como bem lembrado, quando investigamos algum fenômeno sob a perspectiva de tal método, dentre uma série de características presentes nesse tipo de investigação, uma delas é não deixar que as nossas "vontades" influenciem nos resultados. O método científico não deve deixar espaço para subjetividades. Em outras palavras, não podemos "forçar" um resultado para que ele nos agrade e fiquemos satisfeitos por eles terem correspondido às nossas imposições interiores. Acho as onças pintadas das nossas Américas os animais mais majestosos e lindos que já vi em toda minha vida! Seria muito divertido (para mim, claro) que a estrutura do universo tivesse alguma coisa relacionada a elas... os campos gravitacionais seguindo padrões dos desenhos das pintas… tudo seria belo e significativo... Mas somente para mim e, talvez também, para a maioria daqueles que compartilham dessa OPINIÃO! Opinião – essa é a palavra que não vai muito bem com o método científico.

---

[19] Não que a possível teoria da senhorinha não pudesse ser advinda de alguma cosmologia antiga, tal como vimos no segundo capítulo, mas o que estamos realmente discutindo aqui é o fato de um conhecimento ser construído através do método científico ou não. Nesse caso, foram anos de pesquisa científica.

Quando afirmamos saber o que sabemos (em referência às descobertas científicas como um todo e, nesse caso, às ideias de movimentos celestes que Bertrand explicava, ou, pelo menos, tentava explicar em sua palestra) estamos apoiados em investigações meticulosas e detalhadas da natureza e seus fenômenos. Inclusive, investigações realizadas e conferidas por cientistas, muitas vezes, de diversas partes mundo, diferentes culturas, gostos musicais, crenças religiosas, idiomas, hábitos alimentares e torcidas de times de futebol diferentes (dentre outras inúmeras particularidades). Não importa, sempre chegaram nos mesmos resultados e, quando não, surgia uma nova frente de pesquisa e todo o processo se repetia. Um cientista indiano adorador da deusa hindu da sabedoria, Sraswati[20], e que adora comer Samosa[21] encontrará os mesmos resultados ao ser questionado sobre qual a força gravitacional que a Terra exerce na Lua, do que um cientista mexicano, católico, cujo prato preferido é Guacamole[22] com muita pimenta Serrana. Os dados usados nesse cálculo, massas dos dois corpos, constante gravitacional e distância entre eles independem de qualquer opinião ou qualquer outra subjetividade de cada um dos cientistas. As operações matemáticas idem - $23 \times 2 = 46$ seja no México, Índia, Brasil ou qualquer outro lugar do planeta.

O geocentrismo foi uma teoria que durou muito tempo e teve pensadores famosos como Aristóteles e Ptolomeu como defensores. Para eles, a Terra era o centro do que consideravam como o universo, composto pelo Sol, Lua, alguns planetas[23] e as estrelas. Todos giravam em torno dela e, com os aperfeiçoamentos feitos por Ptolomeu, foram descritos os movimentos realizados por eles, seguindo as esferas que tais corpos deveriam ocupar, tendo a Terra em seu centro. O modelo ptolomaico tinha grandes problemas para descrever a órbita lunar. Para adequar o modelo ao movimento observado da Lua, Ptolomeu previa que ela teria que se aproximar duas vezes mais da Terra durante um certo período de tempo. Logicamente, ao fazer isso, ela apareceria duas vezes maior no céu! No entanto, esse modelo durou toda a idade antiga, pois, além de outros fatores, também satisfazia as concepções filosóficas dos gregos que acreditavam que as esferas representavam a perfeição e a forma como os elementos se agrupavam. Com a introdução dos epiciclos, o modelo se adequava à teoria, tornando-se observacionalmente válido. Outra questão que fez o geocentrismo durar boa parte da idade média foi porque também não entrava em conflito com a fé cristã, a qual acreditava ser o universo criado por Deus em nossa função. Dessa forma, ocupar "o seu centro" com tudo girando ao nosso redor, reforçava isso.

Em 1514 uma teoria contrária começou a ser difundida. O padre polonês Nicolau Copérnico defendia uma ideia onde o Sol ocupava o centro do universo e todo o restante o orbitava. Importante relembrar que, como discutido no capítulo 1, Aristarco de Samos já havia proposto essa ideia na Grécia, mais ou menos 250 a.C. Isso viria a ser conhecido como heliocentrismo e começou a ganhar força há aproximados 100 anos depois dos estudos de Copérnico. Dois astrônomos, um italiano, chamado Galileu Galilei, e outro alemão, Johannes Kepler, deram bases mais sólidas à teoria. Em 1609, Galileu abalou o "mundo geocêntrico", através de um telescópio, o qual ele mesmo aperfeiçoou e fez observações importantes, incluindo o planeta Júpiter e quatro corpos que o orbitavam. Estava ali um indício de a Terra não era o centro de absolutamente tudo no universo, pois, em Júpiter, quatro satélites o orbitavam, e não orbitavam diretamente a Terra. Mas isso não era, contudo, uma evidência que derrubava o geocentrismo definidamente, porém, já o colocava em dúvida. O problema das órbitas circulares ainda persistia, pois se supunha que os planetas orbitavam o sol em trajetórias circulares, mas isto não concordava com as observações, e foi Kepler quem introduziu o conceito das órbitas elípticas. Embora suas fortes concepções religiosas não o fizesse "gostar" do movimento elíptico, e sim do movimento circular, seus cálculos e, principalmente, as observações de Marte o levaram a admiti-las, pelo menos, para a validação do modelo. Ele também defendia, erroneamente, que o que se conhecia à época por força magnética era a responsável pelo movimento elíptico. Contudo, a teoria magnética e as órbitas elípticas eram contraditórias entre si (a órbita elíptica de Marte não podia ser conciliada a essa teoria) e Kepler não tinha uma solução para o problema.

---

[20] Saraswati (sânscrito) é a deusa hindu da sabedoria, das artes e da música e esposa de Brahmā, o criador do mundo.
[21] A samosa é um pastel triangular de origem indiana.
[22] Purê de abacate temperado com sal, limão, azeite, cebola, tomate e pimenta.
[23] A palavra planeta significa "andarilho", no grego. Se referiam às "estrelas" que tinham um movimento diferente das demais no céu.

O homem que habita o imaginário coletivo como sendo aquele que foi atingido por uma maçã na cabeça e teve uma brilhante ideia, foi o que pôs luz sobre a questão das órbitas elípticas e o movimento dos corpos celestes. Isso ocorreu no ano de 1666, enquanto ele estava em quarentena, na casa da mãe, devido à Peste Bubônica[24]. Notável mencionar que, ao mesmo tempo que fazia isso, também inventou o cálculo integral[25] e diferencial[26]! O esforçado Isaac Newton, cujo crânio não fora atingido pelo fruto proibido dos cristãos, segundo os relatos do próprio[27], publicou, em 1687, um dos mais importantes tratados das ciências naturais – "Philosophiae Naturalis Principia Mahematica". Nele, através de uma matemática complexa e inovadora, explicava os movimentos dos corpos pelo espaço e tempo, dois conceitos independentes um do outro, uma ideia que seria superada somente 300 anos depois com Albert Einstein. Também introduziu a teoria da gravidade. Basicamente, ela explicava o porquê das coisas caírem, e como todos os corpos do universo se atraem mutuamente. Postulava que quanto maior eram as suas massas e a aproximação entre eles, maior era a força que os atraía – a gravidade. Usando essa teoria, ele demonstrou as órbitas elípticas da Lua ao redor da Terra e dos planetas ao redor do Sol.

Mas essa teoria também o fez refletir sobre as questões da posição das estrelas e o estado do universo. Como a gravidade gerava atração entre os corpos, ele indagou que as estrelas começariam a se atrair mutuamente e, consequentemente, em algum momento, entrariam em choque. Em 1861, ainda intrigado, ele escreve a um amigo cientista dizendo que isso poderia acontecer, porém, somente se o número de estrelas do universo fosse finito e também estivessem confinadas em um espaço igualmente finito. Caso considerasse o universo e o número de estrelas infinitos, isso não seria possível, pois não haveria um centro do Universo onde as estrelas pudessem cair! Na verdade, esse pensamento se torna paradoxal porque em um Universo infinito e com infinitas estrelas, qualquer localidade pode ser considerada o seu centro, uma vez que estará rodeada pelo mesmo número de estrelas. Dessa maneira, o colapso delas gerado pela força gravitacional também ocorreria. Como resolver o paradoxo?

---

[24] Causada pela bactéria Yersinia pestis e pode se disseminar pelo contato com pulgas infectadas. Matou ⅔ da população europeia nessa época.
[25] Utilizado para calcular a área abaixo de uma curva no plano cartesiano. Muito aplicado em problemas de física e engenharia.
[26] Derivar a curva de uma função em um determinado ponto, significa encontrarmos a reta tangente à ela nesse ponto. Também muito utilizado em problemas da física e engenharia.
[27] Ver artigo de Roberto de Andrade Martins – A maçã de Newton – história, lendas e tolices (p.175 e 176)

Uma possível saída seria pensar que o universo pudesse se expandir. Me imagine a um metro de você. Eu estou sobre marca de um "x" e você sobre a de um "o", ambos desenhados no chão em que pisamos. De repente, ficamos mais distantes, por exemplo, há uns 2 metros um do outro. Você olha para os meus pés e eles continuam sobre o "x". Olha para os seus próprios pés e eles continuam sobre a mesma marca do "o". Você conclui que não nos movemos, mas, ainda assim, a distância entre nós aumentou. Progressivamente, o processo se repete. O que está acontecendo, já que nós não estamos nos locomovendo? Uma observação mais criteriosa revela que o espaço é que está se dilatando! Essa é outra ideia que foge das nossas experiências cotidianas e vamos vê-la com mais detalhes nos próximos capítulos. Por hora, pense que em nosso lugar, agora, estão as estrelas ou planetas. Neste caso, o universo estaria se expandindo e as distâncias entre os planetas e estrelas aumentando. Essa foi uma ideia que não apareceu em nenhum momento nessa época e nem durante toda a história escrita da humanidade, pois os filósofos e estudiosos consideravam que o universo deveria ser estático. Tentaram realizar várias modificações na teoria da gravidade, de maneira a colocá-la como repulsiva, caso os corpos estivessem a longas distâncias uns dos outros, surgindo uma espécie de equilíbrio, resultando em um universo estático. Hoje, porém, é sabido que tal equilíbrio, de um universo estático e eterno, seria facilmente quebrado por qualquer pequeno distúrbio nas distâncias entre os corpos.

Essa questão ficou por muitos anos sendo debatida pelos mais variados cientistas e filósofos[28]. Em 1826, um médico e astrônomo alemão chamado Heinrich Olbers, publicou um artigo, o qual continha um problema que mais tarde viria a ser conhecido como o "Paradoxo de Olbers". Ele fazia a seguinte reflexão: ao observarmos o céu noturno, encontramos mais espaços negros do que estrelas. Em um universo infinito, estático e homogeneamente incrustado de estrelas, isso não deveria acontecer pois, para qualquer direção em que olhássemos, nossa linha de visão recairia sobre uma estrela, o que nos faria enxergar o céu noturno tão brilhante quanto o Sol matinal. Implicitamente a essa ideia, estava lançada também a teoria de que as estrelas tivessem se formado em um tempo específico, ou seja, não existiram para sempre.

A teoria do Big Bang é a que melhor explica atualmente o estado inicial do universo (temos um capítulo à frente somente sobre esse tema). Como já brevemente discutido aqui, e também será analisado com mais detalhes à frente, a descoberta da expansão do universo por Edwin Hubble, na década de 20, trouxe uma série de novos paradigmas para a cosmologia. Dentre eles, o que se relacionava à possível origem do universo era o fato de o universo atual estar se expandindo. Alguns cientistas começaram a se indagar, caso pudéssemos retroceder o relógio para tempos passados e ver o que acontecia, qual seria o cenário? Muito provavelmente veríamos o universo se encolhendo, progressivamente, a medida que voltássemos no tempo (o movimento contrário à expansão), até chegar ao ponto em que todo o universo e seu conteúdo estariam confinados em um único ponto. E aqui começa um dos maiores mistérios que a Cosmologia investiga hoje – qual era o tamanho desse ponto? Era infinitamente pequeno? Mas qual o significado desse "infinitamente"? Fisicamente, o que quer dizer infinito? Uma vez que nos propomos a querer entender racionalmente o Universo, conceituar e saber o valor do infinito se torna algo essencial. Seria ali o início do universo ou existia algo antes? O universo surgiu do nada? Essas são perguntas que, além de muito instigantes e interessantes, são a base de inúmeros trabalhos hoje na Cosmologia, inclusive o "trabalho de uma vida" de muitos cientistas, e, mesmo assim, ainda não foram respondidas (talvez algum leitor desse pequeno livro pode vir a ser alguém que trará boas contribuições a esse tema, quem sabe?!). Dessa forma, devemos ter em mente que, quando falamos em teoria do Big Bang, estamos nos referindo a partir de um certo momento em que o universo era muito jovem, pequeno e quente, porém, o estado anterior a esse momento (conhecido como tempo de Planck e veremos sobre ele mais adiante) ainda é incerto, e vem sendo muito estudado. A teoria do Big Bang ainda não descreve o que teria ocorrido antes desse tempo, mas isso não significa que os cientistas não estão pesquisando sobre isso e já possuem algumas ideias (veja bem, ideias, e não comprovações!). Teria o Universo começado ali, ou o Big Bang é apenas a parte de expansão de um processo cíclico muito maior, onde o Universo se expande até um certo ponto, para, e depois se contrai até um valor muito pequeno, para então voltar a se expandir (apenas essa última etapa citada de expansão seria o que conhecemos por Big Bang, ou seja, apenas uma fase em um ciclo)? Um importante cientista que dedicou a sua vida a estudar Cosmologia e tentar responder também a essas perguntas sobre a origem do Universo foi Stephen Hawking. De uma maneira resumida, o cenário que ele propõe para esse possível início do Universo tem mais a ver com a primeira opção. Nele, não faz sentido perguntar sobre o que teria ocorrido antes do Big Bang porque, simplesmente, o tempo não existia, e passou a existir a partir daquele instante. Você pode perguntar: "Mas origem do tempo? Ele não é eterno e sempre existiu?". De acordo com Hawking, não. Como será visto mais adiante, a Teoria da Relatividade Geral prevê que o tempo se desacelera conforme "caminhamos" na direção de um centro de grande massa (nesse caso, o ponto pequeno onde toda a matéria do universo está concentrada). Estamos falando de um cenário de densidade infinita, ou seja, uma quantidade de matéria absurdamente grande de tal forma que o tempo nesse contexto se desacelera a ponto de parar! Isso mesmo, o relógio pararia e o tempo seria estático. Assim, na expansão do Big Bang, tem origem não só o espaço, mas também o próprio tempo (espaço-tempo). Sob essa perspectiva, não faria muito sentido falar sobre algum momento antes desse início porque o próprio tempo não existiria[29]. É interessante também conhecer o que uma outra linha de estudos da Cosmologia diz, como por exemplo, a que o cosmólogo brasileiro Mario Novello também trabalha[30]. Nessa teoria, a história que se passa antes do Big Bang é o cenário de contração e expansão do Universo a qual mencionamos algumas frases atrás. O Universo sempre existiu, e vive em um ciclo de expansão e contração (nesse momento estaríamos vivendo no momento de expansão do Universo). Saber o que de fato aconteceu antes do instante do tempo de Planck ($10^{-43}$ s), é uma tarefa que exige muita dedicação e conhecimento. Como dito, pode ser o trabalho de uma vida inteira de uma pessoa! Neste livro, procuraremos apenas nos ater a teoria do Big Bang, em um capítulo específico, a qual discorre o que aconteceu a partir do tempo de Planck. É importante ressaltar que a teoria do Big Bang não é algo criado como uma história de ficção científica ou enredo para filmes de super-herói. Ela possui uma forte fundamentação teórica e evidências físicas, como a radiação cósmica de fundo, os testes das reações nucleares nos átomos e observações da expansão do universo. Hoje, na comunidade acadêmica, como dizemos no começo, é a melhor teoria que descreve o pssado do Universo, fortemente sustentada por observações. Desde que esse capítulo não é destinado para tratar especificamente do Big Bang, por hora, basta termos esses conceitos introdutórios em mente – o universo, num passado muito distante, era extremamente quente, denso (muita matéria concentrada em um único ponto) e começou a se expandir, evoluindo até o cenário em que vivemos hoje.

---

[28] Nessa época, eram duas ocupações pouco diferenciadas uma da outra.

Como foi mencionado acima a Teoria da Relatividade Geral, torna-se importante fazer uma reflexão sobre como os cientistas tentam descrever a natureza. Ao investigar o universo e os seus fenômenos, eles se deparam com uma gama imensa de complexidades. Assuntos que devem ser detalhados ao máximo possível, com o maior grau de precisão de que possa alcançar. Muitas vezes, um pequeno detalhe de alguma teoria torna-se o trabalho de toda uma vida de gerações de cientistas. Tal complexidade requer de nós a habilidade de saber dividir esses assuntos em "recortes isolados" da realidade, para que seja humanamente possível estudá-los e investigá-los a fundo. É necessário tentar simplificar o problema, as suas condições e criar um "recorte" onde tenhamos controle dos parâmetros envolvidos. Mas não devemos confundir essa habilidade com não sermos dinâmicos e heterogêneos em nosso conhecimento, elegendo apenas algumas coisas como importantes e fechando os olhos para o resto. Não é isto que está sendo dito. Pelo contrário, em se tratando de ser curiosos e adquirir conhecimento, devemos conhecer e aprender o máximo, sobre os mais variados campos de estudos da ciência, em diferentes áreas e com diversos autores! Conhecer outros departamentos do conhecimento que não sejam os que já estamos acostumados, juntamente com as suas teorias e métodos (acredite, eles são imensamente numerosos), torna o nosso próprio conhecimento mais robusto e adaptável ao mundo.

---

[29] Você pode encontrar o próprio Hawking falando sobre isso, no seu livro "Breves respostas para grandes questões", mais especificamente no capítulo "Como tudo começou".
[30] Você pode acessar uma aula do professor sobre a teoria nesse endereço: https://www.youtube.com/watch?v=W21s-blOskY&t=1304s

Hoje existem duas grandes teorias abrangentes que nos ajudam a criar os recortes para o estudo do universo: a Teoria da Relatividade Geral e a Mecânica Quântica. Ambas serão temas de capítulos posteriores e aqui cabe uma pequena introdução.

A Relatividade Geral é a teoria do famoso físico Albert Einstein. Foi apresentada em 1915 e significou uma revolução para o mundo. Ela estuda, principalmente, o universo em grande escala, a origem da gravidade e como os corpos interagem entre si através dela. As dimensões típicas da maioria dos problemas nessa teoria são da ordem de centenas de quilômetros até milhões de milhões dessa unidade. Estamos interessados em campos gravitacionais, distorções no espaço-tempo (os quais, aliás, são dependentes um do outro), interações entre objetos de grandes massas, estrelas e buracos negros dentre outros assuntos.

Já quando nos referimos à Mecânica Quântica, estamos falando do infinitamente pequeno. Essa teoria foi proposta por Max Planck por volta do ano de 1900. Seu conceito fundamental era o de quantização da energia, ou seja, a unidade fundamental de energia, chamada de quantum. Uma analogia útil para abstrair tal conceito é imaginar uma parede composta por tijolos, todos iguais e que não podem ser fragmentados. A parede será considerada como uma quantidade de energia qualquer. Os tijolos, portanto, são a sua unidade fundamental básica, a menor divisão que podemos fazer da energia. Se fossemos "quebrando" a parede, iríamos chegar na unidade do tijolo inquebrável. O Princípio da Incerteza também é outro conceito muito importante dessa teoria e igualmente revolucionário. Basicamente, ele nos diz que no nível subatômico, não podemos conhecer, ao mesmo tempo, a velocidade e a posição de uma partícula. No instante em que conhecermos uma das duas grandezas, a outra apresentará uma incerteza maior na sua medição.

Contudo, existe um fato intrigante sobre as duas teorias: embora ambas funcionam muito bem em seus respectivos campos de estudo, elas são incompatíveis entre si. Muitos trabalhos no mundo científico estão voltados para a tentativa de unir as duas em uma só. Stephen Hawking também foi um dos principais cientistas que dedicaram a sua vida nessa busca. Aliás, como mencionamos acima, a sua busca pelas respostas sobre a origem do Universo tem tudo a ver com a tentativa de unir a Mecânica Quântica com a Teoria da Relatividade em uma única teoria que pudesse explicar tudo. Outro importante cientista, o italiano Carlo Rovelli, também é um explorador da realidade, desenvolvendo seus estudos sob os campos da Gravidade Quântica (uma teoria que, como o próprio nome diz, mescla as outras duas grandes). Dessa forma, eles e mais inúmeros e esforçados cientistas espalhados por todo o mundo vão tentando juntar os recortes de realidade na tentativa de elaborar uma nova teoria mais abrangente. O que o futuro reserva, se isso será possível ou não, não se pode prever. Só é possível continuar investigando com curiosidade e criatividade, cultivando o conhecimento, fazendo-o progredir, checando os resultados, validando nossas hipóteses e tentando aprender o máximo com elas. Também é bom que sigamos admitindo os erros quando acontecem, mantendo a humildade de reconhecer nossas limitações e nunca querendo impor nossas vontades ao universo. Assim, vamos tentando formar a sua imagem, no fundo, a nossa própria imagem, apoiados no nosso saber científico...

*Para conhecer mais:*

*- Uma breve História do Tempo, Stephen Hawking, capítulo 1 "Nossa imagem do Universo".*

*- A maçã de Newton – História, lendas e tolices, Roberto de Andrade Martins (artigo disponível na internet).*

*- Sete Breves Lições de Física - Carlo Rovelli.*

*- A Máquina do Tempo – Mario Novello.*

*Filmes e documentários*

*- A Teoria de Tudo (2014) – Conta a vida e as pesquisas de Stephen Hawking.*

*- Cosmos (2014) – episódio 1 "De pé na Via Láctea".*

*Perguntas de fixação*

*Se você fosse perguntado por alguém, pode ser algum/a amigo/a ou parente, sobre o que os cientistas sabem atualmente sobre como é o universo, o que você responderia?*

1) *Vamos supor que você e seus amigos/as estão superinteressados em ciências e decidem aprender mais sobre ela. Vocês decidem que cada dia irão passar 1h na biblioteca aprendendo física, química, matemática e biologia em dias alternados. No dia da física, vocês se dirigem às estantes de livros de física e começam a procurar por algum livro interessante. Um amigo/a acha um, com um senhor sorridente, de bigode e cabelo meio atrapalhado na capa e o seguinte título "A Teoria da Relatividade". Todos ficam muito intrigados e ainda mais animados quando leem o nome do autor "Albert Einstein". "Esse livro deve ser muito da hora! Bora estudar ele! ", diz um dos seus amigos. "Mas eu nem sei o que é Teoria da Relatividade! O que ela estuda? Será que é legal? ". Seu outro amigo fica mudo, pois também não tem ideia do que se trata. Você é questionado sobre o que é essa teoria. O que você responderia?*
2) *Muitas pessoas, erroneamente, usam a expressão "Isso é mais difícil do que física quântica", para se referir a atividades que acham complicadas. Na verdade, a física quântica nem é tão difícil assim, só precisa de dedicação nos estudos para que seja aprendida. Você poderia conceituar abaixo o que é mecânica quântica? (Esse é um nome um pouco mais específico para se referir à física quântica, uma espécie de nome oficial, física quântica = mecânica quântica)*

# Teoria da Relatividade Especial e Geral

*Objetivos*

*Neste capítulo, apresentamos alguns conceitos básicos da Teoria Geral e Especial da Relatividade. É esperado que, ao final, o leitor saiba as características e principais conceitos de cada uma assim como solucionar os problemas básicos que disponibilizamos nesse mesmo capítulo.*

**Introdução**

A Teoria Especial e Geral da Relatividade marca uma revolução na física e na maneira como a humanidade conhecia o universo e suas leis. Estamos falando do início do século XX, mais precisamente do ano de 1905, quando um recém graduado em física, chamado Albert Einstein, publicou uma série de cinco brilhantes artigos, apresentando e discutindo a Teoria Especial da Relatividade. Nessa época, ele trabalhava em um escritório de patentes e não possuía qualquer tipo de vínculo profissional com nenhuma universidade ou laboratório. Todo seu trabalho foi desenvolvido com uma originalidade e esforço impressionantes, aliados à sua personalidade criativa e curiosa. Dez anos depois, Einstein expandiu sua teoria, combinando-a com a teoria gravitacional de Newton, e dando o nome de Teoria Geral da Relatividade.

O que será apresentado nos próximos tópicos são os principais conceitos de ambas as teorias.

1. **Teoria da Relatividade Especial**

    **1.1. Movimento Relativo**
    Responda rápido, nesse exato momento, você está se movimentando ou parado? E se caso você estiver sentado, confortavelmente, em uma cadeira, enquanto lê o livro, e eu lhe disser que, na verdade, você está percorrendo 30km a cada segundo que passa? Dependendo do ritmo da sua leitura (o da maioria das pessoas é quatro palavras por segundo), enquanto lê somente essa pequena frase, você já terá percorrido 210 km! Pode parecer loucura, mas, de fato, mesmo você estando sentado e estudando agora, também está se movimentando a aproximadamente 30km/s. O que acontece é que estamos usando referenciais diferentes para determinar nossos movimentos e velocidades. Quando digo que o seu movimento é de 30km/s, tenho como referencial o Sol, o qual é orbitado pelo nosso planeta que executa o movimento de translação ao seu redor, aproximadamente, nessa velocidade.[31] Como somos habitantes da Terra, também estamos realizando esse movimento. Se você respondeu no início que estava parado, foi porque tinha como referencial (mesmo que de forma inconsciente), provavelmente, o chão ou a cadeira onde está. Dessa forma, chegamos a uma conclusão de que, para determinar movimentos, velocidades e posições dos corpos, é necessário que tenhamos primeiramente referenciais. Se vou dizer que algo está parado (ou em movimento) é necessário especificar em relação a que referencial. Note, neste próximo exemplo, como todas os movimentos e velocidades possuem seus referenciais:

---

[31] Um fato interessante - o Sistema Solar também se movimenta em relação ao centro da galáxia, aproximadamente, a 240 Km/S.

> *Uma pessoa caminhando dentro de um ônibus, o qual acabou de partir de uma estação rodoviária, e se afasta dela a 80km/h, está se dirigindo à cabine do motorista para solicitar uma informação sobre o horário da sua chegada ao seu destino. Em relação à sua poltrona, a pessoa caminha a 1m/s para chegar ao motorista. Em relação à estação, a pessoa está a 83,6km/h.*

Outro exemplo que pode demonstrar esse conceito de relatividade do movimento é você imaginar que está parado em relação a uma linha ferroviária, há alguns metros dela. Então surge um trem, que vai passando e você nota um amigo dentro de um dos vagões. Ele segura uma maçã e a joga para o alto, deixando-a cair de volta nas mãos dele. Se, nesse momento, você perguntar para ele qual a trajetória a maçã percorreu, ele responderá que se trata de uma linha reta, pois a maçã subiu e desceu, executando uma trajetória retilínea. Mas, para você, que observa **de outro referencial** (fora do trem), a maçã descreveu uma parábola no ar.

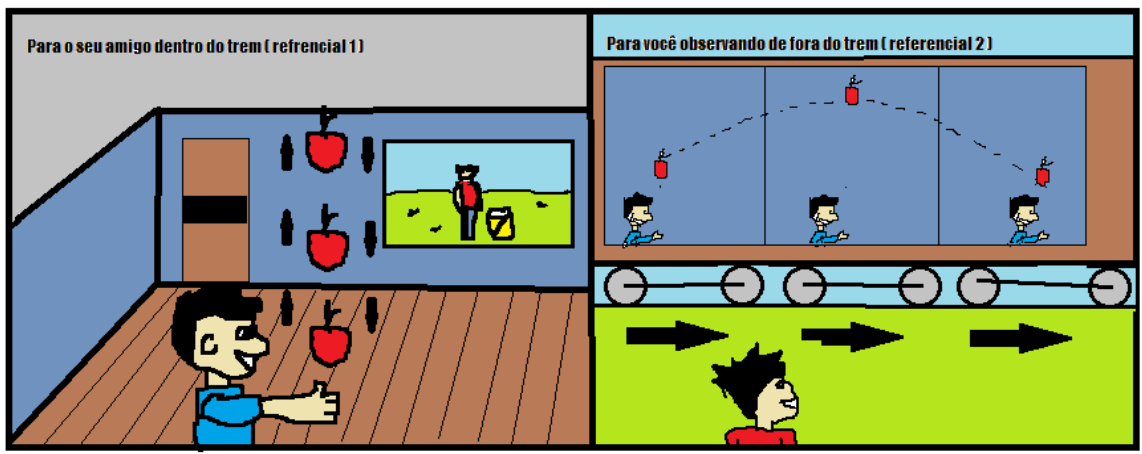

*Fonte: do autor.*

Ambas observações em relação às trajetórias estão corretas em seus respectivos referenciais. Este é mais um exemplo do princípio de relatividade do movimento. Consegue pensar em outro?

O conceito de relatividade não foi proposto por Einstein. Na verdade, era uma ideia que foi desenvolvida primeiramente por Galileu[32] entre os séculos XVI e XVII, e é essencial que a entendamos para compreender a Relatividade Especial. Vamos ao conceito de referencial inercial.

> *Referenciais inerciais são quaisquer sistemas de referenciais, os quis se movem com velocidade constante (movimento uniforme) entre si.*

---

[32] Visite o site da Universidade Federal do ABC para mais detalhes - http://propg.ufabc.edu.br/mnpef-sites/relatividade-restrita/a-relatividade-de-galileu/

Você, parado em reação à linha ferroviária, enquanto o trem passa sobre ela com velocidade constante de 80km/h, em relação a você, forma um sistema de referenciais inerciais juntamente com o trem. Para você, o trem é um referencial inercial (você o vê se movendo com velocidade constante), e para algum passageiro do trem ou o maquinista, você também é um referencial inercial (se algum deles olha pela janela, verá que você é quem parece se afastar deles com uma velocidade constante de 80km/h). Se você estivesse caminhando em relação ao sul, a 1 m/s constantes, enquanto o trem se dirigia para o norte, a 60kmh constantes, também formaria com ele um sistema de referenciais inerciais. Assim, para determinar se referenciais são inerciais ou não entre si, é preciso saber se estão a velocidades constantes, um em relação ao outro.

## 1.2. Os princípios da Teoria Especial da Relatividade

Vamos supor que você está viajando em um ônibus com uma velocidade constante de 100km/h, em relação a um ponto qualquer de uma longa estrada em linha reta. Pense que se trata de uma estrada ideal, ela não possui buracos, ondulações e nem qualquer tipo de irregularidade no asfalto, de modo que sua viagem é suave e você não sente qualquer perturbação vinda da pista ou até mesmo dos motores e dispositivos do ônibus. Pela janela, você observa as inúmeras árvores, lado a lado, que cercam todo o comprimento da estrada. Durante essa observação, de repente, vem uma sensação diferente – as árvores parecem correr para longe de você! Se trata de algo muito comum de acontecer nesse tipo de situação. Seguindo essa lógica, existiria algum experimento que pudesse ser feito para determinar que é você quem está se movimentando ao invés das árvores? Segundo a teoria de Einstein, não! O primeiro postulado nos diz que para referenciais inerciais as observações dos mesmos fenômenos físicos devem ser sempre iguais. Imagine que o ônibus que você está não possui mais janelas. Ele está viajando a 80km/h, mas você não sabe disso. Consegue realizar algum experimento que te mostre que você está parado ou em movimento? Você pode tentar vários, com blocos, pêndulos e o que mais quiser, mas será impossível provar que você está em movimento ou em repouso.

> Primeiro postulado:
>
> *As leis da natureza são as mesmas para todos os referenciais inerciais, e não existe um referencial inercial preferível a outro.*

O segundo postulado diz respeito à velocidade da luz[33]. Desde jovem, Einstein era intrigado com os fenômenos luminosos, e gostava de imaginar como seria se pudesse viajar ao lado de um feixe de luz [34]. Passava horas imaginando e explorando quais as consequências, caso isso fosse possível. Quando já adulto e realizando seus estudos e

---

[33] Representada pela letra *c*. O valor oficial para esta grandeza é 299 792 458 m/s, e é muito comum se utilizar a aproximação de 300.000 Km/h.
[34] Para saber mais, consultar o livro *Física Conceitual* de Paul Hewwit, 11ª edição, p. 625.

cálculos que resultariam em sua teoria, ele postulou que a luz possuía velocidade constante:

> Segundo postulado:
>
> ***A velocidade de propagação da luz no vácuo é a mesma para qualquer observador situado em um referencial inercial.***

Se você estivesse viajando em um foguete pelo espaço e outra nave viesse em sua direção (ambos em velocidades constantes) emitindo pulsos de luz, você mediria *c* como sendo a velocidade dos pulsos. Caso a nave estivesse se afastando, você também encontraria *c*, como valor da velocidade da luz.

Na década anterior, no ano de 1887, dois físicos realizaram um experimento em que haviam encontrado essa mesma constância para a velocidade da luz. O experimento de Michelson-Morley[35] tinha a intenção de investigar a existência do suposto "éter", uma substância em que se acreditava permear todo o espaço. Medindo as velocidades de raios luminosos que vinham de encontro à Terra, enquanto ela se movia de encontro a eles, e também enquanto ela se movia no sentido oposto, eles encontraram o resultado de que a velocidade desses feixes luminosos não se alterava. Em outras palavras, a velocidade da luz era a mesma, não importando se iam de encontro ou se afastavam da fonte emissora. Não é possível afirmar se Einstein sabia da existência do experimento e seus resultados.

### 1.2.1. Simultaneidade

Voltemos à situação em que você está parado em relação a uma linha ferroviária, enquanto um amigo passa em um trem. Imagine que a velocidade constante desse trem é muito grande, superior em mais de dez vezes a de 80km/h que falamos antes. No exato momento em que vocês dois estão alinhados entre si, dois raios caem, simultaneamente, a distâncias iguais do ponto onde vocês estão alinhados:

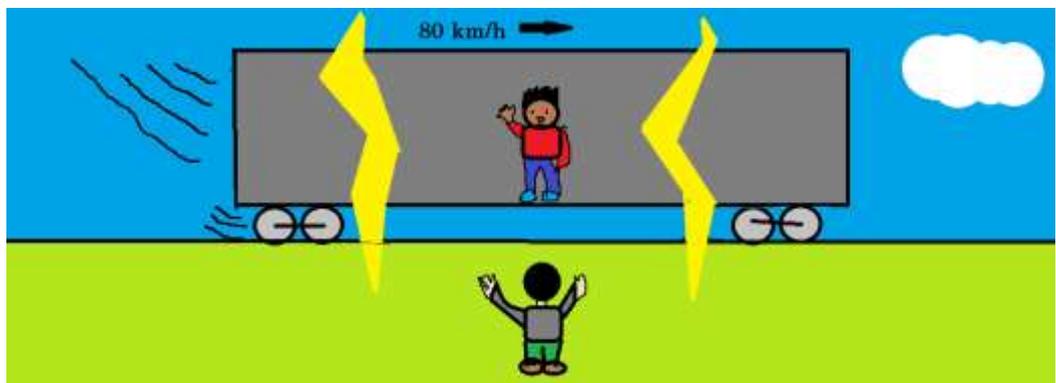

*Fonte: do autor*

Você percebeu os dois raios como acontecendo ao mesmo tempo, pois ambos viajaram à mesma velocidade e por uma mesma distância. Dessa forma, chegaram juntos até seus olhos. Mas, e seu amigo? Ele lhe dirá que o raio que caiu no mesmo sentido em que o trem viaja foi o primeiro a acontecer, e só instantes depois o outro acorreu. Ambos os raios têm a mesma velocidade (eles viajam à velocidade da luz), mas percorreram distâncias diferentes, pois o raio da direita no desenho acima percorreu um caminho menor do que o outro, visto que o trem se desloca em sua direção, diminuindo a distância. Interessantemente, tanto você quanto seu amigo estão corretos em suas observações, a partir dos seus respectivos referenciais, o que nos leva a concluir que:

> *Eventos que são simultâneos para um referencial, não são, necessariamente, simultâneos para outro que se move em elação ao primeiro.*

---
[35] Para conhecer mais, acesse o site da Universidade Federal do ABC - http://propg.ufabc.edu.br/mnpef-sites/relatividade-restrita/o-experimento-de-michelson-morley/

No seu referencial, os raios aconteceram ao mesmo tempo, mas, no referencial do seu amigo, um deles aconteceu primeiro do que o outro. Esse é o conceito da simultaneidade, o qual é uma das consequências da Teoria da Relatividade Especial.

### 1.2.2. Espaço-tempo

A física newtoniana considera espaço e tempo como grandezas separadas na natureza. Einstein nos mostrou em sua teoria que, na verdade, ambos são dependentes e se influenciam. Embora nossa concepção cotidiana é a de um espaço tridimensional (por exemplo, em coordenadas x,y e z), na verdade, habitamos um *espaço-tempo* quadridimensional, sendo o tempo a quarta dimensão. Ao nos referirmos à localização de um objeto, então, seria correto dizer , por exemplo, que a sua localização está em (x,y,z,t) sendo "t" o instante de tempo naquele referencial. Como veremos mais adiante, devido à velocidade da luz ser constante, independente do referencial inercial, surgem consequências nas medições de distâncias e tempos. Basicamente, como a definição de velocidade é *distância percorrida / tempo decorrido*, e estamos nos referindo a uma velocidade que não se altera (c), quando uma variação ocorrer em quaisquer uma das outras duas grandezas, a outra grandeza também deverá sofrer mudanças. Temos um *espaço-tempo*! (Note que é a velocidade da luz que estabelece a relação entre os dois)

Esses efeitos serão mais significativos para nós quando estivermos lidando com objetos que possuem velocidade extremamente altas, próximas à velocidade da luz (também conhecidas como velocidades relativísticas). Em situações do nosso cotidiano, não estamos acostumados a lidar com esses tipos de situações. Para entendermos como o espaço-tempo se altera entre diferentes referenciais, é necessário aprender dois outros conceitos importantes: a dilatação do tempo e a contração dos comprimentos – consequências diretas dos postulados da Relatividade, mostrados acima.

### 1.2.3. Dilatação temporal

Imagine que você está a bordo de uma nave, viajando pelo espaço, com velocidade *v*. Nela, existe um dispositivo que mede o tempo, baseado na emissão, reflexão e absorção de pulsos de luz. Um pulso é emitido de um equipamento que está localizado no chão da nave, chega até o teto, onde é refletido para o chão de e o mesmo equipamento o absorve, lançando novamente o pulso, de forma sincronizada e contínua. O sistema todo funciona como um relógio de luz, sendo o tempo entre a emissão do pulso e a sua recepção pelo mesmo equipamento que o lançou, a unidade básica de tempo (em um relógio comum, a unidade básica são os segundos). Você está dentro da nave, observando todo o processo, e chamaremos esse conjunto de referencial 1 (seu referencial, note que você está em repouso em relação à nave).

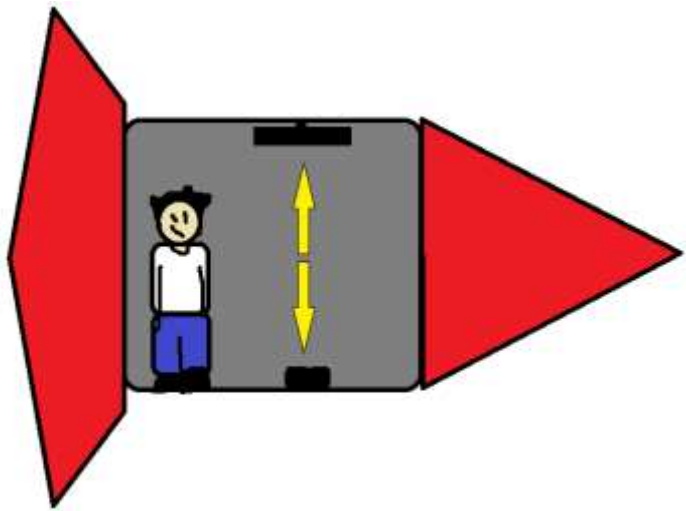

Fonte: do autor

Sobre a superfície de um planeta qualquer, está seu amigo, que o observa passando a bordo da nave. Ele, no planeta (referencial 2), é um referencial inercial em relação a você e vice-versa. Quando ele observa o seu "relógio de luz", enxerga um fenômeno um pouco diferente do que você está enxergando. No seu referencial, o pulso de luz descreve uma linha reta, com subida e descida (figura acima). Vista a partir do referencial do seu amigo (referencial 2), a luz percorre uma distância maior, uma vez que a nave se move em relação a ele (figura a baixo):

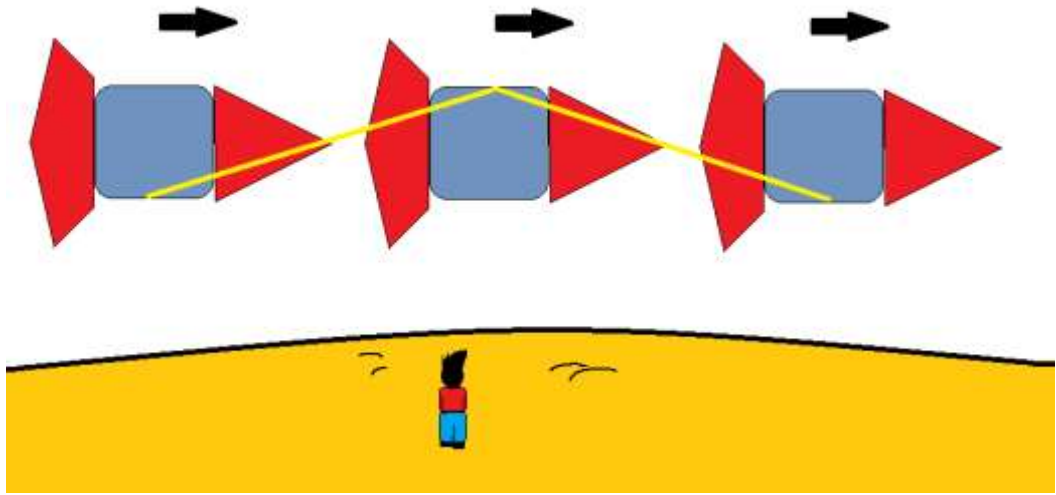

Fonte: do autor.

Como vimos no segundo postulado, não importa qual seja o referencial, a luz sempre terá velocidade igual para ambos (estamos considerando os dois referenciais no vácuo, portanto a velocidade da luz é *c)*. Quando a observação foi realizada a partir do referencial 2, a luz é vista como percorrendo uma distância maior da percorrida no referencial 1. Portanto, a *medição do tempo,* feita a partir do referencial 2, também encontrará um valor **maior** para essa grandeza:

**Referencial 1:**

$$c = \frac{d1}{t1}$$

Onde d1 é o valor da distância medido por você que está no referencial 1. Igualmente é t1, valor do tempo decorrido medido no referencial 1.

**Referencial 2:**

$$c = \frac{d2}{t2}$$

Onde d2 é o valor da distância percorrida medido pelo seu amigo que está no referencial 2. Igualmente t2, valor medido de tempo decorrido a partir do referencial 2.

Mas como vimos:

$$d2 > d1$$
$$e$$
$$c\,não\,varia$$

Portanto:

$$t2 > t1$$

Isso quer dizer que seu amigo mede intervalos de tempo muito maiores para o seu referencial do que os que você mede. Você poderia dizer a ele que o tempo para um pulso de luz ir do chão até o teto, ser refletido e voltar para o aparelho no chão é de 1 segundo; enquanto que seu amigo, dependendo da velocidade da nave em que você está, lhe dirá que o tempo, na verdade, é de 2 segundos. Resumindo, ele vê o tempo passar para você em um ritmo mais lento do que o tempo que você mede no seu referencial.

Através de álgebra e geometria podemos entender mais a fundo esse conceito:

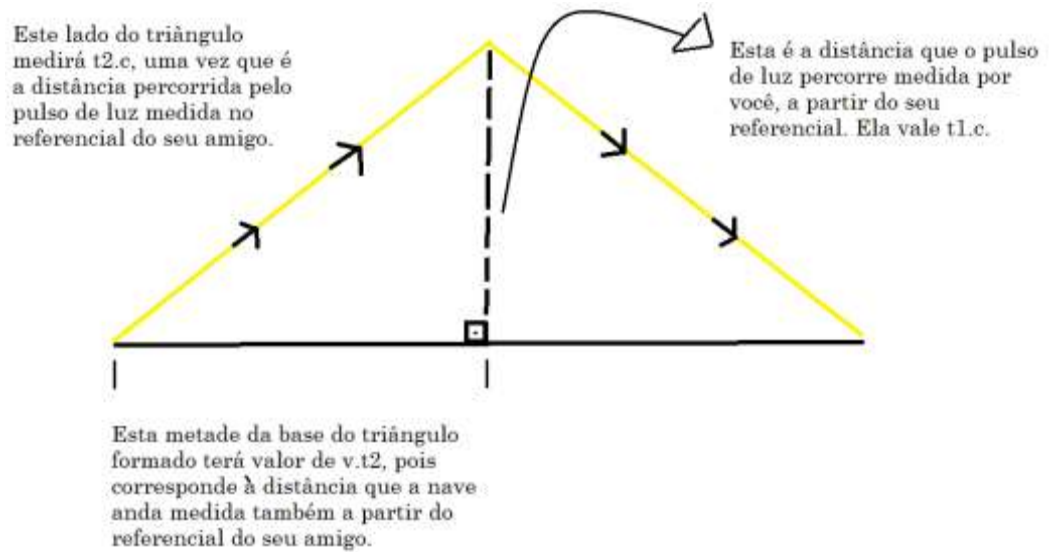

Utilizando o Teorema de Pitágoras para o triângulo retângulo formado na figura acima:

$$t_2^2 \cdot c^2 = v^2 \cdot t_2^2 + t_1^2 \cdot c^2$$

$$t_1^2 \cdot c^2 = t_2^2 c^2 - v^2 \cdot t_2^2$$

$$t_1^2 = \left[t_2^2 \cdot (c^2 - v^2)\right]/c^2$$

$$t_1^2 = t_2^2 \cdot \left(1 - \frac{v^2}{c^2}\right)$$

$$t_2 = \frac{t_1}{\sqrt[2]{1 - v^2/c^2}}$$

Ou:

$$t_2 = \gamma \cdot t_1 \text{, onde } \gamma = 1/\sqrt[2]{1 - v^2/c^2}$$

Para você, dentro da nave, a medição do valor do tempo é t1 (também chamado de tempo próprio). Note como ele difere de t2 (tempo medido a partir do referencial 2) pelo fator γ, conhecido como fator de Lorentz. Esse nome é em homenagem a Hendrick Lorentz, o cientista que encontrou essa relação matemática ao tentar explicar o experimento de Michelson-Morley. Embora ele tenha chegado ao valor primeiro do que Einstein, ele não tinha uma explicação física para o fenômeno. Einstein chegou no mesmo fator tempos depois, juntamente com a sua interpretação física.

Ao analisar o fator de Lorentz, é fácil perceber como ele nunca será menor do que 1 e, assim, no caso do tempo, provocará a sua dilatação (quando γ >1). Como *v* é a velocidade do objeto que está no outro referencial (no exemplo, a velocidade da sua nave em relação ao seu amigo), a fração ($v^2/c^2$) contida em γ, será sempre menor do que 1, devido à velocidade da luz ser muito superior a qualquer outra que conhecemos (não existe velocidade de informação/matéria maior do que *c*,

---

**Caso 1**

*A Fórmula 1 é um esporte conhecido por seus carros ultra potentes e tecnológicos. Os recordes de velocidades chegam muito próximos a 400km/h . Imagine um carro a 400km/h! Pois bem, nosso carro de Fórmula 1 viajará com essa velocidade v= 400km/h (constante) por uma longa pista em linha reta. A bordo dele, o piloto marca o tempo $t_0$. Você está parado em relação ao carro, também marcando o tempo t. Como o tempo t e $t_0$ se relacionam?*

$t_0$ é o tempo próprio, ou seja, o tempo medido por quem está no referencial que se move.
t é tempo medido por você a partir do seu referencial.
**Vimos que:**

$$t = \gamma \cdot t_0$$

**Onde:**

$$\gamma = \frac{1}{\sqrt[2]{1 - \frac{v^2}{c^2}}}$$

**Resolvendo $v^2/c^2$ para v = 400km/h:**

$$\frac{400^2}{300000^2} = 0.0000017$$

**Resolvendo $\gamma$:**

$$\gamma = \frac{1}{1 - 0.0000017} = \frac{1}{0.9999} = 1.0001$$

**Ou seja: $t = 1.0001 t_0$, que significa, por exemplo, quando o piloto marcar 1 segundo no relógio dele, você verá o mesmo relógio marcar 1.0001 segundos. Uma diferença pouco significativa de 0.0001 segundo. Resolvemos minuciosamente apenas para mostrar que o efeito da dilatação temporal acontece, só que a velocidades baixas, que não se comparam a da luz, ele é insignificante para nossa percepção. Perceba que quando resolvemos a fração $v^2/c^2$, por v apresentar um valor menor que c, o resultado foi praticamente 0. Quando isso acontece, usamos o arredondamento "para 0", em outros termos, dizemos que o resultado *tende* a zero. Sempre que for aplicar alguma fórmula que tem o fator de Lorentz, fique atento para esse termo $v^2/c^2$, quando v for muito inferior a c, substitua a fração por zero.**

---

como veremos adiante). Dessa forma, γ nunca possuirá um valor negativo, fazendo com que o t1 (tempo próprio) sofra dilatação. Cabe ressaltar, mais uma vez, que essa dilatação temporal será significativa para nós quando o objeto possuir velocidade *v* relativística (próxima da luz)! Vejamos um exemplo:

---

**Caso 2**
*Estamos em um futuro onde a tecnologia evolui a ponto de sermos capazes de construir naves espaciais que chegam muito próximas à velocidade da luz. Um amigo embarco em uma dessas e viaja à velocidade constante de 87% de c ( 0,87c, ou , aproximadamente, 261.000km/h ). Ele marcará o tempo em um relógio ($t_0$ tempo próprio), e você marcará o seu (t), enquanto fica em repouso em relação ao referencial do seu amigo. Os relógios são idênticos e começam a marcar o tempo sincronizadamente. Como é a relação entre os valores de tempo medidos por vocês?*

**Resolvendo mais diretamente:**

$$t = \gamma \cdot t_0$$

$$\gamma = \frac{1}{\sqrt[2]{1 - \frac{(0,87c)^2}{c^2}}}$$

$$\gamma = \frac{1}{\sqrt[2]{1 - 0,7569}} = \frac{1}{0,4930} = 2,0283$$

**Portanto:**

$$t = 2,0283 \cdot t_0$$

**Quando seu amigo marcar 1 minuto, no seu relógio já se passaram, aproximadamente, 2. Vocês se falam através de um comunicador e, ao ser questionado por quanto tempo já está viajando, seu amigo responde que são 2h. Você olha o seu relógio e ele marca apenas 1h.**

---

Esse é um dos efeitos gerados pela Relatividade Especial. Se você pudesse, **a partir do seu referencial,** observar as coisas acontecendo dentro da nave, veria tudo em câmera lenta, pois o tempo no referencial do seu amigo se *dilatou* em relação ao seu. Mas como seriam as coisas para ele? Ele veria tudo dentro da nave acontecendo lentamente, como em um replay sendo reproduzido em baixa velocidade? Não! Para ele, tudo acontece normalmente e não teria absolutamente nada de anormal. As diferenças só seriam notadas quando fossem **comparadas entre os dois referenciais.**

### 1.2.4. Paradoxo dos gêmeos

Baseados na ideia da dilatação temporal, surge um paradoxo conhecido como "Paradoxo dos gêmeos". Pense em dois irmãos gêmeos univitelinos[36] de 20 anos de idade. Eles decidem fazer um teste utilizando a dilatação temporal da Relatividade Especial. Combinam que, um deles ficará na Terra, enquanto o outro fará uma viagem a bordo de uma nave, com velocidade constante, pelo espaço e retornará tempos depois. Quando se reencontrarem, estarão com a mesma idade ou alguém terá envelhecido mais do que o outro? Antes de responder, lembre-se do primeiro postulado que estudamos! Para o gêmeo que ficou na Terra, a nave do seu irmão está em movimento (com velocidade constante) e ele a vê se afastando do planeta gradativamente, de forma que afirmará que o seu irmão retornará mais jovem, pois ele estava em movimento. Porém, de acordo com o primeiro postulado, é perfeitamente possível o irmão da nave afirmar que quem estará mais novo é o que ficou na Terra, pois, para ele, é o planeta que vai se afastando na mesma velocidade constante, enquanto a nave está em repouso. Meio doido tudo isso, mas seguindo os postulados e leis da teoria (é como um cientista deve agir) são suposições perfeitamente aceitáveis e que não violam nenhum princípio da teoria. No entanto, apresentam resultados que divergem! Isso é um paradoxo! Quem está correto? A menos que prestemos atenção aos detalhes dessa situação, não conseguiremos uma resposta correta. A chave para a solução está em compreender que o gêmeo que está na nave não só *vai*, mas também *retorna* para a Terra. Durante esse processo, existem dois acontecimentos que devem ser levados em conta, pois interferem diretamente em nossa análise. O primeiro é o fato de que para retornar, é imprescindível que a nave sofra aceleração, nem que por alguns poucos instantes, para mudar seu sentido e viajar de volta para a Terra. No instante em que isso ocorre, o referencial muda, e deixa de ser inercial. A medida em que retorna e sua velocidade é constante de novo, ele está em outro referencial de espaço-tempo diferente do que ocupava quando fazia o caminho de ida. Assim, ele, primeiramente, se afasta da Terra, sofre aceleração ao mudar de sentido, e volta em direção ao planeta. Como a velocidade da luz não se altera, independente do movimento uniforme do observador ou da fonte, um fenômeno irá acontecer durante a viagem de ida e volta desse gêmeo.

---

[36] Gêmeos idênticos, gerados a partir de um único óvulo e têm a sua gestação na mesma placenta. Possuem genomas iguais.

Imagine que na Terra ficou um emissor de pulsos luminosos muito potente. Ele emite um pulso a cada 10 minutos segundo o relógio do irmão que ficou no planeta. Na nave, existe um receptor específico que detecta quando um pulso chega até ele. Vamos supor que a velocidade da nave é tal que ela recebe um pulso com o dobro do período de emissão, ou seja, o pulso luminoso chega à nave 20 minutos depois que foi emitido (contados no relógio da nave). De acordo com o **relógio da nave,** a viagem irá durar 2 horas, sendo desprezível o tempo de manobra para virar e retornar. Dessa forma, durante a primeira hora de ida, a nave recebeu 3 pulsos de luz (1 a cada 20 minutos). Quando ela retorna, a situação se inverte, pois, ela passa a receber 1 pulso a cada 5 minutos (sendo 12 nesse tempo). Quando o gêmeo retorna à Terra, ele confere o contador de pulsos e vê que, ao todo, recebeu 15 pulsos de luz. Mas, de acordo com o relógio terrestre, o tempo para emitir 15 pulsos é 2 horas e 30 minutos! Sendo assim, enquanto que a viagem para o gêmeo da nave durou 2 horas, para o seu irmão se passaram 2 horas e 30 minutos e ele é quem será o mais velho dos dois[37].

Para a resolução do "Paradoxo dos gêmeos" é essencial entender que o irmão que permaneceu no planeta Terra se manteve todo o tempo em apenas um referencial. O outro alternou entre dois referenciais, separados pela aceleração (o ato de frear para alterar o movimento) na mudança de sentido ao retornar. Podemos dizer, com outras palavras, que o gêmeo viajante experimentou **duas regiões diferentes do espaço-tempo** e seu irmão, na Terra, apenas uma. Em cada uma dessas regiões, o ritmo do tempo se passou de uma maneira diferente. Incrível!

---

[37] A dilatação temporal acontece em "todos os relógios", sejam eletrônicos, biológicos ou químicos. As células do corpo do gêmeo que ficou na Terra realmente estarão mais envelhecidas do que as do seu irmão.

### 1.2.5. Contração dos comprimentos

Outra consequência muito importante da Relatividade Especial é a contração dos comprimentos (também chamada de contração dos espaços). Ela é expressa na seguinte equação:

$$L = L_0/\gamma$$

Onde, L é o comprimento do objeto que está sendo observado quando ele está se movimentando com velocidade v e $L_0$ é o seu comprimento quando parado. O fator de Lorentz é $\gamma$ (note que agora ele divide $L_0$, provocando diminuição do seu valor). Essa equação mostra a relação matemática entre as medidas de **comprimento** que um observador, em um determinado referencial inercial, faz em relação a um objeto, o qual está em outro referencial inercial em relação ao observador. Quando o objeto está parado em relação ao observador, este mede o seu comprimento como sendo $L_0$. Se o objeto se movimenta com velocidade constante v em relação ao observador, este medirá um comprimento menor L do objeto. Assim, como acontece na dilatação temporal, os resultados dessa contração só serão expressivos para nós quando a velocidade do objeto v for próxima à velocidade da luz. Vejamos um exemplo:

*Exemplo:*
*O comprimento medido de uma nave espacial em repouso é 100 metros. Qual será o seu comprimento medido por um observador em repouso em relação a ela, enquanto ela se movimenta a uma velocidade constante de 87% do valor de c (velocidade da luz)?*
**Resolvendo:**
$$L = L_0/\gamma$$
$$\gamma = \frac{1}{\sqrt[2]{1 - \frac{(0,87c)^2}{c^2}}}$$
**Como já resolvido no Caso 2 da dilatação temporal, $\gamma$ terá resultado igual a 2,0283.**
**Portanto:**
$$L = L_0/2,0283$$
$$L = \frac{100m}{2,0283} = 49,30m$$
**Ou seja, o observador em repouso enxergará a nave como tendo apenas a metade do seu comprimento original quando estava parada.**

Duas ressalvas são importantes nesse assunto de contração dos comprimentos. A primeira é que esse fenômeno acontecerá apenas na **direção** em que o movimento do objeto se desenvolve, ou seja, se ele se move horizontalmente, a contração se dará apenas nessa direção.

A outra diz respeito à *natureza* dessa contração. Não é o comprimento em si do objeto que se contraiu sozinho, mas, na verdade, é o espaço-tempo em que ele está contido que diminuiu. O espaço tempo no referencial do objeto fica reduzido em relação ao espaço-tempo do observador em repouso. Uma analogia pode ser feita se pensarmos em uma criança que desenha um carrinho na superfície de uma bexiga murcha. Depois de desenhá-lo, ela assopra o balão até ele ficar bem cheio e grande. O desenho, naturalmente, também aumentou. *Não foi a criança que o desenhou um carrinho maior*, mas a superfície em que ele estava contido que se distendeu, fazendo-o também aumentar. No caso da contração dos comprimentos, o objeto tem o seu diminuído porque o próprio espaço-tempo se contraiu.

### 1.2.6. Momentum relativístico

A expressão da física clássica para o momentum pode ser escrita como:
$$p = m.v$$

Em que o p é o momentum (quantidade de movimento) do corpo em questão, m a sua massa e v a velocidade.
Também temos a relação entre p e o impulso I no mesmo corpo:
$$I = F.\Delta t = \Delta p$$

Onde F é força necessária, juntamente com o período de tempo $\Delta t$, para causar a variação do momentum $\Delta p$ do corpo em movimento.

Analisando essa equação, podemos supor que, dadas as circunstâncias, aumentando indefinidamente o impulso, podemos aumentar indefinidamente também a variação do momentum, o que significaria o aumento indefinido da velocidade (visto que a massa não sofreria alteração). Dessa forma, atingir velocidades espetaculares, como a da luz, seria uma tarefa apenas de engenharia e muita tecnologia, certo? Na verdade, Einstein mostrou que não é bem assim. O fator de Lorentz também aparecerá na equação do *momentum relativístico $p_r$*:

$$p_r = \gamma.m.v$$

Novamente, as consequências serão mais significativas para nós, quando o objeto estiver desenvolvendo velocidade relativística. Em nosso cotidiano, por exemplo, na engenharia de automóveis e aviões, esse efeito é desprezível.

*A medida que a velocidade de um corpo aumenta, de acordo com a Teoria da Relatividade Especial, seu momentum também aumenta, ficando cada vez mais difícil atingir velocidades superiores.*

## 2. Teoria da Relatividade Geral

### 2.1. Sistemas de referência acelerados e não acelerados

Nos exemplos dos tópicos anteriores, em que usamos trens, carros e naves, sempre atribuímos a eles velocidades constantes durante toda a análise da situação. Isso fazia parte para se determinar *referenciais inerciais*, os quais são essenciais no desenvolvimento dos conceitos da Relatividade Especial. Trabalhando agora as ideias da Relatividade Geral, nossos referenciais vão deixar de ser não-acelerados e passarão a sofrer mudanças de velocidade (referenciais acelerados). Estaremos a bordo de uma nave, viajando pelo sistema solar. A gravidade mínima faz você flutuar pelo compartimento interior. A nave está equipada com um motor de propulsão que, se acionado, a impulsionará no sentido desejado. Nossa viagem está tranquila, a nave não possui janelas e a movimentação está tão suave (velocidade constante e sem perturbações) que você flutua e tem a sensação de que a nave está parada. De repente, o motor é acionado e você é impulsionado contra uma das paredes do compartimento, a qual se transforma *na seu chão*. Você já não está mais voando solto pelo compartimento, mas está de pé, apoiado no que antes era a *parede* do fundo da nave, sentindo o seu peso. A aceleração é tal que você se sente como se estivesse na Terra, com o mesmo peso que tem no planeta. Qualquer objeto que você soltasse das mãos cairia da mesma maneira que outro objeto, em condições iguais de altura, cairia em outra experiência realizada na Terra (desconsideramos possíveis diferenças entre as duas atmosferas). Como eu disse no início que a nave não possuía janelas, como você poderia afirmar que ela, na verdade, não está estacionada na Terra ao invés de estar viajando pelo sistema solar? Einstein nos mostrou, através da Teoria Geral da Relatividade, que não existem diferenças entre os resultados de experimentos realizados em um campo gravitacional newtoniano[38] dos resultados para os mesmos experimentos, caso estes fossem realizados em referenciais acelerados. Um campo gravitacional provoca os mesmos efeitos do que um referencial acelerado, ou seja, gravidade e aceleração são equivalentes.

**2.2. A deflexão da luz**

Um dos fatos que torna a Teoria Geral da Relatividade revolucionária é que os fenômenos descritos anteriormente sobre a gravidade, aplicados a situações mecânicas (objeto caindo), podem ser estendidos a todos os fenômenos de outra natureza, assim como o eletromagnetismo e a óptica. No exemplo anterior, o objeto cai de maneira idêntica nos dois experimentos. Como vimos, no referencial acelerado, foi devido à aceleração do motor, o qual impulsionou o *chão* para cima, que objeto e *chão* se encontraram. Na Terra, foi devido à *força da gravidade* (teoria newtoniana) que o objeto caiu. O que aconteceria se ao invés de um objeto qualquer, estivéssemos falando de um feixe de luz? Seguindo o exemplo da nave acelerando para cima, um feixe de luz que fosse atravessar a sala em que você está, de lado a lado, descreveria uma parábola descendente (o comprimento e/ou velocidade da nave teriam que ser *exageradamente* grandes para você perceber o fenômeno!). À medida que o raio vai avançando, sua distância com o chão que está sendo impulsionado para cima vai diminuindo, formando uma trajetória parabólica. Assumindo que não existem diferenças entre as observações entre um referencial acelerado e um campo gravitacional newtoniano, portanto, o mesmo deve acontecer com a luz se submetida à gravidade.

As constatações experimentais para o fenômeno da deflexão (curvatura) da luz submetida a um campo gravitacional vieram no ano de 1919, durante um eclipse solar. Quando acontece esse fenômeno, a luz solar fica ofuscada no planeta Terra, e se torna mais fácil observar as estrelas que estão posicionadas "atrás" do Sol (a luz solar é tão intensa que causa interferência. Somente com o eclipse, a região do *céu profundo*, adjacente ao Sol, ficaria visível). Dessa forma, seria possível comparar a posição delas neste momento (quando a sua luz passasse perto do sol) com o de quando a trajetória da luz proveniente

---

[38] Campo de força gerado entre um corpo com massa $m_1$, com outro de massa $m_2$. Chamamos $F$ de *força da gravidade* –> $F = G.m_1.m_2r_2$ , em que G é a constante gravitacional de valor $6,74 \times 10^{-11}$ $m^3$ $kg^{-1}$ $s^{-2}$ e r é a distância entre os dois corpos.

delas não sofreria influência alguma da gravidade causada pela grande massa solar. De acordo com a teoria, seria possível observar o fenômeno da deflexão da luz pela gravidade caso ela fosse submetida a um campo gravitacional intenso. O eclipse solar de maio de 1919 representava essas condições ideais – o céu profundo da região das estrelas estudadas ficaria com boa visibilidade e, ao mesmo tempo, a luz proveniente delas passaria perto do Sol. O resultado experimental que comprovaria que a luz sofreu deflexão seria se houvesse uma determinada diferença entre as posições medidas dessas estrelas em dois momentos diferentes, um com o eclipse e em outro, em outra época, em que não houvesse a influência solar. O desenho abaixo mostra esse raciocínio:

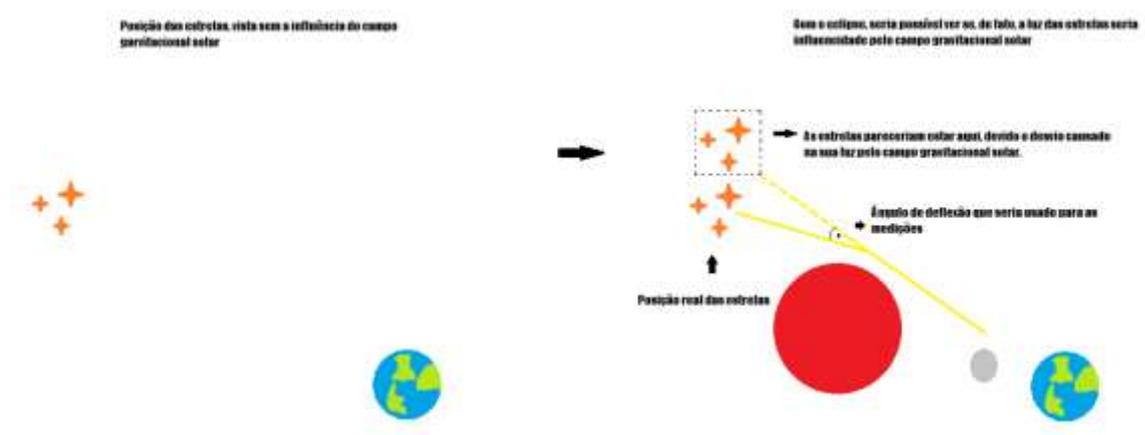

Fonte: do autor

Os cálculos de Einstein previram que aconteceria um desvio de 1,75 segundos de arco (valor do ângulo). As observações aconteceram no dia 29 de maio, nas cidades de Sobral, no Ceará, e na Ilha de Príncipe, no continente africano[39]. Tempos depois, com as fotografias reveladas e descartadas aquelas que não poderiam ser usadas por causa da má qualidade, os cálculos foram feitos e revelaram o mesmo valor encontrado por Einstein, e sustentaram a teoria.

Einstein também forneceu a explicação para o fenômeno. A gravidade não deveria ser interpretada como uma força, mas a própria *curvatura* do espaço-tempo. Qualquer massa *deforma* o espaço-tempo à sua volta, tornando a sua geometria curva. Corpos mais massivos produzem maior curvatura e os efeitos da sua gravidade são mais sentidos do que de corpos menos massivos. No caso do experimento envolvendo o eclipse, podemos entender que a luz se desviou não em razão de uma força que o Sol fez sobre ela, mas porque passou em uma região de *espaço-tempo* curvo, devido à massa solar. Você pode se perguntar, "então porque não é possível ver esse efeito aqui na Terra, já que o planeta *também é muito grande"?* Mesmo que *grande* para nossos padrões, a curvatura que ele cria ao seu redor não se compara com a do Sol, cuja massa é mais de 300.000 vezes superior à da Terra! De fato, a Terra causa desvio na luz, mas é muito pequeno. Você mesmo causa uma deformação do espaço-tempo ao seu redor!

---

[39] Para saber mais dessa história acesse o artigo " Einstein e o eclipse de 1919", no site da Sociedade Brasileira de Física - http://www.sbfisica.org.br/fne/Vol6/Num1/eclipse.pdf

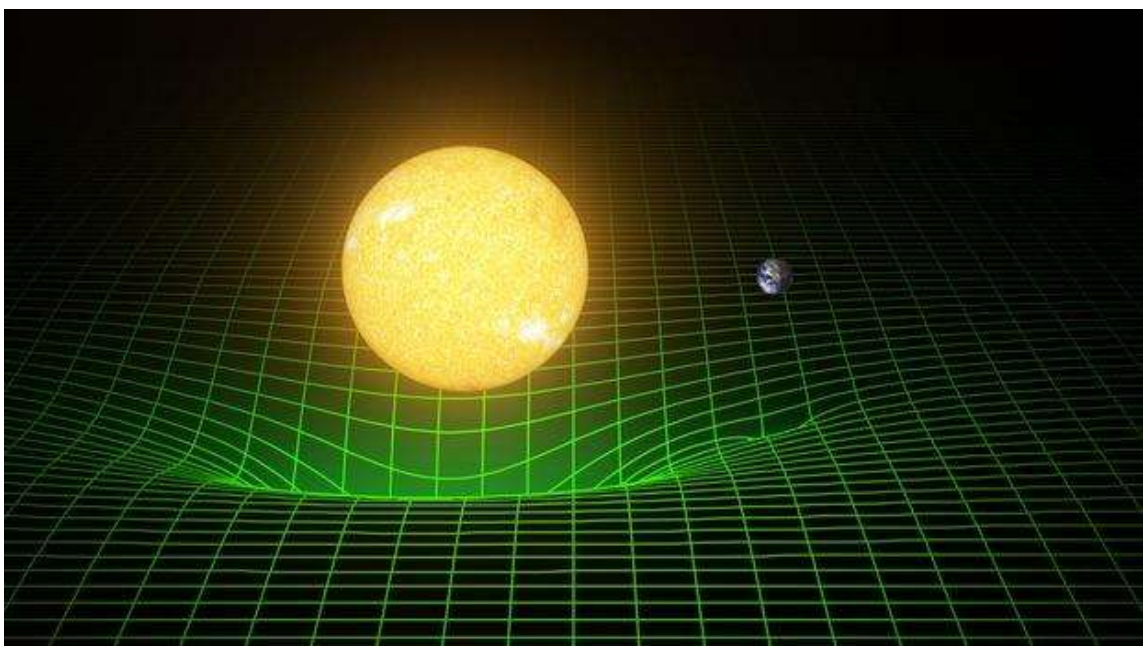

(Fonte:LIGO–Caltech - https://www.ligo.caltech.edu/image/ligo20160211e)

Representação gráfica mostrando o espaço-tempo deformado pela massa solar com a Terra percorrendo a geometria curva, o que resulta na sua órbita.

## 2.3. Desvio gravitacional para o vermelho

Para entendermos os efeitos da gravidade sobre o tempo, previstos pela Teoria Geral da Relatividade, vamos recorrer a um exemplo. Pense em três relógios idênticos e sincronizados no momento do início do experimento, conforme a figura abaixo:

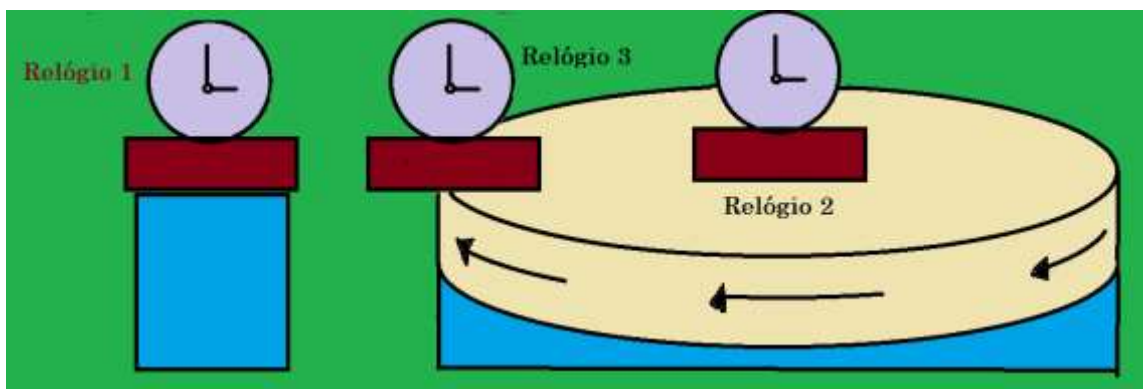

O disco sobre o qual estão os relógios 3 e 2 gira, sendo que o 2 ocupa o centro. Dessa forma, o relógio 3 está em movimento em relação ao 1 mas em repouso em relação ao relógio 2 (a distância entre eles não varia). O relógio 1 também está em repouso em relação ao 2. Um observador em repouso, ao lado de 1, analisa todo o sistema. Como vimos na Relatividade Especial, ele vê os relógios 1 e 2 funcionarem no mesmo ritmo. Já o relógio 3 ele vê funcionar com um ritmo mais lento. Um observador colocado no centro do disco, junto ao relógio 2, começa a observar o sistema e também percebe que o relógio 3 funciona mais lentamente em relação ao 2, embora, entre ele e aquele relógio, não exista movimento relativo. Depois de analisar toda a situação, ele percebe que a única diferença entre o seu referencial e o do relógio 3 é que sobre o relógio 3 age uma força centrifuga[40]. Ao caminhar *no sentido que age a força* sobre o relógio 3, ele nota que o tempo vai se passando mais lentamente naquele relógio. Lembrando que a Relatividade Geral nos diz que não há diferenças observacionais entre estar num referencial acelerado e em um campo gravitacional newtoniano, a conclusão que se pode chegar é que **quando nos deslocamos no sentido que age a força gravitacional, o tempo flui mais lentamente.** Sendo assim, como seria uma situação entre dois irmãos gêmeos idênticos, os quais moram um ao nível do mar e o outro no topo de uma grande montanha? Como é a passagem do tempo para os dois? Algum deles envelhece mais rápido? A resposta é sim! Embora seja uma diferença insignificante, muito menor do que 1 segundo, o tempo passa mais lentamente para o gêmeo que mora ao nível do mar em relação ao seu irmão, pois ele está mais próximo do centro gravitacional do planeta.

---

[40] Em situações de movimento circular, é força que age do centro para fora da trajetória.

O desvio gravitacional da luz para o vermelho é um fenômeno observado que deriva do fato da gravidade desacelerar o tempo. Os átomos são como relógios, sendo a vibração dos elétrons em seu interior como os ponteiros. Quando uma estrela emite luz, ela o faz em frequências específicas (que resultarão nas suas cores). Essas frequências nada mais são do que a frequência de vibração dos elétrons nos seus átomos. Como a gravidade *torna o tempo mais lento*, ela desacelera os elétrons diminuindo a sua frequência de vibração. Uma estrela massiva teria a frequência da sua luz emitida com algum desvio, visto que seu campo gravitacional é intenso. Essa frequência de deslocamento resulta na cor vermelha, a qual possui as frequências mais baixas no espectro de cores[41], por isso o nome *desvio gravitacional da luz para o vermelho*. Os mesmos átomos da estrela na superfície de um planeta, por exemplo, com menos massa, emitiriam luz sem a sua frequência alterada, portanto não desviada para o vermelho. Na década de 1920 o fenômeno já havia sido observado, mas por problemas técnicos e dificuldades a sua comprovação não foi confiável. Na década de 60, um laboratório da universidade de Harvard conseguiu comprovar o efeito[42].

## 2.4. O espaço-tempo e a gravidade

Quando nos referimos a corpos que estão submetidos a campos gravitacionais relativamente fracos entre si, a teoria newtoniana da gravitação é suficiente para realização de cálculos e aplicações de maneira satisfatória[43]. No entanto, para campos gravitacionais intensos, a teoria de Newton não é suficiente e se torna necessária a aplicação da Teoria Geral da Relatividade. Um exemplo disso é o que acontece com o planeta Mercúrio. Por ser o mais próximo do Sol, ele é mais fortemente afetado pelo campo gravitacional do que o restante dos planetas. Um fenômeno que chamava a atenção há tempos era a precessão das órbitas dos planetas. As órbitas são elípticas, tendo o Sol como um dos focos. A precessão é quando o periélio (ponto da órbita em que o planeta passa mais próximo ao sol) é deslocado em alguns graus. Mercúrio tinha uma precessão de sua órbita relativamente grande e os cálculos, utilizando a teoria newtoniana, não eram suficientes. Para adaptar a teoria com a observação, chegaram a supor a existência de um planeta desconhecido nas redondezas de Mercúrio para tentar explicar as incoerências dos cálculos[44]. Einstein aplicou sua teoria ao problema e, realizando os cálculos, encontrou o valor exato que correspondia à observação do fenômeno (Mercúrio tinha uma precessão de 42 segundos de arco a cada século). Esse também é um dos fatos que comprovam experimentalmente a Teoria da Relatividade Geral.

---

[41] O espectro de cores da menor frequência para a maior – vermelho, laranja, amarelo, verde, azul e violeta.

[42] Ver em *Física Conceitual – Paul Hewwit, 11ª edição, p.660.*

[43] Como curiosidade, pesquise como é a física de satélites e outros veículos espaciais. Você verá que a física newtoniana é a disciplina fundamental dessas aplicações.

[44] Veja aqui a incrível história de Vulcano - https://g1.globo.com/ciencia-e-saude/noticia/vulcano-o-planeta-procurado-por-mais-de-meio-seculo-e-que-einstein-expulsou-do-ceu.ghtml

Portanto, devemos entender a gravidade como sendo o efeito de um espaço-tempo curvo devido à presença de matéria. Os planetas orbitam o sol não em função de uma *força* que os atrai, mas por estarem em um espaço-tempo que foi *deformado* pela massa solar. Sua órbita nada mais é do que seu caminho por um plano inclinado curvo. Toda massa deforma o espaço ao seu redor. Ao seu redor o espaço também está deformado, mas não suficiente para que algum objeto *o orbite.*

Apenas como curiosidade e motivação, colocaremos aqui a equação da gravitação de Eintein. Não se preocupe em entende-la na forma que está aí, pois ela utiliza partes mais avançadas da linguagem matemática, as quais não fazem parte da nossa abordagem:

$$\underbrace{R_{\mu\nu}}_{\#1} - \frac{1}{2}\underbrace{g_{\mu\nu}}_{\#2}\underbrace{R}_{\#3} + g_{\mu\nu}\underbrace{\Lambda}_{\#4} = \frac{8\pi G}{c^4}\underbrace{T_{\mu\nu}}_{\#5}$$

#1 = Tensor de Curvatura de Ricci

#2 = Tensor Métrico ("métrica")

#3 = Tensor Curvatura Escalar

#4 = Constante Cosmológica

#5 = Tensor Momento-Energia

Conceitualmente, os termos que estão à esquerda da equação (Tensores de Ricci, Métrico e de curvatura escalar) mais a constante cosmológica (será muito importante no contexto da expansão do Universo, a qual veremos mais à frente) estão relacionados à geometria do espaço-tempo. A parte da direita, a qual contém o tensor momento-energia, como o próprio nome diz, está relacionada a energia-massa (são equivalentes na Teoria Geral da Relatividade) existente no universo. Nessa equação, fica claro a ligação entre a energia-massa e a geometria, sendo que uma está sobre o efeito da outra. Como dito, essa parte é apenas uma curiosidade, não se preocupe em entendê-la a fundo nesse momento.

## 2.5. Ondas gravitacionais

Em setembro de 2015 o Observatório de ondas Gravitacionais por Interferômetro (Laser Interferometer Gravitational-Wave Observatory - LIGO), com laboratórios nos estados de Lousiana e Washington, nos Estados Unidos, detectou outras evidências observacionais da Teoria Geral da Relatividade[45]. Os dos fenômenos previstos por Einstein, era o das ondas gravitacionais. Essas ondas se formam quando uma massa se movimenta ou gira pelo espaço-tempo, criando perturbações que se propagam através dele, igual a ondas causadas em um lago, quando uma pedra é jogada em suas águas. Ao invés da água, o meio de transporte das ondas gravitacionais é o próprio espaço-tempo. No entanto, elas são muito difíceis de serem detectadas, por serem fenômenos de baixa intensidade (valores com magnitude baixa) e também por ser muito fácil a interferência de quaisquer outros movimentos nos equipamentos de detecção. Qualquer vibração mínima seria uma perturbação significativa nos equipamentos, e invalidaria o experimento. Cem anos após a publicação da Teoria Geral da Relatividade, o projeto LIGO, através de uma tecnologia muito desenvolvida e um trabalho de engenharia preciso, conseguiu detectar ondas gravitacionais provenientes da fusão de dois buracos negros[46], o que aconteceu há mais de 1,2 bilhões de anos-luz[47] de distância.

---

[45] Acesse o site da Sociedade Brasileira de Física para mais detalhes em http://www.sbfisica.org.br/v1/index.php?option=com_content&view=article&id=964:2017-10-05-17-17-18&catid=151:destaque-em-fisica&Itemid=315

3. **Considerações finais**

A Teoria Geral e Especial da Relatividade é um marco na história científica da humanidade. A partir dela, muito conhecimento foi gerado e nossa compreensão acerca do universo aumentou. A cosmologia moderna de que estudaremos é totalmente dependente dela.

Outro ponto a ser comentado é sobre a teoria newtoniana. Muitos que veem, pela primeira vez, a teoria de Einstein, tomam uma iniciativa precipitada de avaliar a física de Newton como errada. Não é o caso. Como discutimos no capítulo "*Algumas palavras sobre ciência*", o conhecimento científico é algo construído por geração em geração, num contínuo processo de aperfeiçoamento. A mecânica newtoniana se torna *um caso especial* da Teoria Especial da Relatividade para baixas velocidades. Vejamos um exemplo:

A relação entre tempos em referenciais inerciais, na Relatividade Especial, é dada por:

$$t = \frac{t_0}{\sqrt[2]{1-v^2/c^2}} = \gamma t_0$$

Mas, como visto, quando v é muito inferior ao valor de c (situações do nosso cotidiano[48]), temos que $\gamma = 1$. Dessa forma:

$$t = t_0$$

Ou seja, os tempos entre os referenciais são iguais, tal como é na mecânica newtoniana. Faça o mesmo processo acima para as equações de comprimento e momentum.

**Para aprofundar mais:**

- ***Einstein, sua vida, seu universo*** – Walter Isaacson, Companhia das Letras, 2007.

- ***Genius*** – Série de televisão Da *National Geographic* sobre a vida Albert Einstein.

---

[46] Buracos negros são assuntos do tema de Teoria da Relatividade Geral mas serão tratados em um capítulo posterior específico.

[47] Unidade de distância da luz viajando, pelo vácuo, durante um ano. Em quilômetros, 1 ano-luz equivale a 9.460.730.472.580,8 km.

[48] Estamos no ano de 2020 e ainda não existem meios de transporte que viagem a velocidades relativísticas e as viagens espaciais não são comuns para nós. Quando assim for, lidaremos com a Relatividade rotineiramente. Um exemplo de como uma tecnologia atual é afetada pela Relatividade Especial vem dos Sistema de Posicionamento Global (GPS, em inglês). Veja este artigo da Revista Exame -https://exame.com/ciencia/sem-teoria-da-relatividade-de-einstein-gps-nao-existiria/

**Perguntas para fixação**

**a) Para um determinado sistema de referência, uma explosão x aconteceu primeiro do que outra y. É possível que para outro sistema de referência, y aconteça primeiro do x? Explique.**

*Sim. Dependendo do sentido de movimentação do segundo referencial da pergunta em relação ao primeiro referencial, o qual viu x acontecer primeiro do que y, é possível que para esse segundo referencial y aconteça antes do que x, como mostra a Teoria Especial da Relatividade (rever exemplo do trem no tópico que trata desse assunto).*

**b) Muitos equipamentos utilizados na topografia (estudo e descrição de terrenos geográficos) são baseados na tecnologia da luz. Eles direcionam feixes de luz em linha reta para fazer medidas da topografia de uma determinada área. Mas, é correto dizer que a luz viaja em linha reta? Podemos confiar nas medidas desses equipamentos?**

*Como vimos, a gravidade afeta também a luz, curvando-a. Como existe gravidade em todo planeta Terra, a luz é afetada por ela em toda a sua extensão. Porém, como também vimos, para que essa deflexão seja realmente significativa para nós, é necessário um campo gravitacional extremamente intenso para que isso ocorra. O exemplo do eclipse de Sobral mostrou bem esse fato. Como o campo gravitacional da Terra não é suficientemente forte para curvar a luz **de uma maneira que seja significativa para nós, aqui na Terra,** esse efeito não interfere no nosso cotidiano a ponto de causar nos causar problemas, como, por exemplo, nas medidas dos equipamentos que utilizam a luz em topografia.*

<p style="text-align:center"><span style="color:#4a86e8">**Capítulo 5**</span></p>

# Conceitos básicos da Teoria Quântica

*Este é um capítulo de apoio, contendo alguns princípios básicos da Teoria Quântica. Pode ser trabalhado diretamente com os alunos. O professor deve ficar atento às notas de rodapé, pois elas contêm muitas sugestões de material complementar às explicações contidas no capítulo. A teoria quântica é apresentada apenas em conceitos básicos, no entanto indispensáveis à sua iniciação no aprendizado.*

**Introdução**

No capítulo anterior, sobre Teoria Especial e Geral da Relatividade, vimos que as consequências mais significativas dela aconteciam quanto maiores fossem as velocidades e/ou mais massivos os corpos envolvidos. De forma geral, ela é uma teoria que lida diretamente com a estrutura macroscópica do universo, por exemplo, os campos gravitacionais de corpos celestes como estrelas, planetas, buracos negros e outras estruturas massivas. Porém, para se ter um entendimento mais profundo da Cosmologia Moderna, é necessário também que aprendamos como matéria e energia são e se relacionam em escalas muito pequenas. A teoria quântica cuida desses assuntos. Ela lidará com a estrutura da matéria, principalmente, em escala atômica, subatômica e molecular[49]. Estaremos interessados em desvendar as leis que regem estruturas muito menores que a espessura de um fio de cabelo e, no entanto, são cruciais para interpretamos como de fato o Universo é.

**A natureza ondulatória da luz e o experimento da dupla fenda**

Para a física clássica, um dos grandes desafios era lidar com a natureza da luz[50]. Alguns físicos, como Newton, a consideravam como um conjunto de partículas. Outros, como Huygens, atribuíam o caráter ondulatório à luz. Foi no ano de 1801 que o físico Thomas Young realizou um experimento, o qual ficou conhecido como "experimento da dupla fenda", onde ele conseguiu demonstrar que a luz se comportava como onda[51], analisando o que ocorria quando um feixe de luz passava por uma espécie de parede com duas fendas e incidia sobre outra sem fendas, posicionada logo atrás dela. Basicamente, se a luz fosse composta por partículas, era esperado que surgissem pontos luminosos apenas na direção das fendas, indicando que somente as partículas que passaram por elas conseguiram chegar até a parede posterior, e as demais ficaram impedidas pela região sem fendas. No entanto, o padrão encontrado por Young foi de várias "franjas" luminosas alternadas com franjas escuras, indicando que a luz, ao passar pelas fendas, sofreu o fenômeno de difração e interferência (fenômenos característicos de ondas), resultando naquele padrão de reflexão apresentado (consultar o quadro explicativo "Difração e interferência de ondas" nas páginas posteriores). No mesmo século, James Clark Maxwell introduziu o conceito de "onda eletromagnética" que classificava a luz como uma radiação transportando energia. Heinrich Hertz, um físico alemão, demonstrou experimentalmente a existência da radiação eletromagnética através da construção de emissores e receptores de ondas de rádio[52].

---

[49] O tamanho médio dos átomos é $10^{-10}$m. Em um milímetro é possível alinhar, aproximadamente, 10 milhões de átomos. A escala subatômica é menor do que a atômica, e abrange subpartículas como elétrons, prótons e nêutrons dentre outros tipos.

[50] Consultar " Física Conceitual ", de Paul Hewwit, 11ª edição, p. 554.

[51] Acessar o site da Universidade Federal do Rio Grande do Sul para mais informações em: http://www.if.ufrgs.br/historia/young.html.

[52] O experimento de Hertz pode ser visto aqui: https://sites.ifi.unicamp.br/lunazzi/files/2014/03/FredericoC-Monica_RF2.pdf

**A origem da Teoria Quântica e a radiação do corpo negro**

No final do século XIX, havia um problema envolvendo a radiação emitida por um corpo negro[53] (alguns exemplos do nosso cotidiano se aproximam do conceito de corpo negro tais como a lâmpada de filamento e o carvão aquecido) e a elaboração de um modelo matemático que descrevesse esse processo corretamente. Os cientistas não conseguiam uma solução que correspondesse ao que era observado experimentalmente, ou seja, as teorias clássicas existentes não conseguiam propor uma explicação que concordasse inteiramente com o fenômeno[54]. O corpo negro, quando aquecido, absorve toda radiação incidente sobre ele e também a emite em sua totalidade. Ao se aquecer, sua coloração vai mudando conforme a temperatura aumenta. Pela teoria clássica, se acreditava que quanto mais se aquecesse um corpo negro, maiores seriam as frequências de emissão da radiação. Porém, um problema acorria quando, basicamente, a observação do fenômeno mostrava que essa emissão acontecia em apenas algumas frequências específicas. Os valores medidos eram diferentes dos valores do modelo matemático construído utilizando as teorias clássicas. Apenas para valores de ondas de baixas frequências os modelos clássicos funcionavam, ficando o restante dos dados em total desacordo. Foi Max Planck quem propôs o modelo que correspondia aos dados empíricos e explicava o fenômeno da radiação do corpo negro[55]. As abordagens clássicas assumiam que a energia (radiação) existia em qualquer frequência, sendo assim um fenômeno contínuo. Planck introduziu a hipótese que, na verdade, a energia só poderia ser armazenada na matéria em "pequenos pacotes", quantidades discretas (em oposição aos valores contínuos das teorias anteriores). Pense em um muro muito extenso. De longe ele parece ser contínuo, sem interrupções ou divisões em seu comprimento. No entanto, ao nos aproximarmos, vemos que ele é constituído de tijolos, os quais são as suas pequenas *unidades elementares*. A hipótese de Planck (em 1901) propunha que a energia armazenada na matéria só existiria sob a forma de unidades elementares, as quais ele chamou de *quantum*. A lei de Planck foi descrita matematicamente como:

$$E = h.f$$

Ela expressa o quanto de energia possui um *quantum* na frequência *f* de vibração. Chamamos o termo $h$[56] de constante de Planck. É importante destacar que a **energia** será **diretamente proporcional à *frequência***. A energia de um corpo era um múltiplo de *h!* A Teoria Quântica nasceu com os estudos de Max Planck e se desenvolveu com a participação de outros importantes físicos.

---

[53] Detalhes sobre a radiação do corpo negro, radiação térmica e o experimento no artigo "Radiação de Corpo Negro" de Mônica Bahiana, Universidade Federal do Rio de Janeiro – UFRJ – disponível em https://www.if.ufrj.br/~marta/cederj/quanta/mq-unid2-textocompl-1.pdf

[54] Mais sobre o problema do corpo negro em

[55] HUSSEIN, Mahir S.; SALINAS, Sílvio Roberto Azevedo. 100 anos de física quântica. Editora Livraria da Fisica, 2002, p.1.

[56] *h*= 6,6260 x 10⁻³⁴m⁻²Kg/s. Observe que a constante de Planck tem um valor muito pequeno.

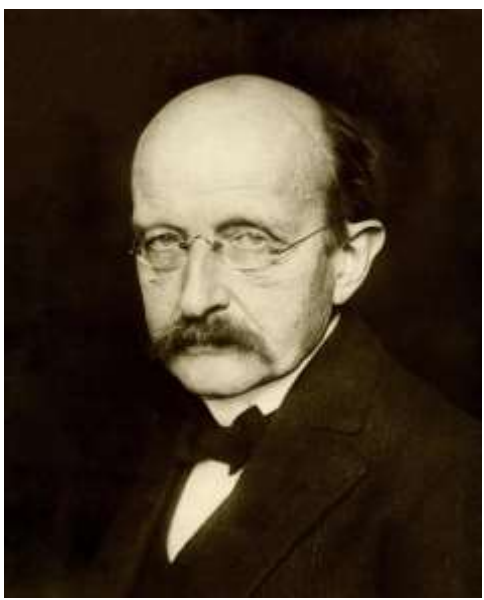

*Max Planck        Fonte:Wikipedia*

**Efeito fotoelétrico**

Pouco tempo depois da publicação da Planck, o jovem Albert Einstein estendeu a ideia do quantum para a luz, baseado no efeito fotoelétrico. O efeito fotoelétrico consiste na emissão de elétrons de uma superfície metálica, negativamente carregada, quando uma luz incide sobre ela. Até 1905, o fenômeno não era explicado satisfatoriamente, pois as abordagens que consideravam a luz como sendo uma onda não conseguiam encontrar a solução para algumas peculiaridades que aconteciam no efeito[57]. As principais eram que não existia um intervalo de tempo entre a incidência da luz na placa e a emissão do elétron, sendo um efeito que ocorria quase que instantaneamente. Outro efeito não explicado era que, independente da frequência ou intensidade da luz incidida, os elétrons eram emitidos sempre a partir do mesmo instante de tempo e, caso a luz possuísse uma frequência baixa (por exemplo, a luz vermelha) o fenômeno de emissão não ocorria. A quantidade de elétrons ejetados em um determinado intervalo de tempo era proporcional à intensidade da luz. Ao considerar a luz como também sendo composta por *pequenos pacotes* de energia (ficariam conhecidos por *fótons,* nome proposto por Einstein) Einstein conferiu o caráter de *conjunto de partículas* à natureza da luz, as quais obedeciam à lei de Planck. Basicamente, a sua explicação para esse fenômeno era que a luz, no momento em que interagia com a placa metálica carregada, tinha o comportamento de partícula, ou seja, funcionava como uma saraivada de pequenas partículas (fótons) bombardeando a placa. Os elétrons da placa, por sua vez, absorviam esses fótons imediatamente ao contato com eles e, à medida que mais fótons eram absorvidos, mais esses elétrons se excitavam, culminando em um limite de excitação (energia) que, quando ultrapassado, o elétron se *soltava* da placa e era ejetado. O fato de os elétrons absorverem instantaneamente os fótons, resolvia a questão de que não havia um tempo de retardo entre a incidência da luz na placa e a emissão do elétron. Depois de onze anos, a comprovação experimental para a explicação de Einstein veio do físico Robert Milikan, o qual ganhou o prêmio Nobel por esse feito (curiosamente, ele queria demonstrar que Einstein estava errado). A relação diretamente proporcional entre a energia do fóton e a sua frequência foi evidenciada. Uma curiosidade sobre Einstein é que ele ganhou o prêmio Nobel em 1921 pelas suas contribuições à explicação do efeito fotoelétrico e não pela Teoria Especial e Geral da Relatividade.

---

[57] Consultar "Física Conceitual", de Paul Hewwit, 11ªedição, p. 557.

**Dualidade na natureza da luz: ora onda, ora partícula.**

O experimento da dupla fenda de Thomas Young revelou que a luz se comportava como uma onda, sofrendo os processos de difração e interferência. Maxwell e Hertz, anos depois, contribuíram com essa ideia através do conceito e da demonstração da radiação eletromagnética.

No entanto a explicação do efeito fotoelétrico mostrava características corpusculares da luz (partículas), sendo ela composta de minúsculas partículas (características comuns às partículas são massa e velocidade, seu produto é o momentum) energéticas que viriam a ser conhecidas por fótons.

Como é o comportamento da luz, então? Onda ou partícula? Onda e partícula ao mesmo tempo! Assim, dizemos que a natureza da luz é dual, apresentando em certos momentos comportamentos de onda e em outros de partícula. Os fótons, ao serem emitidos por uma fonte ou absorvidos pelo receptor, desenvolvem o comportamento de partícula. Já enquanto estão fazendo esse caminho fonte-receptor, se propagam como ondas.

**A teoria de Louis de Broglie**

Louis de Broglie foi um físico francês que também contribui para o desenvolvimento da Teoria Quântica. Seu principal trabalho, com o qual ganhou o prêmio Nobel de Física em 1929, foi expandir o conceito da dualidade onda-partícula do fóton para o elétron. Ele chegou à seguinte equação:

$$\lambda = \frac{h}{p}$$

Onde $\lambda$ é o comprimento da onda gerada pelo corpo enquanto esse tem o momentum p. O termo *h* é a constante de Planck. Essa equação mostra que qualquer corpo dotado de momentum possui uma função de onda associado a ele. Eu, você, seu cachorro, um carro ou um pássaro, nos movimentando, o fazemos *como uma onda*. Um avião *Boeing 747*, um dos maiores existentes no mundo, o qual pode viajar pesando até 440.000kg, e atingir 900km/h, sob essas condições, teria uma onda relacionada ao seu momentum de 6,02 x $10^{-42}$m. Assim, não experimentamos essa dualidade com corpos do nosso dia-a-dia, pois as ondas geradas têm grandeza insignificante para nós. O termo momentum (m.v) está no denominador e *h* é um valor muito pequeno comparado a eles, de forma que $\lambda$ possui um valor muito pequeno.

**O princípio da incerteza**

O princípio da incerteza é um dos principais conceitos da Teoria Quântica. Ele foi proposto pelo físico alemão Werner Heisenberg, o qual também ganhou o Nobel de Física em 1932. Esse

> **Difração e Interferência de ondas**
>
> O fenômeno da difração pode ser entendido como sendo o que acontece quando uma onda passa por um orifício de tamanho semelhante ao seu comprimento. Aquele ponto se torna um *novo emissor* de onde, de onde surgem novas ondas que se propagam.
>
> Já a interferência consiste quando mais de uma onda se sobrepõem umas às outras causando ou interferência construtiva ou destrutiva. Se for construtiva, as ondas estão em fase (picos e vales estão sincronizados) e suas amplitudes se somam, formando uma nova onda de amplitude maior. Já a destrutiva é o contrário, as ondas defasadas se somam, porém, diminuindo a amplitude e podendo até mesmo se anularem completamente.

princípio diz que, a nível quântico, quanto mais conhecemos a posição de uma partícula, menos conhecemos o seu momentum e vice-versa. A sua equação é:

$$\Delta x . \Delta p \geq \frac{h}{2\pi}$$

Onde, $\Delta x$ é a incerteza, relacionada à posição medida da partícula, e $\Delta p$ está relacionado ao seu momento. Como $h$ e $2\pi$ são valores constantes, quanto mais diminuímos uma das incertezas, mais a outra aumenta. O ato de medir uma dessa grandezas, nesse contexto, altera o seu próprio valor. Pense, se quisermos medir a posição de um fóton, em um determinado momento, precisamos incidir luz sobre ele. Ao fazer isso, esse fóton estará sujeito a colisões com outros fótons (provenientes da luz), o que altera sua velocidade e, portanto, o momentum. O princípio da incerteza de Heisenberg nos mostra que existe um limite para a exatidão das medidas no mundo quântico.

É preciso ressaltar que, esse princípio é válido apenas para o nível quântico (por exemplo, fótons e elétrons dentre outras partículas). A partir daí ele já não pode ser aplicado. As incertezas das medidas que, por exemplo, a engenharia trabalha, nada têm a ver com o princípio da incerteza de Heisenberg. Elas são causadas ou por limitações dos instrumentos de medida ou por erros esperados da pessoa que executa o processo. O mundo quântico deve ser entendido em seu contexto e muito das suas leis não se aplicam à matéria e escala maior.

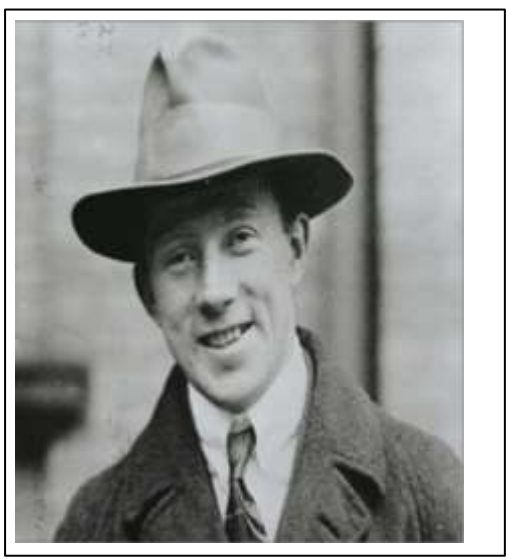

Fonte: Wikipedia

Werner Heisenberg (1901 - 1976) nasceu em Wurtzburgo, na Alemanha. Desde muito jovem era um ávido leitor de filosofia e ciências, sendo Platão uma grande influência. Estudou física e matemática na Universidade Munique. Foi um físico prolífico, tendo se dedicado a desenvolver as bases da Mecânica Quântica. Ganhou o prêmio Nobel de física em 1932.

 **"Um especialista é alguém que sabe quais os piores erros que podem ser cometidos na sua área e os evita."**

**Comentários Finais**

Este breve capítulo da Teoria Quântica serviu para introduzir os conceitos mais básicos do grande e complexo corpo de conhecimento que o assunto abrange. O que muito resta foge ao propósito da apresentação desta teoria aos alunos do ensino médio como parte suplementar a um curso de introdução à Cosmologia Moderna.

**Referências Bibliográficas:**

HEWITT, P. **FÍSICA Conceitual**. 12ª edição, p.553-570, 2011.

**Para aprender mais:**

*- Uma breve história do tempo – Stephen Hawking, capítulo "O princípio da incerteza".*

*- A parte e o todo – Werner Heisenberg.*

**Perguntas de fixação:**

a) ***Se formos medir a temperatura de uma água quente, contida em um recipiente, colocando um termômetro dentro dela, certamente estaremos interferindo na medida, pois, se o termômetro estiver mais frio que a água, ele causará uma pequena mudança na temperatura original da água. Isso tem a ver com o princípio da incerteza de Heisenberg? Explique.***
*Não. Embora o fenômeno de interferir na temperatura da água quente do exemplo com o termômetro seja um fenômeno que também ocorrerá no nível microscópico da matéria, essa interferência não será significativa a ponto de conferir tal incerteza à medida de temperatura. É importante lembrar que a temperatura é medida do grau de agitação dos átomos e moléculas que formam um determinado corpo, sendo que quanto maior essa agitação, maior é a temperatura. Como vimos, o princípio de incerteza de Heisenberg é significativo para escalas ainda menores do que átomos e moléculas e, dessa forma, ele não interfere nessa medida.*

b) ***Qual cor de luz possui maior energia: vermelha (f = 480THz) ou violeta (f=789THz)? Justifique.***
*Para resolver essa questão, não é nem necessário fazer conta, uma vez que não se pede o valor da energia, mas sim qual dos dois é o maior. Devemos lembrar que aprendemos na aula como a **energia de um feixe de luz é diretamente proporcional à frequência dessa luz**. Como a frequência da luz violeta é superior à frequência da vermelho, concluímos que a cor violeta é mais energética que a vermelha. Se quiserem calcular o valor, basta aplicarem a fórmula já discutida no capítulo $(E = h.f)$.*

<p align="center" style="color:blue">**Capítulo 6**</p>

# A expansão do universo, um cocheiro e um atleta – uma grande história científica (pouco provável) do século XX

*Atenção, professores! A leitura desse capítulo é recomendada, primeiramente, a vocês, pois está escrito numa linguagem acadêmica do ensino superior, na modalidade de artigo.* Ele servirá de base para que montem suas aulas sobre a expansão do universo. Caso queiram trabalhar essa leitura diretamente com seus alunos, recomendamos que os auxiliem durante todo o processo e façam a leitura juntos. As biografias de Edwin Hubble e Milton Humason[58] podem constituir uma boa maneira de inspirar e motivá-los para a ciência, principalmente os fatos da vida de Milton Humason, o qual iniciou seu trabalho no Observatório do Monte Wilson como um guia de mulas e, através do seu interesse e esforço, participou ativamente dos estudos conduzidos por Hubble sobre as distâncias e velocidades das galáxias. Outra sugestão é que os professores de história e/ou geografia podem auxiliar na introdução do texto, uma vez que ela trata do cenário geopolítico do início do século XX como um imoportante pano de fundo para o nascimento da Cosmologia Moderna.*

## RESUMO

O modelo de expansão acelerada do Universo, tal como conhecido hoje, é fruto de uma história de pesquisas e desenvolvimentos, tanto no campo teórico quanto prático da física e astronomia. Representa uma das maiores descobertas científicas, não só do século XX, mas de toda a história da humanidade, e também uma das primeiras da Cosmologia Moderna. Ao longo de aproximadamente 30 anos, teorias e modelos do universo, propostas por físicos e cosmólogos, juntamente com observações astronômicas, andaram lado a lado, ora em sintonia, ora conflitantes, guiando-os no caminho em busca da compreensão de como funciona parte da dinâmica do Cosmos em larga escala. **Dinâmica, essa, negada durante séculos**, inclusive por Einstein, mas que começou a ser aceita a partir de uma publicação do matemático e cosmólogo russo Aleksander Friedmann, em 1922. Em 1929, Hubble e Humason realizam observações que seriam determinantes para as conclusões sobre a expansão do universo[59]. O presente texto tem como tema principal contar um pouco da história compreendida em todo esse período, descrevendo suas principais ideias e acontecimentos, dando destaque para o principal personagem, Edmund Hubble.

---

[58] Existem duas biografias recomendadas. Uma, mais atual, de Milton Humason, ainda sem tradução para o português, chamada "The Muleskinner and the Stars" (VOLLER, 2016) escrita por Ronald Voller. Outra, sobre Hubble, escrita em 2003, pelo astrônomo e astrofísico brasileiro Augusto Damineli, com o nome de "Hubble e a expansão do universo" (DAMINELI, 2003).

[59] Como dito, as observações de ambos os astrônomos seriam cruciais para o modelo de um universo não-estacionário (dinâmico), ou seja, não estamos afirmando serem os mesmos os responsáveis por tal proposta de modelo. Inclusive Hubble, como um ótimo "experimentador", foi muito cauteloso ao publicar seus resultados e nunca defendeu publicamente a ideia de um universo que se expandia (consultar "Cosmologia e Ensino da Ciência: Problemas e Promessas", de Helge Kragh, p.5). A ciência é um processo de pesquisa contínua, que se renova sob a dúvida e trabalho das gerações posteriores a uma ideia ou teoria. Ressaltar aos alunos esses conceitos, como também a importância de Hubble e Humason para todo o processo.

**1. Introdução – Contextualização da época**

O século XX foi marcado por profundas mudanças políticas, sociais e econômicas. Em um cenário turbulento de guerras e conflitos de âmbito mundial, já no seu início, esse novo século mostrava o caráter de ruptura com o passado e transformações radicais, nas mais variadas esferas das organizações humanas. A primeira Guerra Mundial estouraria nos anos 10; as transmissões de rádio e telégrafo diminuíram as distâncias e apresentaram novos mundos para as pessoas; a Revolução Russa pretendia distribuir riquezas e poderes entre os proletários; e os carros inavdiram as cidades que se tornaram maiores e mais presentes. Nas palavras do historiador Eric Hobsbawn, " [...] para os que cresceram antes de 1914, o contraste foi tão impressionante que muitos – inclusive a geração dos pais deste historiador, ou pelo menos de seus membros centro-europeus – se recusaram a ver qualquer continuidade com o passado. " (HOBSBAWN,1995, p.30). No campo científico não foi diferente. Com a publicação da Teoria da Relatividade Especial (1905), Geral (1915) e da Teoria Quântica nos meados da década de 20, o mundo científico sofreu um grande impacto ao ver a Física Clássica, proposta por Isaac Newton, três séculos antes, ser confrontada por essas novas teorias. No domínio quântico, ou o chamado infinitamente pequeno, as medições simultâneas de *momentum* e posição, antes bem determinadas pelas equações clássicas quando aplicadas a corpos de maior dimensão, mostravam-se falhas nas escalas microscópicas devido ao Princípio da Incerteza, proposto por Werner Heisenberg em 1927 (SILVA, 2014). Nele, ao se aferir a posição de uma partícula, não é possível medir, simultaneamente e com precisão, o seu *mometum* (produto da velocidade pela massa) e posição, e vice-versa. Já no contexto do macrocosmo, o infinitamente grande, a teoria clássica era posta à prova e refletida por outro ponto de vista através da Teoria da Relatividade de Albert Einstein. Na Teoria Especial, a noção de espaço e tempo, separados e únicos, dá lugar a uma relação espaço-tempo inseparável, onde o tempo e as dimensões dos corpos podem ser relativos para os observadores, dependendo dos seus referenciais. Nas palavras do próprio Einstein, sobre a relatividade galileana, base para a física clássica, a qual era vigente em seu tempo, "[...] Enquanto todos estavam convencidos de que os fenômenos da natureza podiam ser representados com auxílio da mecânica clássica, a validade deste princípio da relatividade nunca foi posta em dúvida. Mas, os novos desenvolvimentos da eletrodinâmica e da óptica foram tornando cada vez mais claro que a mecânica clássica era uma base insuficiente para a descrição de todos os fenômenos físicos. Com isto, também passou a ser discutida a questão da validade do princípio da relatividade [...]" (EINSTEIN, 1916, p.19). Partindo dos alicerces sobre a constância da velocidade luz e a validade das leis física para quaisquer observadores no universo, ele postulou, dentre outras leis, que a gravidade, na verdade, era consequência da curvatura do espaço, causada pela existência de massa que deformava o "tecido espaço-tempo" ao seu redor (STACHEL,2001). Dessa forma, por exemplo, o movimento dos planetas ao redor do sol não deveria mais ser entendido como em função de uma força, assim descrita por Newton, mas devido a curvatura criada pela grande massa do sol no centro do sistema solar, fazendo com que os planetas o orbitassem dentro da porção de espaço tempo deformada por ele mesmo. Era sob essas condições que o mundo científico estava no início do século XX. Esse clima de surgimento de novas ideias e teorias científicas, instabilidade política e desenvolvimento tecnológico foi o contexto para outra grande revolução envolvendo o conhecimento da dinâmica e estrutura do Universo. .

**2. Hubble e o descobrimento das Galáxias**

Edwin Powel Hubble nasceu no dia 20 de novembro de 1889, em Marshfield no estado de Missouri nos Estados Unidos. Seu fascínio pela Astronomia começou por volta dos 8 anos de idade, motivado pelo seu avô materno. Porém, seguindo os passos de seu pai, concluiu o curso

de Direito em 1910 pela Universidade de Chicago, exercendo a profissão por pouco tempo. Doutorou-se em Astronomia em 1914. Se alistou para servir durante a Primeira Guerra Mundial como major e comandou um batalhão na França. Em 1919, retornou aos Estados Unidos e foi trabalhar no Observatório de Monte Wilson, Califórnia, onde realizaria quase todo seu trabalho até o ano de sua morte, em 1953.

Quando Hubble vai trabalhar no Observatório Monte Wilson, uma das maiores questões que rondavam a Astronomia da época era a natureza das nebulosas espirais[60] (WAGA, 1998). No dia 26 de abril de 1920, um debate, que mais tarde viria a ser conhecido como "O Grande Debate da Astronomia", aconteceu no Museu de História Natural Smithsonian em Washington, Estados Unidos (HOSKIN 1976). Nele, dois cientistas discutiram ideias opostas sobre o tema "Escalas do Universo". Heber Curtis, astrônomo do Observatório Lick, defendia que as nebulosas espirais, na verdade, eram outras galáxias semelhantes à nossa em dimensão e número de componentes. A ideia da existência de outras galáxias já havia sido proposta pelo astrônomo amador Thomas Wright, o qual acreditava na existência de "outras vias Lácteas", e Immanuel Kant, que influenciado por Wright, propõe em seu "História Natural e Filosofia do Céu" que as nebulosas seriam outras galáxias comparáveis à nossa (OLIVEIRA FILHO; SARAIVA, 2017, p.475). Curtis tinha como principal argumento que as novas[61] observadas nas nebulosas deveriam ser similares às observadas em nossa galáxia, sendo, portanto, essas espirais outros sistemas estelares. Já Harlow Shapley, astrônomo do mesmo observatório de Hubble, argumentava que essas nebulosas não seriam outras galáxias, mas apenas objetos nebulosos comuns. Shapley carregava a reputação de já ter medido a posição do Sol na Via Láctea e mensurado a melhor medida do seu tamanho para a época. No entanto, se baseava em dados errados sobre o movimento próprio de algumas estrelas presentes nessas nebulosas, calculados pelo astrônomo holandês Adriaan van Maanen. Seguindo os valores de Maanen, Shapley concluiu que, caso as nebulosas fossem grandes como a Via Láctea, a velocidade dos corpos em suas bordas seria superior a velocidade da luz, o que entrava em conflito com a Teoria da Relatividade de Einstein. Concluía, portanto, que as nebulosas espirais deviam ser objetos menores e estavam localizadas dentro do raio da Via Láctea. Ao final do debate, não foi possível chegar a uma conclusão devido à falta de dados observacionais, pois, os astrônomos dessa época tinham muitas dificuldades para medir a distância de objetos mais longínquos. Hubble, por sua vez, subordinado de Shapley no Observatório do Monte Wilson, estava mais inclinado em aceitar a ideia de que as nebulosas estavam além dos limites da Via Láctea (extragalácticas), porém, ao mesmo tempo, pensava que faltavam "medidas confiáveis das distâncias" (DAMINELI, 2003, p.52) não sendo possível, naquele momento, uma afirmação contundente sobre essa teoria. Outro fator o preocupava – opor-se abertamente à teoria de Shapley, seu chefe, poderia lhe trazer problemas profissionais significativos. Dessa forma, ele conduziu seus estudos sem chamar a atenção do restante dos astrônomos do observatório.

---

[60] No contexto dessa época, as nebulosas eram quaisquer corpos de aparência difusa (mesmo aspecto de nuvens, daí o nome) que os astrônomos observavam no céu noturno e, devido a essas características, sabiam que não se tratavam de apenas uma estrela ou planeta. As grandes questões envolvendo as nebulosas nessa época era em relação às suas possíveis composições (se eram aglomerados de estrelas, planetas ou outros elementos) e se eram objetos celestes localizados dentro ou fora dos limites da Via Láctea. A ideia vigente era que todo o universo se consistia de apenas uma galáxia, a nossa Via Láctea. Dessa forma, medir distâncias tanto dos limites da nossa galáxia quanto da posição das nebulosas eram grandes desafios da época.

[61] Quando uma estrela explode e libera no espaço seu conteúdo chamamos esse fenômeno de "supernova", o qual acarreta na desintegração total da estrela ou da maior parte de sua massa. Já o fenômeno das "novas" ocorre em sistemas formados por duas estrelas (binários), onde uma consome o material da outra até o ponto em que também explode. A diferença, nesse caso, é que o processo não leva a destruição da estrela que explodiu, voltando essa a ter dimensões menores.

Três anos se passaram do debate e, em outubro de 1923, Hubble começa a fazer observações da nebulosa de Andrômeda. Nela, ele identificou uma estrela variável e duas novas. A estrela variável seria de grande importância para o seu trabalho. Estrelas variáveis apresentam oscilações periódicas em sua luminosidade (LEDOUX;WALRAVEN,1958). Baseado nos trabalhos de Henrietta Leavitt[62], astrônoma americana responsável por descobrir a relação proporcional entre o período de pulsação de uma estrela variável cefeida e a sua luminosidade absoluta (energia por unidade de tempo vinda de uma fonte emissora de luz), Hubble a classificou como uma cefeída. Cefeidas são estrelas que apresentam picos de pulsação em seu brilho da ordem de dias, e têm massa de aproximadamente entre 5 e 10 vezes a massa solar (OLIVEIRA FILHO; SARAIVA, 2017, p.370). Como descoberto por Leavitt, em 1912, quanto maior fosse o período de oscilação da luminosidade aparente[63] de uma estrela (energia por unidade de área por unidade de tempo recebida de uma fonte luminosa), maior seria a sua luminosidade absoluta (LEAVITT; PICKERING,1912). Como a intensidade da luz (fluxo de energia em uma direção) cai com o inverso do quadrado da distância entre a fonte emissora (estrela) e a receptora (planeta Terra), Hubble conseguiu calcular a distância da cefeída até a Terra. A medida encontrada foi de 300.000 parsecs, ou seja, 1 milhão de anos-luz. Essa medida era muito superior à medida conhecida na época do raio da Via Láctea, que era de aproximadamente 15000 parsecs. Na noite do dia 5 outubro de 1923, Hubble desceu a montanha do Observatório de Monte Wilson e escreveu na página 156 do seu diário de bordo: " [...] nesta chapa (H335H), três estrelas foram encontradas, duas das quais eram novas, e 1 provou ser uma variável, que mais tarde foi identificada como uma Cefeida – a primeira a ser encontrada em M31" (CHRISTIANSON,1996, p.156). Aproximadamente cinco meses depois, escreve uma carta para Shapley relatando suas observações: " Você se interessará em ouvir que encontrei uma variável cefeída na nebulosa Andrômeda (M31)... Em anexo segue uma cópia da carta de luz que, mesmo grosseira, mostra de forma inquestionável as características de uma cefeída. ... Usando o valor de Seares... a distância obtida possui um valor acima de 300.000 parsecs." (WAGA,1998). Para um valor como esse, que situava Andrômeda em uma posição muito além das fronteiras da Via Láctea, as conclusões teriam que ser no sentido de que Andrômeda estava muito além da galáxia e seria um outro sistema estelar independente (outra galáxia). Relatos da cientista Cecilia H. Payne, primeira doutora em Astronomia do Observatório de Harvard, a qual estava presente no momento em que Shapley havia lido a carta em seu escritório, contam que o mesmo exclamou: "[...] Aqui está a carta que destruiu o meu universo. " (CHRISTIANSON,1996, p.159). Diante de tais observações e cálculos, concluiu-se que realmente Andrômeda estava muito além dos limites da Via Láctea e que, portanto, podia ser considerada como outra galáxia. Hoje sabe-se que a distância entre Andrômeda e a Via Láctea é 2,2 milhões de anos-luz, o dobro calculado por Hubble (OLIVEIRA FILHO; SARAIVA, 2017, p.476). Como menciona Waga (1998, p.164), essa discrepância foi devido ao fato de que somente na década de 50, um astrônomo alemão chamado Walter Baade, mostrou que existem 2 categorias diferentes de Cefeidas, o que levou a uma revisão de todos os cálculos de distâncias que já haviam sido feitos utilizando esses objetos como referência. Em 1926, Hubble ainda propõe um sistema de classificação das galáxias, que ficaria conhecido como Diagrama de Hubble, em que as dividia de acordo com o seu formato – elíptco, espiral e espiral barrado (HUBBLE, 1926).

---

[62] Ela e outras astrônomas como Williamina Fleming, Annie Cannon, Cecilia Gaposchkin e Antonia Maury fizeram parte do que ficou conhecido como as mulheres "computadoras de Harvard" (DAMINELI, 2003, p.61). No final do século XIX e início do XX elas foram recrutadas por Edward Pickering, astrônomo e físico de Harvard, para trabalharem no observatório dessa universidade, classificando e catalogando estrelas. Em uma época de muito preconceito em relação às mulheres nas posições de trabalho, principalmente nos ambientes científicos, a história delas e das outras que compuseram esse quadro de astrônomas pode servir como inspiração e motivação para a carreira científica. A quantidade de dados relacionados à classificação das estrelas que elas produziram foi muito significativa para os estudos em astronomia que seguiriam nas décadas posteriores. Por exemplo, Williamina Fleming, que era empregada doméstica na casa de Pickering antes de ser recrutada, catalogou e classificou mais de 10.000 estrelas em relação ao seu brilho. Fragmentos da sua história assim como das outras astrônomas também podem ser encontrados nessa matéria do jornal *El país*:
https://brasil.elpais.com/brasil/2015/10/28/ciencia/1446051155_519282.html

[63] Em palavras simples, é o brilho que você vê quando observa uma estrela daqui da Terra (com os efeitos da atmosfera já corrigidos). Já o brilho intrínseco (ou absoluto) é o brilho aparente que a estrela tem se for observada há uma distância padrão de 10 parsec (aproximadamente 32 anos-luz). Imagine você observando o brilho de uma vela que esteja há 10cm de você. Depois, essa vela é colocada há 100 metros e você observa seu brilho novamente. Na primeira situação, se imaginarmos que nessa escala 10 cm são iguais há 32 anos-luz, você está observando o brilho intrínseco (ou absoluto) da fonte luminosa (nesse caso, a vela, no caso dos astrônomos, são as estrelas). Na segunda situação, você observa o brilho aparente.

**3. A Expansão do Universo**

A descoberta da Lei de expansão do Universo foi feita por Edwin Hubble com a colaboração de outro astrônomo, chamado Milton Humason (OLIVEIRA FILHO; SARAIVA, 2017, p.505). Porém, passos importantes nesse processo foram realizados antes, por outros cientistas (WAGA, 2000). A história se inicia em 1901, quando o astrônomo americano Vesto Slipher é contratado para trabalhar no Observatório Lowell, no estado do Arizona, Estados Unidos. Durante dez anos, utilizando um espectroscópio, analisou a luz proveniente de estrelas e nebulosas. No ano de 1912, constatou que a luz proveniente da, até então, *nebulosa* de Andrômeda apresentava um desvio para o azul. Foram realizadas 8 observações, entre os meses de setembro e dezembro de 1912. Utilizando o efeito Doppler, no qual a frequência de uma onda tem seu valor alterado se a sua fonte emissora ou o receptor se movem em relação ao seu meio de propagação, calculou a velocidade de Andrômeda para cada uma dessas observações, como mostra a figura 1:

Figura 1 - Tabela original do trabalho de Slipher, publicado em 1913.

| 1912, September | 17, | Velocity, | —284 km. |
|---|---|---|---|
| November | 15–16, | " | 296 |
| December | 3–4, | " | 308 |
| December | 29–30–31, | " | —301 |
| Mean velocity. | | | —300 km. |

Fonte: Slipher(1913)

O valor médio calculado para a velocidade foi de 300 Km/s. Quando o Efeito Dopler é aplicado para a luz, se a fonte emissora se aproxima do receptor o espectro de cor é desviado para o azul, pois o comprimento de onda fica menor e a frequência maior. Se a fonte está se afastando, o desvio do espectro vai para o vermelho (baixa frequência). A equação (1) descreve o fenômeno:

$$z = \frac{(\lambda 1 - \lambda 2)}{\lambda 1} = \left(\frac{Ve}{c}\right) equação (1)$$

Onde, $\lambda 1$ representa o comprimento da onda no referencial da fonte, $\lambda 2$ o comprimento da onda medido pelo observador, $Ve$ a velocidade da fonte em relação ao observador e c a velocidade da luz. O termo z recebe o nome de desvio para o vermelho, em inglês, redshift. Valores positivos para z representam afastamento da fonte em relação ao observador. Valores negativos, aproximação. Nas palavras de Slipher, suas conclusões foram: " Consequentemente podemos concluir que a Nebulosa de Andrômeda está se aproximando do Sistema Solar com uma velocidade de, aproximadamente, 300 quilômetros por segundo. " (SLIPHER,1913, p.56). Em 1923, Slipher já havia mesurado a velocidade de 41 nebulosas (THOMPSON, 2011). Somente cinco apresentavam valores de z negativos, ou seja, se aproximavam da Via Láctea. As velocidades encontradas estavam numa faixa entre 200 e 1000 Km/s, valores muito superiores aos de velocidades encontrados para estrelas comuns (WAGA, 2000).

Concomitantemente, os avanços teóricos sobre a cosmologia estavam acontecendo, principalmente na Europa. Em 1917, Albert Einstein publica o seu " Considerações Cosmológicas na Teoria Geral da Relatividade", onde discute as consequências da sua teoria na estrutura do universo e propõe o primeiro modelo cosmológico relativístico (EINSTEIN, 1917). Uma característica marcante do seu modelo era o fato de ser estático e conseguir relacionar a matéria do universo à essa característica. Um universo estático era uma crença comum entre os cientistas da época (WAGA,2000). Desde o início da ciência moderna, por volta do século XVI, as concepções sobre a ordem do Cosmos estavam alinhadas com esse princípio. Como o campo gravitacional exerce atração entre os corpos, Einstein adicionou uma constante ($\Lambda$) em sua equação, a qual ficou conhecida como constante cosmológica, desempenhando a função de anular a contração do universo resultante devido à gravidade através de um efeito de expansão. No mesmo ano, um astrônomo holandês chamado Willem de Sitter publica 3 trabalhos aplicando a "Teoria da Gravitação de Einstein" (Relatividade), assim chamada por ele, na cosmologia (DE SITTER,1917). Seus resultados mostraram que era possível encontrar soluções para o modelo de um universo estático com a constante cosmológica, porém sem a existência de matéria, o que entrava em conflito com uma das principais características do modelo de Einstein. Após ler uma carta recebida de Einstein, em 24 de março de 1917, a quem de Sitter chamava de "professor", na qual ele concordava não ser possível propor uma solução nessas condições, de Sitter relata no seu "Sobre a Relatividade da inércia": "Ele [Einstein], portanto, postula o que chamei acima da impossibilidade lógica de supor que a matéria não existe. Podemos chamar isso de "postulado material" da relatividade da inércia." (DE SITTER, 1917, p.1225). De Sitter ainda propõe que a velocidade de afastamento entre objetos espalhados aleatoriamente pelo universo, aumentaria com a distância. Como seu universo é vazio, desprovido de massa, eles representam apenas partículas hipotéticas de testes.

Aleksandr Friedmann foi um físico russo que primeiramente propôs soluções matemáticas de um universo em expansão para as equações de Einstein[64] (SOARES, 2012). Num primeiro momento, contrariado pelas novas ideias de Friedmann, Einstein publica uma nota, onde diz estarem erradas as soluções encontradas pelo russo (WAGA, 2000). Somente um ano mais tarde, ele admitiu que o trabalho de Friedmann estava correto. O modelo proposto por Friedmann é considerado o ***modelo padrão cosmológico*** até os dias atuais (FAGUNDES, 2002). Ele está de acordo com os princípios de homogeneidade e isotropia (respectivamente, o universo tem a mesma constituição em qualquer parte; para qualquer direção que olharmos, a partir de qualquer ponto no Universo, ele terá a mesma aparência, não existindo **lugares privilegiados.** Os dois princípios são válidos em escalas muito grandes, aproximadamente, 1 bilhão de anos-luz) do universo, e data a sua idade como sendo de 10 bilhões de anos, um valor muito próximo ao que se considera correto hoje (aproximadamente 13,8 bilhões de anos). Estava publicado alí o primeiro modelo teórico que postulava um universo em expansão, um marco importante na história da cosmologia. Seu trabalho foi muito pouco divulgado, permanecendo desconhecido por muito tempo. Friedmann morre pouco tempo depois de tifo[65]. Tempos depois da morte de Friedmann, um padre belga chamado Georges Lemaître, propõe, em 1927, um modelo de universo inflacionário (LEMAÎTRE,1927). Neste modelo, num instante primordial, toda a matéria e espaço estavam concentrados em um ponto infinitamente pequeno que sofre uma expansão abrupta e dá origem ao universo. Sua teoria é a base do que se chama "Teoria do Big Bang", a qual é atualmente aceita, pela maioria da comunidade científica internacional, como sendo a que melhor explica o passado do universo[66]. Na teoria de Lemaître aparece pela primeira vez uma relação linear derivada das suas equações, relacionando a distância entre as galáxias e sua velocidade de afastamento. Também é creditado a Lemaître o primeiro cálculo do valor para o que viria ser conhecido como constante de Hubble, com valor de 625 km/s/Mpc (BLOCK,2011). O trabalho de Lemaître também teve pouquíssima divulgação, permanecendo desconhecido por muitos anos.

---

[64] Essas soluções matemáticas que estamos tratando aqui, tanto as de Friedman, quanto de De Sitter e outros que virão, são desenvolvimentos teóricos matemáticos desses cientistas tentando encontrar soluções para as equações e modelos. Basicamente, um modelo matemático é um conjunto de equações que tentam descrever um processo qualquer. No nosso caso, resumidamente, esses modelos estão tentando representar, através da matemática, como é o funcionamento do Universo em larga escala. Não é nossa intenção aqui mergulhar em detalhes sobre como e quais são essas soluções, até porque seriam necessários muitos conhecimentos avançados de Cálculo e Física para apenas introduzir o assunto. No ponto em que estamos, uma breve introdução à Cosmologia Moderna, basta que entendamos conceitualmente e historicamente alguns aspectos básicos sobre *o que estava acontecendo,* mas sem querer entrar muito no *como isso estava acontecendo.* Aprender ciências, em geral, é uma tarefa que exige paciência para construir o conhecimento pouco a pouco e de maneira sólida. Respeitar o *tempo* de assimilação conhecimento é importante.

Voltando um pouco no tempo, em 1922, através de observações, o astrônomo alemão C. Wirtz encontra uma relação logarítmica entre velocidade e distância das nebulosas. Usou como padrão o diâmetro aparente delas para inferir sua distância em relação a Terra. Quanto menor o diâmetro aparente, mais afastada estava a nebulosa. Seus resultados, no entanto, não foram levados à diante (WAGA, 2000). No ano de 1928, segundo Fagundes (2002), H. Robertson encontra a relação linear entre distância e velocidade das galáxias que, mais tarde, pelas mãos de Edwin Hubble, viria a ser conhecida como lei de Hubble. Já, segundo Waga (2000), na verdade, Robertson encontrou uma relação aproximada de linearidade entre essas duas grandezas, mas foi Hubble quem "coloca sobre uma base firme a validade da lei que indica que a razão entre a velocidade de afastamento de uma galáxia e sua distância é uma constante"(WAGA, 2000, p.165). Não é possível afirmar, categoricamente, se Hubble já havia lido o trabalho de Robertson e foi influenciado por ele.

## 4. A Lei de Hubble

As observações que serviram de dados, para que Edwin Hubble chegasse nos resultados que dariam origem à lei que leva o seu nome, foram realizadas também pelo astrônomo Milton Humason (OLIVEIRA FILHO; SARAIVA, 2017, p.505). Humason nunca teve educação formal, abandonando a escola ainda muito jovem. Começou trabalhando como guia de carroças (cocheiro) puxadas por mulas, as quais transportavam as partes desmontadas do telescópio do Monte Wilson na fase inicial de construção e montagem. Depois de pronto, se interessou pelo funcionamento do telescópio e aprendeu a operá-lo. Foi contratado para ser auxiliar noturno, dando assistência para o trabalho dos astrônomos. Foi lá que começa a trabalhar com Hubble. Catalogando o desvio para o vermelho de 24 galáxias, Hubble e Humason publicam, em 1929, seus resultados, os quais apontavam uma relação de linearidade entre distância e desvio para o vermelho (HUBBLE, 1929). Quanto maior era a distância de uma galáxia em relação à Terra, maior era o seu desvio para o vermelho (z). Como visto na equação (1), $Ve$ = cz. Portanto, quanto maior a distância d, maior era a velocidade de afastamento $Ve$:

---

[65] Doença infecciosa causada pela bactéria *Rickettsia prowazekii*. Normalmente é transmitida por piolhos. Causa febres altas e dores no corpo.

[66] Cuidado! Dizer que é a mais aceita não significa que ela já seja comprovada e definitiva. Iremos abordar a Teoria do Big Bang no próximo capítulo.

$$Ve = Ho.d \quad \text{equação 2}$$

Onde, $Ho$ é a constante de Hubble, com unidade de Km/s.Mpc[67], que está relacionada com a taxa de expansão do universo. A figura 2 mostra o diagrama original do trabalho:

Figura 2 – Diagrama original, relacionando a velocidade (Km/s) de afastamento com a distância (Mpc).

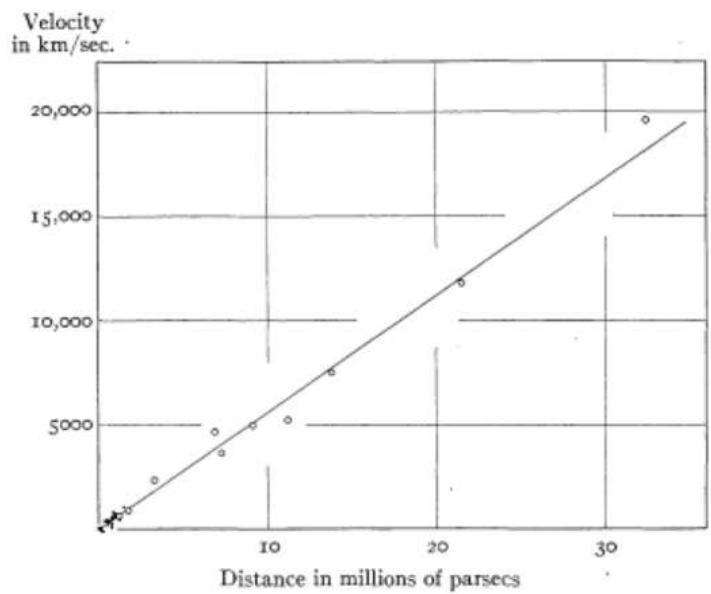

Fonte:Hubble (1929)

Dois anos após a publicação, Hubble e Humason estenderam sua pesquisa para avaliar a relação desvio/distância, ou velocidade/distância, de 40 galáxias. A mesma relação de linearidade foi encontrada (HUBBLE; HUMASON,1931). A constante $Ho$, calculada na época, foi de 500Km/sMpc. Hoje, sabemos que o valor é de aproximadamente 74 Km/sMpc (RIESS,2018), e essa discrepância é devida aos métodos ainda rudimentares de análises de distâncias e velocidades da época de Hubble, comparados com os atuais. O trabalho de Hubble e Humason foi considerado a evidência empírica para da expansão do universo. Hubble, o qual não se aventurava em interpretar seus próprios resultados como sendo em função da expansão do universo, nunca defendeu publicamente essa teoria (DAMINELI, 2003). Se limitou a apenas demonstrá-los e publicar os dados. O fato é que seu trabalho foi muito significativo para o contexto da época e teve grande alcance entre os cientistas. Albert Einstein ficou muito interessado pelas novas observações e foi pessoalmente visitar o Observatório do Monte Wilson, chegando a afirmar que " a descoberta da expansão do universo foi obra de dois californianos " (DAMINELI, 2003, p.101). Embora todo o alcance que o trabalho teve na época, a aceitação da ideia da expansão ainda demorou alguns anos. Passados esses acontecimentos, Humason continuou trabalhando como astrônomo. Em 1961, descobriu um cometa que foi batizado em sua homenagem (ROEMER,1962). Morreu em 18 de junho de 1972 na cidade de Mendocino, California. Durante os anos 30 e 40, Hubble se dividiu entre uma vida social agitada e animada, devido à sua forte tendência pessoal em *gostar da fama* (DAMINELI, 2003) e o trabalho, investigando, principalmente, galáxias e suas formações. Descobriu um asteroide no dia 30 de agosto de 1935, o qual recebeu o nome de 1373 Cincinnati (WARNER,2005). Trabalhou para o governo americano, durante a Segunda Guerra Mundial, desenvolvendo pesquisas com trajetórias de mísseis. Faleceu em 1953, pouco antes do Comitê do Prêmio Nobel aprovar que astrônomos pudessem concorrer ao Nobel de Física, o qual ele tinha muitas chances de receber, juntamente com Humason. Seus trabalhos, além de revolucionarem a maneira de como se entendia o universo, constituíram a base sólida das pesquisas posteriores que buscaram entender a dinâmica do Cosmos.

---

[67] "Mpc" leia-se "mega parsec", uma unidade de distância, muito comum em Cosmologia, para grandes distâncias. 1Mpc = 3.260.000 anos-luz.

**Atualidade**

Em 2011, três físicos receberam o prêmio Nobel de Física[68] por comprovarem e expansão acelerada do universo, ou seja, além de se expandir, o Universo o faz com velocidades cada vez maiores[69]. Embora saibamos que o universo se expande de maneira acelerada, alguns estudos atuais têm encontrados resultados que podem apontar que a taxa de expansão pode variar conforme a região do Universo[70]. No entanto, se trata de algumas controvérsias existentes, porém, sem nenhuma comprovação, apenas levantamento de hipóteses.

Alguns pontos devem ser estudados e frisados para que os alunos tenham clareza em certos conceitos, os quais, muitas vezes, não são de fácil assimilação nem por profissionais da área[71]. Dentre eles:

- os alunos devem compreender que, na expansão do universo, não são os objetos que viajam, se afastando uns dos outros pelo espaço, mas o próprio espaço que se expande. A analogia da régua pode ser útil – uma régua se expandindo, mas não por causa dos números que "se movimentam" para longe uns dos outros, mas porque o próprio espaço entre eles aumenta. Os números permanecem nas mesmas posições, porém, é o espaço existente entre eles que se expande, fazendo-os ficarem mais distantes uns dos outros, conforme o tempo passa.

- Edwin Hubble não "descobriu" a expansão do universo. Foi quem observou, juntamente com Milton Humason, e relacionou a velocidade de afastamento das galáxias às suas distâncias relativas à Terra, encontrando uma relação linear, onde as velocidades de afastamento aumentavam conforme mais distantes as galáxias. Anterior a isso, Georges Lemaître já tinha proposto teoricamente um universo com matéria que se expandia. Ainda no campo teórico, não podemos nos esquecer de Friedmann e De Sitter, os quais também desenvolveram modelos do universo em expansão. No campo observacional, anteriores a Hubble e Humason, os astrônomos Slipher, Landmark e Silberstein já observavam as velocidades de afastamento das nebulosas, porém, utilizavam métodos pouco confiáveis e decisivos na medição das distâncias, deixando suas observações com níveis de incertezas elevados (DAMINELI, 2003). Reforçar o caráter de continuidade da ciência e do tempo que se leva para averiguar hipóteses e teorias é necessário.

**5. Palavras finais**

Este artigo mostrou uma breve história sobre os acontecimentos que permearam as três primeiras décadas do século XX, referentes á descoberta da expansão do universo. A narrativa histórica, aliada aos conhecimentos de cunho um pouco mais técnico, procurou transmitir o conhecimento

---

[68] Uma notícia da época do prêmio: http://g1.globo.com/ciencia-e-saude/noticia/2011/10/tres-cientistas-recebem-nobel-por-estudos-sobre-aceleracao-do-universo.html

[69] Mais informações nessa notícia:
https://revistagalileu.globo.com/Ciencia/Espaco/noticia/2019/04/universo-esta-se-expandindo-muito-mais-rapido-do-que-o-esperado.html

[70]Mais informações aqui: https://canaltech.com.br/espaco/novo-estudo-sobre-expansao-do-universo-pode-mudar-o-que-sabemos-sobre-o-cosmos-163267/#:~:text=De%20acordo%20com%20a%20pesquisa,Hubble%20para%20determinar%20a%20velocidade.

[71] Recomenda-se a leitura do artigo de Helge Kragh "Cosmologia e ensino de ciências: Problemas e Promessas"

de uma forma completa, visando a imersão do professor no contexto dos eventos. Compreender a história desses acontecimentos é importante para entender o significado da mudança de visão sobre a estrutura dinâmica do universo, impulsionada pelas pesquisas de todos os cientistas aqui citados, principalmente, as observações de Edwin Hubble e Milton Humason, e os trabalhos de Georges Lemaitre e Aleksander Friedman. Durante quase toda a história da humanidade, as concepções sobre o universo estavam todas bem distantes de modelos que admitissem a sua dinâmica, muito menos a sua expansão. Por isso, a descoberta da expansão do universo deve ser um fato reconhecido como uma das grandes revoluções científicas da história e, portanto, deve ser estudada a fundo sobre seus aspectos históricos e técnicos. Ela também marca o nascimento da Cosmologia Moderna.

**Perguntas de fixação**

a) **Imagine se você, enquanto conversa com sua avó (ou alguém mais velho que seja próximo a você) sobre as coisas que aprende na escola, precise explicar sobre a expansão do universo. Como você explicaria a ela? Faça um pequeno texto explicando os conceitos principais.**

*As atividades de escrita "cartas ou mensagens a um parente/amigo mais velho" podem parecer simples demais ou repetitivas, mas não são. Nelas podemos analisar como, de fato, o aluno "interiorizou" o conteúdo que foi trabalhado com ele. Ao pedir para "explicar a alguém mais velho", na verdade, estamos estimulando que ele use a maneira mais clara e objetiva de se expressar em relação ao que viu e refletiu.Richard Feynman utilizava um mecanismo de aprendizagem similar, montando "apresentações" sobre algum assunto que ele tinha cabado de estudar.*

<p style="text-align:center">Capítulo 7</p>

# A história térmica do Universo – Do Big Bang até hoje

*"Neste capítulo apresentaremos conceitos a teoria do Big Bang. Espera-se que, ao final, os alunos possam ter uma maior compreensão sobre essa teoria, a qual é hoje a mais aceita no mundo acadêmico como a explicação do **passado do Universo**. É muito importante deixar claro que a teoria do Big Bang trata de assuntos até um tempo específico da história do Universo. O que veio antes de desse tempo ainda não possui comprovações observacionais e é tema de profundas e complexas investigações. No entanto, a teoria do Big Bang, em si, é fortemente apoiada em evidências observacionais, ou seja, sabemos como era o Universo primitivo, porém, ainda não sabemos ao certo o que veio antes desse tempo"*

**Introdução**

É atribuído a Georges Lemaître, o padre e cosmólogo belga, o qual também propôs um dos primeiros modelos do universo em expansão, como sendo um dos criadores da teoria que viria a ser conhecida como Big Bang[72]. Em 1931, a sua hipótese sobre a possível *origem* do universo defendia que este teria evoluído a partir de um "átomo primordial", em estado estacionário, extremamente quente e denso, não fazendo menção a qualquer espécie de "explosão" e nem a uma possível idade do universo. Apenas cunhava a expansão contínua (tal como vimos no capítulo anterior) a partir das condições iniciais. Foi uma hipótese que não foi levada a sério por muitos anos e, somente no final da década de 40, o físico russo George Gamow, juntamente outros cientistas, a retomaram, propondo um novo modelo, o qual também recebeu pouca atenção[73]. Gamow utilizava as ideias de novos campos de estudos da física, tais como a física quântica e a nuclear. Depois de realizar alguns cálculos, encarregou seu aluno, Ralph Alpher, de terminá-los[74]. Nesses cálculos, Gamow utilizou as equações da física nuclear para tentar entender quais seriam as condições das reações no "átomo primordial" de Lemaître. Ele argumentava que, se a hipótese de Lemaître estivesse correta, a extrema temperatura e densidade do início favoreceriam as reações nucleares. Os resultados dos cálculos mostraram que a composição do universo deveria ser de valores próximos a 75% para o hidrogênio, 25% para o hélio e o restante para todos os outros elementos químicos.

No entanto, outra teoria alternativa mais aceita no meio acadêmico da época era a do Estado Estacionário, proposta por Fred Hoyle, Thomas Gold e Herman Bondi em 1948[75]. Nela, o universo teria sempre existido da maneira como é atualmente, não tendo um início determinado. A teoria não negava a expansão do universo, fato esse já comprovado na época, mas supunha a criação de matéria espontânea, de acordo com alguns princípios quânticos, como forma de manter a densidade constante do universo, uma vez que ele se expande. Porém, uma das histórias mais surpreendentes e improváveis da recém-nascida Cosmologia Moderna iria ajudar a contestar essa visão.

---

[72] "Big Bang" é um termo inglês que significa grande explosão. Foi usado em 1949 por Fred Hoyle, um cosmólogo inglês, ao se referir pejorativamente à teoria a qual ele não aceitava como correta. É um termo que pode criar uma falsa impressão de que o universo tenha começado com uma "explosão", quando na verdade essa não é uma interpretação correta, como veremos a seguir.

[73] Ver em "Cosmologia e ensino de Ciências: Promessas e Problemas", de Helge Kragh.

[74] Ver em "Depois do Big Bang – Da origem ao fim do Universo", p.42, de Alberto Fernandéz Soto.

[75] Pode ser acessado em: <http://articles.adsabs.harvard.edu/cgi-bin/nph-iarticle_query?bibcode=1948MNRAS.108..372H&db_key=AST&page_ind=0&data_type=GIF&type=SCREEN_VIEW&classic=YES>

**Os cocôs de pombos e a radiação cósmica de fundo**

A teoria do Big Bang prevê um início muito quente e denso para o universo, de onde ele vai se esfriando, conforme o tempo passa e se expande. Alguns cientistas acreditavam ser possível medir essa temperatura. Os modelos do final da década de 40 e 50 estimavam um valor entre 1K (graus Kelvin) e 50K, sendo esse último o limite superior máximo calculado por Gamow. Ralph Alpher e Hans Bethe, outro físico que começou a trabalhar com Gamow, chegaram numa faixa menor entre 1K e 10K. Robert Dicke, que era um astrônomo de outra equipe, calculou uma faixa mais precisa, algo entre 5K e 10K. Concomitante a esses desenvolvimentos teóricos, havia equipes trabalhando para desenvolver equipamentos que fossem capazes de detectar a radiação e medir o valor da temperatura. Dentre elas, a equipe do próprio Robert Dicke e outra soviética, de Andrei Doroshkevich e Igor Novikov. Juntamente com Dicke, também estavam Peter Rool, David Wilkinson e James Peebles[76].

É aí que a história tem um acontecimento muito inesperado e que mais uma vez nos mostra como a ciência se constrói em um processo contínuo, o qual, muitas vezes, não é linear (lembrar sobre o que vimos no primeiro capítulo "Uma conversa sobre Ciência"). Há alguns quilômetros, afastados da equipe de Dicke, dois físicos e rádio-astrônomos[77,] chamados Arno Penzias e Robert Wilson, enfrentavam um problema nos sinais das antenas dos Laboratórios Bell, onde trabalhavam com comunicação de rádio, na implementação de um novo sistema. As antenas destinadas ao novo e promissor sistema de comunicações, baseado em sinais transmitidos por satélites, estavam captando um sinal de interferência que parecia ser constante. Eles já haviam calibrado muito bem o equipamento, inclusive, utilizavam uma fonte padronizada, com foco frio de hélio para filtrar qualquer interferência. Chegaram a pensar ser o cocô de pombos presente na estrutura metálica das antenas. Depois de limpá-las e expulsar os pombos que haviam feito morada lá, perceberam que o sinal ainda era o mesmo - constante, vindo de todas as direções e não sofria qualquer variação nos meses do ano. O sinal era muito fraco e a estimação da sua temperatura era de, aproximadamente, 3,5 K. O fato da interferência vir de todas as direções também invalidava a explicação de que ela provinha de uma fonte terrestre, como, por exemplo, transmissores de onda de outras estações. Durante meses, trabalharam para entender de onde vinha a tal "interferência'. Comentaram com vários colegas e profissionais da área, até que Bernard Burke, um astrônomo do Massachusetts Institute of Technology (MIT), sugeriu que eles deveriam entrar em contato com um tal pesquisador, chamado Robert Dicke... Estava feito o elo! Burke já conhecia o trabalho de Dicke e não hesitou em pedir que os dois fossem procurá-lo, pois ele poderia ter a explicação. Como é contado no livro de Alberto Soto (Depois do Big Bang – da origem ao fim do universo, p.54), Dicke estava comendo um sanduíche quando recebeu o telefonema e seus colegas o ouviram repetir as palavras "antena", "sinal contínuo", "todas as direções" e "fundo constante". Quando ele desligou o telefone e olhou para a sua equipe, disse: "Bem, rapazes, parece que fomos superados".

---

[76] SOTO, Alberto Fernadez. **Depois do Big Bang – Da origem ao fim do Universo.** Lisboa: Atlântico Press, p. 52, 2016.
[77] A radioastronomia é um ramo da astronomia que estuda o universo a partir de ondas eletromagnéticas vindas dos objetos celestes.

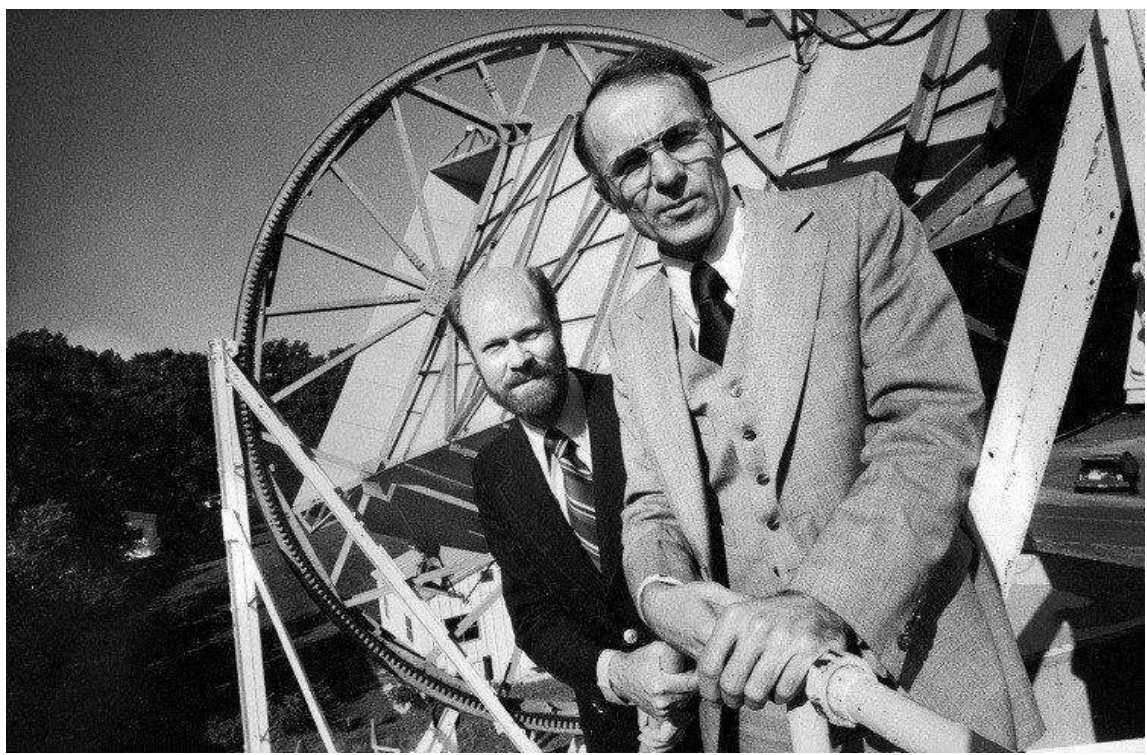

*Arno Penzias (primeiro plano) e Robert Wilson e a antena onde detectaram a radiação cósmica de fundo. Fonte: Sophiaphysics.ie ([http://sophiaphysics.ie/index.php/arno-penzias-robert-wilson-and-the-story-of-the-origins-of-the-universe/](http://sophiaphysics.ie/index.php/arno-penzias-robert-wilson-and-the-story-of-the-origins-of-the-universe/))*

As duas equipes começaram a trabalhar juntas, publicando dois artigos em julho de 1965, na revista *Astrophysical Journal*. O primeiro estipulava as bases teóricas da radiação captada[78], e o segundo o trabalho de detecção feito por Penzias e Wilson[79]. Somente Penzias e Wilson receberam o prêmio Nobel de física em 1978. Hoje, devido às melhores condições tecnológicas do que da época do achado, temos um mapa mais específico dessa radiação de fundo[80] e sabemos, com altíssima precisão, que sua temperatura é de 2,7K. Em 1992, o satélite COBE, da NASA, foi o instrumento utilizado para mapear a radiação de fundo. Foram detectadas diferenças sutis em sua disposição pelo espaço, nos revelando que o universo primitivo possuía algumas pequenas regiões com diferentes densidades de energia.

---

[78] DICKE, Robert H. et al. Cosmic Black-Body Radiation. **The Astrophysical Journal**, v. 142, p. 414-419, 1965.

[79] PENZIAS, Arno A.; WILSON, Robert Woodrow. A measurement of excess antenna temperature at 4080 Mc/s. **The Astrophysical Journal**, v. 142, p. 419-421, 1965.

[80] A radiação cósmica de fundo pode ser entendida como a medida da temperatura do Universo, um resquício da luz primordial que começou a viajar quando ele estava em seu início e continua viajando até hoje, ocupando predominantemente de forma homogênea todas as suas regiões. Veremos mais à frente.

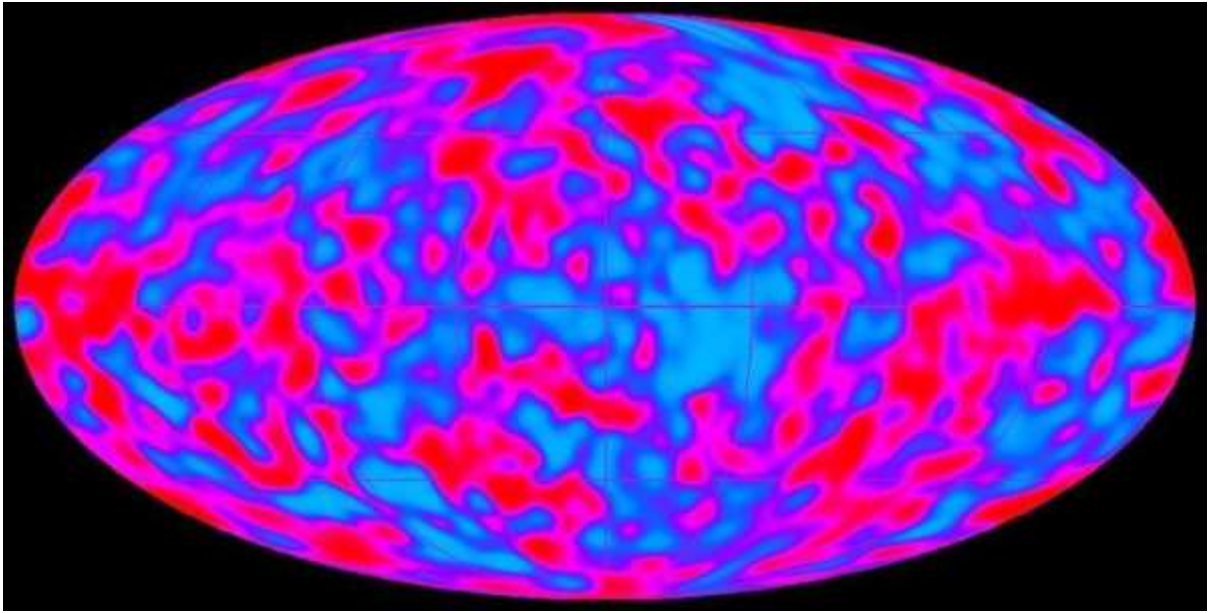

Mapa gerado pelo satélite COBE, entre os anos de 1990 e 1992. As diferentes tonalidades representam as ligeiras variações de temperatura e densidade da radiação cósmica de fundo. Fonte: Nasa (https://www.nasa.gov/topics/universe/features/cobe_20th.html)

Mas por que a teoria do Big Bang passou a ser aceita com a detecção da radiação cósmica de fundo?

**Entendendo alguns princípios físicos da radiação cósmica de fundo e do Big Bang**

A teoria do Big Bang é válida para explicar o universo **a partir de um instante de tempo específico**. Esse instante é chamado de tempo de Planck[81] e equivale a $10^{-43}$s. Antes desse instante, chamamos tudo de "Era de Planck", e não há ainda uma teoria que explique, definidamente, as leis que regem esse período. O universo nessa fase é infinitamente pequeno e denso, de forma que as quatro forças fundamentais da natureza[82] estavam unidas. No tempo de Planck, a temperatura era de $10^{32}$K e a força gravitacional se separou das outras três. Aos $10^{-35}$ a força nuclear forte se separa, deixando apenas a força eletro-fraca (eletromagnética + nuclear fraca) unida. Quando a temperatura já estava em $10^{15}$K no tempo de $10^{-10}$s, a última separação das forças ocorreu, prevalecendo no universo as quatro forças separadas.

---

[81] Existem algumas constantes fundamentais na Física: velocidade da luz (c = 299.792.458m/s), constante de gravitação universal (G=6,6742 x $10^{-11}$m$^3$Kg$^{-1}$.s$^{-2}$) e constante de Planck (h=1,054571x$10^{-34}$J.s).
[82] Resumidamente: gravitacional (entre as massas dos corpos), eletromagnética (entre as cargas elétricas dos átomos), nuclear fraca (entre elétrons e núcleo do átomo) nuclear forte (entre as partículas que formam o núcleo).

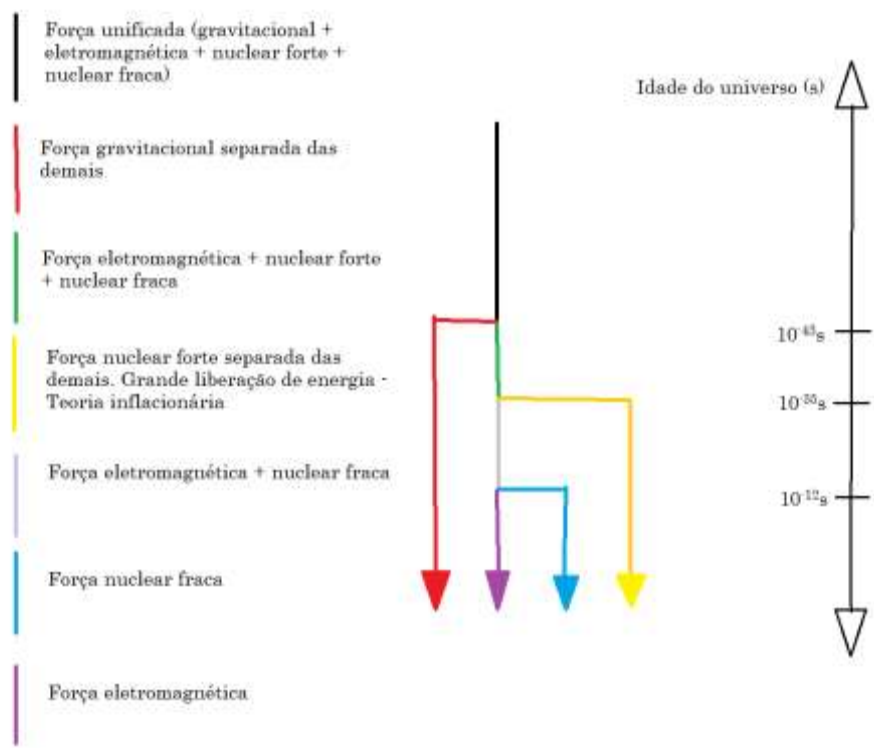

*Diagrama representando a separação das forças no Universo primordial. (Fonte: do autor)*

O universo desse instante está repleto de fótons que colidem, produzindo partículas e antipartículas[83], as quais se aniquilavam, formando novamente a radiação. Quando a temperatura baixou para $10^{14}$K, começa a formação do que conhecemos por prótons e nêutrons (e suas antipartículas). No instante em que elas colidiam com suas antipartículas, eram aniquiladas, produzindo novamente energia, porém, essa proporção não era igual. Existia um número superior de partículas ao de antimatéria, o que contribuiu para que mais partículas restassem ao final das reações e, assim, fosse possível existir matéria no Universo. Aos $10^{-4}$ segundos, a produção de nêutrons e prótons cessa, permanecendo esses apenas em colisão com suas respectivas antipartículas, num processo de aniquilação e produção de fótons, e apenas os elétrons e suas antipartículas estão sendo gerados. Quando estava com 1 segundo de idade, o universo possuía temperatura tal que a produção de elétrons também cessou. As partículas e antipartículas continuaram se aniquilando mutuamente, produzindo elétrons para o Universo, mas a energia já não era suficiente para que o choque de fótons produzisse matéria. O Universo, nesse instante, é um aglomerado de prótons, nêutrons, elétrons livres e muitos fótons, todos colidindo entre si. Quando a idade do Universo era de 3 minutos, os primeiros núcleos leves conseguem se formar, sendo o hidrogênio e hélio os primeiros em maior quantidade, e um pouco de lítio. Essa fase é o que chamamos de nucleossíntese primordial e ela vai durar até, aproximadamente, 380.000 anos. Durante todo esse tempo, as colisões entre fótons e matéria acontecem quase que incessantemente, fazendo com a luz ficasse aprisionada pela matéria. O Universo tinha um aspecto opaco, pois os raios de luz não conseguiam viajar livremente e seu conteúdo era uma mistura de núcleos de hidrogênio e hélio, juntamente com fótons e elétrons livres[84].

---

[83] A grosso modo, antipartículas tem a mesma massa que a sua partícula correspondente, porém cargas opostas. Quando elas se chocam, são aniquiladas, produzindo energia novamente.

[84] SOTO, Alberto Fernadez. **Depois do Big Bang – Da origem ao fim do Universo.** Lisboa: Atlântico Press, p.47-48, 2016.

Quando o Universo estava com temperatura de 3000K (380.000 anos após seu início), a energia já baixou suficientemente para que os núcleos formados possam aprisionar os elétrons e os primeiros átomos neutros de hidrogênio e hélio são formados. Dessa forma, a luz (fótons) já não mais colide com a matéria e pode viajar em linha reta, livremente, pelo espaço. E é justamente essa luz, à temperatura inicial de 3000k, a radiação cósmica de fundo. Alberto Soto faz uso, em seu livro aqui já citado (p.49), da analogia de um cosmólogo americano chamado Charley Lineweaver. Vamos imaginar que estamos em um campo infinitamente grande, completamente cheio de pessoas. Iremos todos começar a gritar, com toda força, juntos e, em um determinado instante, iremos parar também no mesmo instante. O jogo começa e estamos gritando muito alto, por determinado período de tempo. Quando chega o momento combinado, cessamos de gritar. O que acontece? Supondo que a velocidade do som seja de 340m/s, quando todos paramos de gritar a intensidade do som diminuiu, mas não há ainda silêncio absoluto. Ao passar um segundo após esse instante, o som das pessoas que estão há 340 metros à sua volta distantes chegam até você. Quando se passam 3 segundos, chega o som dos que estavam há, aproximadamente, 1km afastados. É claro que conforme o tempo passa, precisaremos de instrumentos mais precisos para captar as ondas sonoras que vêm de mais longe, porém, elas continuam chegando, mesmo em frequências que nosso ouvido não pode captar. Se, por exemplo, para de chegar som de alguma direção, podemos concluir que naquela direção e distância não haviam pessoas gritando no instante em que todos nós estávamos bradando nossos pulmões. Ao passar 1 hora, nos chegam sons das pessoas que estavam a mais ou menos 1200 km. Elas já podem até terem ido embora para suas casas, mas a partir dos sons que nos chegam sabemos que, há uma hora, estavam naquela posição, afastadas de nós a 1200Km, gritando.

Nessa analogia, o som representa a luz que viaja e se choca conosco, que fazemos o papel das partículas. Enquanto todos gritamos juntos, é como se fosse o estado antes dos 380.000 anos de idade do Universo. O momento em que paramos representa quando a luz começa a viajar livremente em todas as direções e conseguimos captar o sinal com maior nitidez pois ela não interage com mais nada. Hoje, quando detectamos a radiação cósmica de fundo, estamos recebendo um sinal gerado há aproximadamente 13,5 bilhões de anos atrás[85], uma foto do seu estado inicial.

A teoria do Big Bang precisou de muitas refinações e aprimoramentos e é a mais aceita nos dias de hoje como a teoria que explica a *infância* do Universo[86]. A detecção da radiação cósmica de fundo foi um marco para a teoria, pois mostrou como suas predições estavam corretas. Alguns problemas surgiram na teoria do Big Bang em relação à homogeneidade da radiação cósmica de fundo e à planicidade tridimensional do universo (por observações, sustenta-se que o universo seja plano, mas pelo modelo do Big Bang ele deveria apresentar certa curvatura, no entanto, não é o que se observa). Uma teoria que se soma a do Big Bang é a da Inflação Cósmica. Resumidamente, a essa teoria prediz que quando a força nuclear forte se separou das demais, ela liberou uma quantidade imensa de energia que provocou uma inflação violenta no universo, indo este do tamanho de um núcleo atômico para o do sistema solar em $10^{-36}$s. A Inflação Cósmica ou teoria inflacionária resolve as duas questões apresentadas aqui, pois, uma expansão rápida e brusca como essa manteria a homogeneidade das regiões e favorece uma geometria tridimensional plana para o Universo. Um pequeno resumo:

---

[85] Tempo estimado em que aconteceu o Big Bang, baseado nos modelos e observações.

[86] Cuidado! Essa afirmação não significa "teoria definitiva" ou "já comprovada". Por exemplo, sugere-se a consulta de vídeos e textos do cosmólogo brasileiro Mário Novello, o qual possui muitas considerações e reflexões sobre a Teoria do Big Bang. Recomenda-se https://www.youtube.com/watch?v=88FG4v885GA

| Idade do Universo | Temperatura (K) | |
|---|---|---|
| $10^{-43}$ s | $10^{32}$ | *Tempo de Planck. A gravidade se separa das outras 3 forças.* |
| $10^{-35}$ s | $10^{25}$ | *A força nuclear forte se separa das forças magnética e nuclear fraca.* |
| $10^{-32}$ s | $10^{27}$ | *Inflação Cósmica causada pela energia liberada na etapa anterior.* |
| $10^{-10}$ s | $10^{15}$ | *Força eletromagnética e nuclear fraca se separam. Era da radiação.* |
| $10^{-7}$ s | $10^{14}$ | *São formados os prótons e suas antipartículas. Era das partículas pesadas.* |
| $10^{-1}$ s | $10^{12}$ | *Formação dos elétrons e suas antipartículas. Era das partículas leves.* |
| 3 min | $10^{10}$ | *Nucleossíntese. Formação dos nêutrons e dos primeiros núcleos de hidrogênio, hélio e um pouco de lítio.* |
| $3,8 \times 10^5$ anos | $10^3$ | *Recombinação. Núcleos atômicos capturam os elétrons formando os primeiros átomos neutros. A radiação vija livremente sem colidir com partículas.* |
| $1 \times 10^9$ anos | 20 | *Formam-se as primeiras galáxias.* |
| $1 \times 10^{10}$ anos | 3 | *Formação do sistema solar.* |

**Palavras finais**

É importante não concebermos a ideia do Big Bang como uma explosão, pois uma explosão *acontece dentro de um espaço e num determinado instante do tempo*. O Big Bang é a própria origem do espaço e do tempo, não havendo nenhuma localiação ou instante antes do seu acontecimento. Dessa forma, ao pensarmos em uma "explosão", inevitavelmente, pensaremos em algo explodindo em um certo local e instante já existentes quando, na verdade, o Big Bang representa o início do espaço-tempo. Muitos astrônomos e cosmólogos também não acham fácil visualizar esse conceito. Mais importante do que isso é a busca do conhecimento e se aprofundar no assunto.

Como um pequeno resumo das provas observacionais do Big Bang, as quais, em conjunto, sustentam a teoria com alto grau de confiabilidade, apresentamos:

- Medições da distribuição de matéria visível[87] no universo como sendo, aproximadamente, 75% de hidrogênio, 25% de hélio e o restante dos outros elementos.
- A radiação cósmica de fundo homogênea em todas as direções a uma temperatura de 2,7K, a qual bate com o modelo e condições iniciais do Big Bang. Abaixo, colocamos as palavras de James Peebles[88], um envolvido direto nos trabalhos de Robert Dicke na época da descoberta, escritas em seu livro "Principles of Physical Cosmology" (Princípios de Cosmologia Física):

> "*A descoberta da radiação cósmica de fundo há 25 anos atrás teve um profundo efeito nos rumos e andamento da pesquisa em cosmologia física. Sua presença torna a imagem envolvente de Friedmann – Lemaître mais credível, porque é difícil conceber como o espectro térmico característico poderia ter sido produzido no universo como ele é em uma época presente. Isto é, a radiação cósmica de fundo é <u>considerada uma evidência quase tangível de que o universo se expandiu de um estado denso.</u> Isto é o que conduziu a maioria das pessoas a abandonar a cosmologia do estado estacionário.*"
>
> James Peebles, Principles of Physical Cosmology, p. 134.

- A expansão do Universo, fazendo com que as galáxias se afastem conforme o tempo passa. Ao *invertermos o relógio*, o Universo culmina em um ponto infinitamente pequeno com densidade e temperatura muito elevadas.

**Pra conhecer mais:**

- **Os Três Primeiros Minutos do Universo – Steven Weinberg.**

**Perguntas de fixação:**

**Explique a teoria do Big Bang e suas evidências observacionais. Ela fala sobre a origem do Universo?**

---

[87] Iremos estudar também a matéria escura.
[88] Vencedor do prêmio Nobel de física em 2019 por seus esforços e descobertas em cosmologia. Também não gosta do termo "Big Bang", pois isso remete a um local e instante de tempo.

<p style="text-align:center"><span style="color:#6FA8DC">**Capítulo 8**</span></p>

# Buracos Negros, carecas ou cabeludos?

*"Este capítulo contém informações sobre o que são buracos negros e como se formam. Também tratamos da primeira imagem de um, divulgada em 2019. O texto pode ser trabalhado diretamente com os alunos. Os buracos negros são um dos fenômenos mais surpreendentes e instigantes da atualidade e esse aspecto deve ser ressaltado. O prêmio Nobel de 2020 também está ligado diretamente ao assunto, sendo os três ganhadores cientistas que desenvolveram trabalhos com buracos negros".*

**Introdução**

Os buracos negros são objetos de estudos recentes em nossa história científica, e a primeira imagem real de um, reconstruída a partir de ondas de rádios, foi obtida somente em 2019. Até então, sua existência era indicada por implicações matemáticas, oriundas de equações da Teoria da Relatividade Geral e os primeiros indícios físicos vieram através da captação de ondas de rádio nos anos 70[89]. Não se sabe ainda hoje se o que cai em seu interior poderá sair e, caso positivo, como saíra. Será a *informação* toda perdida? Os buracos negros emitem algum tipo de sinal? Todas essas questões hoje são muito controversas e não se tem uma resposta em definitivo. Stephen Hawking foi um dos cientistas que estudaram as questões envolvendo os buracos negros e deixou alguns trabalhos[90] que necessitam de mais avanços e testes. Para sintetizar um pouco o quão exóticos são os buracos negros, fiquemos com a fala de Stephen Hawking, em uma palestra[91] dada sobre o assunto:

> *"Dizem que às vezes a realidade é mais estranha que a ficção. Em nenhum lugar isso é mais verdadeiro que no caso dos buracos negros. Os buracos negros são mais estranhos que qualquer coisa já sonhada por escritores de ficção científica."*

**Evolução estelar**

Antes de falarmos especificamente sobre os buracos negros, primeiro é necessário entender como é o ciclo de vida das estrelas, pois é a partir delas que podem se formar os buracos negro.

As estrelas nascem quando nuvens de poeira e gás entram em colapso, por causa da atração gravitacional, concentrando o seu material em um único ponto. O gatilho que pode levar essas nuvens se agruparem dessa maneira é a atração gravitacional, causada pelos movimentos rotacionais das galáxias, interação com campos gravitacionais de estrelas ou a explosão de uma supernova[92]. A nuvem colapsada entra em movimento circular, se tornando um disco achatado. Nesse estágio ela é chamada de *protoestrela*. A medida que o material vai se acumulando em seu centro, o que dará origem ao *núcleo* da estrela, ela vai ficando mais densa e quente, até que o estado de energia é tal que as reações nucleares (ver quadro à direita) comecem e a estrela passa a brilhar.

---

[89] FRANCHI, CMGG; NETO, Manoel F. Borges. Breve História dos buracos negros. p.48-49, 2014.
[90] Um resumo sobre o último artigo envolvendo buracos negros pode ser encontrado aqui: https://canaltech.com.br/espaco/ultimo-artigo-de-stephen-hawking-sobre-buracos-negros-e-publicado-124662/
[91] HAWKING, Stephen. Buracos Negros: Palestra da BBC Reith Lectures. Editora Intrinseca, 2017.

### Fusão Nuclear

A principal reação que ocorre no núcleo de uma estrela é a fusão nuclear. Basicamente ela consiste da reação entre dois átomos de menor massa que fundem para a formação de um átomo de maior massa mais a liberação de uma grande quantidade de energia. No caso das estrelas, o principal combustível gerador de energia é o hidrogênio (especificamente o deutério, que contém 1 próton, 1 nêutron e 1 elétron). Quando dois átomos de deutério se fundem, acontece a formação de um átomo de hélio mais a liberação de energia. Quanto maiores as estrelas, mais elementos, além do hélio, serão produzidos, pois, o núcleo será capaz de comprimir mais matéria e se aquecer, possibilitando novas fusões de elementos pesados.

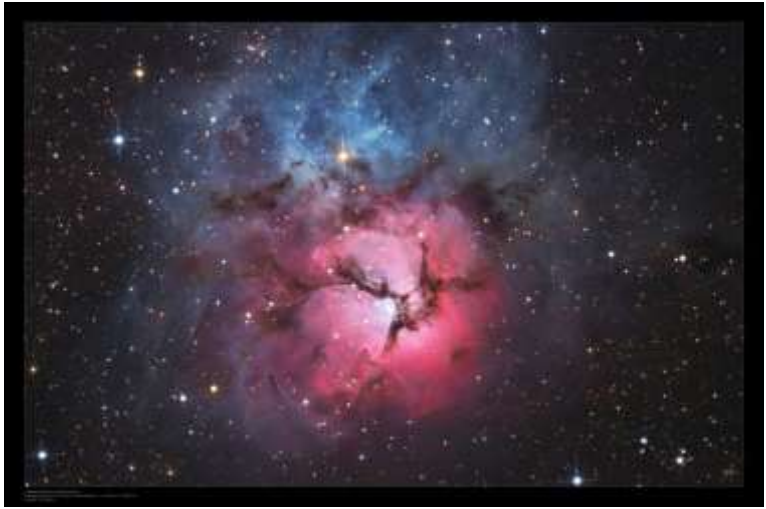

*Nebulosa Trifida – região de formação de estrelas. (Fonte: SpaceToday -*
https://spacetoday.com.br/a-bela-nebulosa-trifida-2/*)*

Depois de formadas, as estrelas jovens continuam cercadas por nuvens de gás e poeira, sendo parte desse material absorvido por elas e o restante ejetado.

A fusão nuclear do hidrogênio no núcleo da estrela é a sua principal forma de gerar energia. Quando o hidrogênio do núcleo se esgota, ela passa a queimá-lo em suas camadas mais externas (pode-se fazer uma analogia estrutural entre as camadas da estrela e as de uma cebola). Essas camadas vão inflando, devido à radiação que vem de dentro, e também se esfriando. Estrelas com a massa aproximada a do Sol tornam-se gigantes vermelhas, e quando o seu núcleo se desintegra (devido ao esgotamento do hidrogênio), ele se torna quente o suficiente (energética) para começar a queimar o hélio que produziu. Estrelas como o Sol conseguem fundir apenas hidrogênio e hélio. A medida que as camadas vão se exaurindo e começa a queima de elementos de uma camada mais externa, a instabilidade da estrela aumenta e as camadas internas do seu núcleo vão se desintegrando. A estrela vai liberando material em forma de uma nuvem (nebulosa) ao seu redor. Esse fenômeno se chama *nebulosa planetária*, pois se assemelha a planetas orbitando a estrela. O que resta é uma pequena estrela branca, chamada *anã branca,* a qual brilhará por alguns milhões de anos e se tornará uma *anã negra*[93]. Caso a anã branca possua alguma outra estrela ou material estelar em suas redondezas, sua gravidade é capaz de capturá-los e ela ainda viverá por muito mais tempo.

---

[92] Fase final da vida uma estrela "supergigante". Imensa explosão que libera material estelar.

[93] Nunca foram detectadas, mas os modelos preveem que elas possam existir. Além da dificuldade de *enxergá-las*, pois emitiriam muito pouca radiação, discute-se que ainda não haveria tempo para formação de anãs negras.

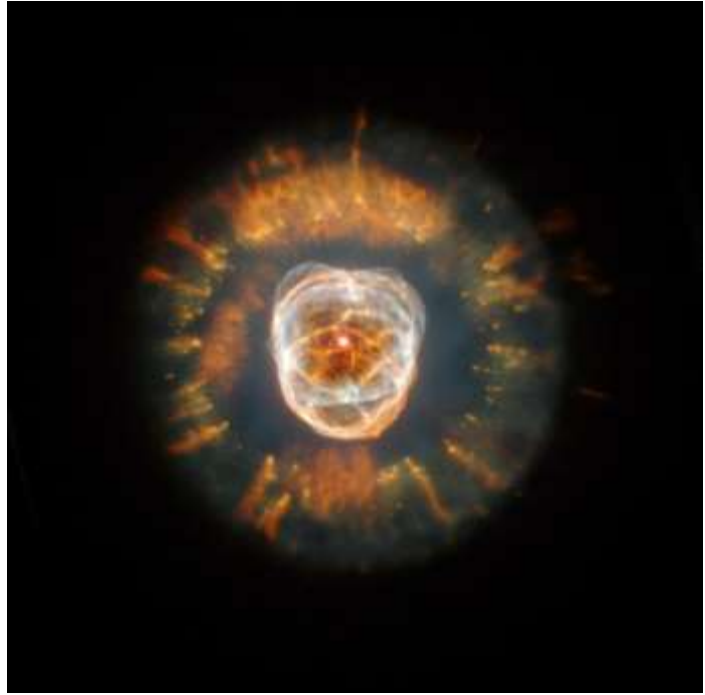

Nebulosa do Esquimó – Um exemplo do fenômeno de nebulosa planetária. (Fonte: Wikipedia https://pt.wikipedia.org/wiki/Nebulosa_do_Esquim%C3%B3)

Para estrelas muito maiores que o Sol, com aproximadamente oito vezes a sua massa, a queima do hélio ainda não representará o fim. Isso se deve, basicamente, ao fato de que, possuindo mais massa (então mais combustível), a estrela é capaz de atingir temperaturas maiores e assim fundir elementos mais pesados em suas camadas externas. Ela também incha, atingindo tamanhos gigantescos (por isso são chamadas *supergigantes*). Entretanto, chega um instante em que a pressão da energia que emana de dentro da estrela diminui a ponto de ela perder sustentação das camadas e então ocorre o colapso. Ela se desintegra e uma explosão de matéria e energia ocorre, chamada *supernova*. Uma *supernova* pode ter o brilho 1 bilhão de vezes maior que a estrela original.

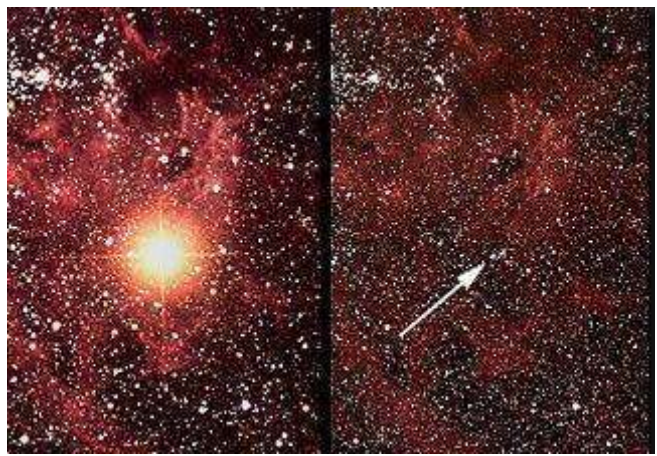

A comparação da explosão da supernova SN1987A (esquerda) com a mesma região do céu antes (direita), na Grande Nuvem de Magalhães. A supernova brilhou mais que toda a galáxia. (Fonte: Universidade Federal do Rio Grande do Sul - http://www.if.ufrgs.br/cref/camiladebom/Aulas/Pages/10.html )

O que resta da explosão de uma supernova é uma grande nuvem de poeira e energia, contendo os elementos químicos formados durante as fusões nucleares da vida da estrela[94] e um núcleo de matéria muito massivo. Esse núcleo é uma estrela de nêutrons e, se a sua massa equivale a, aproximadamente, 1,44 a massa do Sol ou menos, ela permanecerá sendo uma estrela de nêutrons. Mas se esse valor for maior, é quando começa nossa viagem rumo aos buracos negros...

**Buracos negros**

Embora a ideia que viria a ser o conceito de buracos negros já figurasse em interesses de alguns cientistas desde o século XVIII, com John Michell e Pierre Laplace, foi pouco tempo depois que Albert Einstein publicou a Teoria Geral da Relatividade, em 1915, que um físico e astrônomo alemão, chamado Karl Schwarzschild, encontrou uma solução para a equação de campo de Einstein[95]. Basicamente, essa solução implicava, dentre outras coisas, que, em um corpo esférico com massa grande o suficiente, a velocidade de escape na sua superfície seria igual à da luz. Caso essa matéria fosse concentrada em um volume de raio rs (conhecido como raio de Schwarzschild,o qual limita o horizonte de eventos, de onde nem a luz pode escapar), existiria no seu centro um ponto onde toda a matéria pareceria se concentrar, criando uma singularidade no espaço-tempo de densidade infinita.

$$rs = 2Gm/c^2$$

Onde *rs* é o raio de Schwarzschild, *G* a constante gravitacional[96], *m* a massa do corpo e *c* a velocidade da luz no vácuo.

Subrahmanyan Chandrasekhar foi um físico indiano que mostrou através de equações, aplicando a Teoria da Relatividade Geral, que estrelas anãs brancas podem ter um final diferente caso sua massa fosse superior a 1,44 massas solares[97]. Sob essas condições, quando o combustível da estrela começa a se esgotar, a atração gravitacional em seu núcleo se torna muito maior do que as forças que sustentam a estrutura, fazendo-a colapsar sobre si mesma, criando uma singularidade no espaço-tempo. Essa singularidade seria responsável por uma deformação grandiosa no tecido espaço tempo, porém, ela não é observável. Conseguimos observar apenas o horizonte de eventos, um limite de onde nada poderia *escapar,* devido à gravidade, nem mesmo a luz. Tempos depois, essas estruturas celestes viriam a ser conhecidas por buracos negros, e constituem um dos tantos exemplos na ciência em que o modelo matemático perdurou por anos antes das comprovações observacionais[98]

---

[94] Todos os átomos que compõem você e tudo mais que existe, foram forjados dentro das estrelas. Daí vem a famosa frase de Carl Sagan "somos formados de poeira das estrelas".

[95] Equação da Teoria Geral da Relatividade que mostra como a matéria gera a gravidade e, ao mesmo tempo, como a gravidade afeta a matéria.

[96] $G = 6,67408 \times 10^{-11}$ m$^3$ kg$^{-1}$ s$^{-2}$

[97] Esse valor limite ficou conhecido como "Limite de Chandrasekhar" - FRANCHI, CMGG; NETO, Manoel F. Borges. Breve História dos buracos negros. 2014.

[98] Lembrar dos casos das ondas gravitacionais da Relatividade Geral.

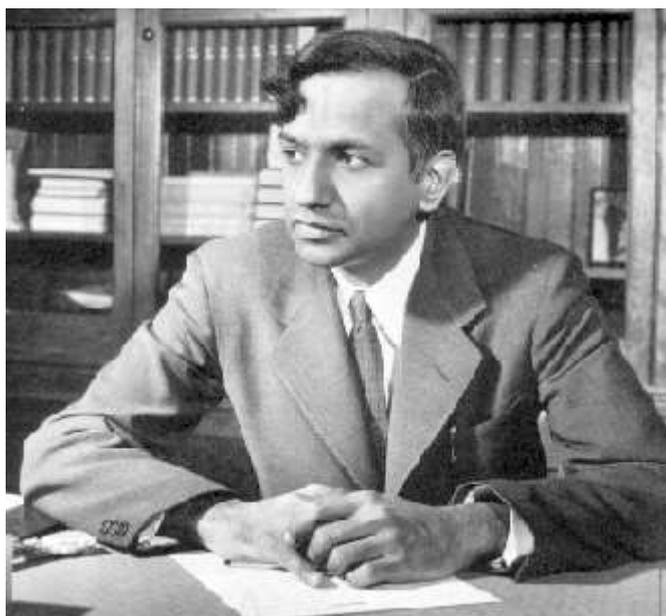

Subrahmanyan Chandrasekhar (1910 - 1995) nasceu na cidade indiana chamada Lahore. Desde criança era curioso e gostava de aprender. Estudou em casa até os 12 anos, onde aprendeu matemática e física com seu pai. Em 1930, formou-se físico no Presidency College, em Madras e, logo em seguida, foi estudar na Inglaterra, através de uma bolsa de estudos concedida pelo governo indiano.

Fonte: Universe Today (https://www.universetoday.com/40852/chandrasekhar-limit/)

Foi nas décadas de 60 e 70 em que os estudos envolvendo esse assunto receberam muita atenção e desenvolvimentos por parte da comunidade acadêmica. Cientistas como Stephen Hawking, Roger Penrose, James Bardeen e Jacob Bekenstein foram alguns dos que empreenderam esforços para ampliar mais a teoria sobre buracos negros nesse período. Um dos trabalhos de destaque é o trabalho de Hawking, publicado em 1974, no qual ele demonstra matematicamente que é possível um buraco negro emitir algumas quantidades de radiação. A quantidade de radiação emitida seria inversamente proporcional à massa do buraco negro, assim, buracos negros menores emitiram quantidades detectáveis de radiação. A radiação que seria emitida por um buraco negro ficou conhecida como *radiação Hawking,* mas, no entanto, até hoje não foi detectada. Ele revelou, com bom humor, em eu dos seus últimos livros[99] algumas reflexões sobre o assunto:

> *"Um buraco negro com a massa do sol vazaria partículas a uma taxa tão reduzida que não seríamos capazes de detectar o processo. No entanto, poderia haver miniburacos negros bem menores com a massa de, digamos, uma montanha. Um buraco negro do tamanho de uma montanha emitiria raios x e raios gama a uma taxa de 10 milhões de megawatts [...]. Os cientistas têm procurado miniburacos negros com essa massa, mas até o momento não encontraram nenhum. O que é uma pena, pois se alguém tivesse conseguido eu ganharia um prêmio Nobel"*

Em torno de um buraco negro, existe uma região chamada *horizonte de eventos*. Ela é o limite até onde pode-se *chegar perto* sem *cair* para dentro dele. É o ponto limite do raio de Schwarzchild. Ao ultrapassar o horizonte de eventos, o campo gravitacional é tão intenso que nada consegue escapar, e cai em direção ao centro do buraco negro. A matéria que é *sugada* para dentro forma uma espécie de halo em volta do horizonte de eventos, um fenômeno conhecido como *disco de acreção*. Uma analogia a esse fenômeno é quando uma grande quantidade de água escoa pelo ralo da pia, formando um halo em volta do orifício, enquanto escorre para dentro. No caso dos buracos negros, essa matéria escoando emite luz na frequência de raios x, e a detecção deles é uma das ferramentas empregadas por astrônomos para estudá-los.

---

[99] Buracos Negros – Palestras da BBC Reith Lectures, 2017, editora Intrínseca, p.44 – 45.

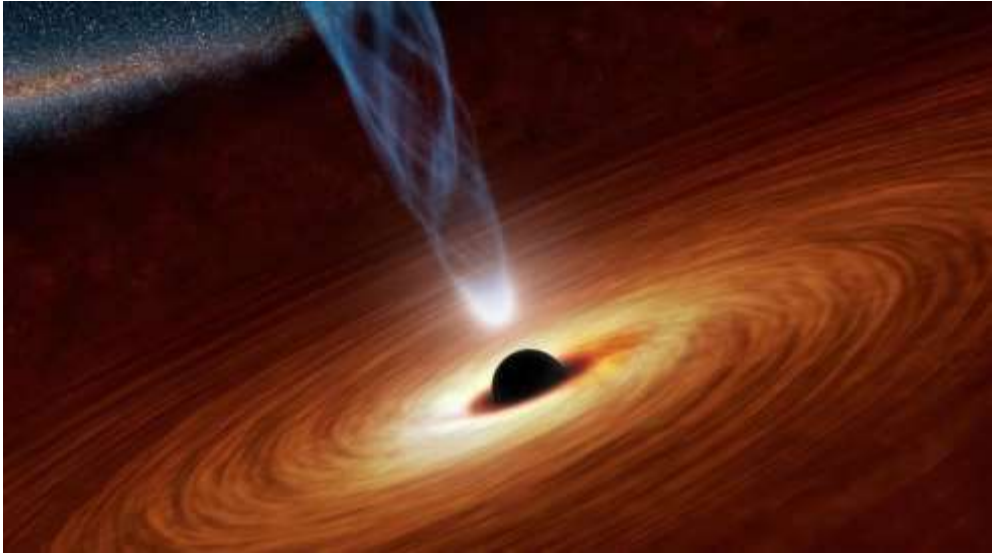

Esta figura é uma representação de um buraco negro. Vemos o *buraco*, em si, ao centro, o que na verdade é um corpo, e, ao seu redor, o disco de acreção, formado pela matéria que está sendo sugada por ele. Pode parecer que ele esteja emitindo luz, mas não é. A luz que aparenta sair do buraco negro, na verdade está sendo expelida pela própria matéria antes de ultrapassar o horizonte de eventos. Esse fenômeno acontece em buracos negros muito massivos, os chamados supermassivos[100].

Mas o que acontece com a matéria que cai dentro de um buraco negro? Ainda não se sabe ao certo, porém, a teoria mais aceita é que as únicas informações preservadas dessa matéria, depois que ela cruza o horizonte de eventos, são: massa, carga elétrica e momento angular[101]. Todo o resto seria perdido. Assim, *pouco importa para um buraco negro* se é você ou um navio gigantesco que ele está engolindo. De acordo com as teorias, somente as suas massas, cargas elétricas e momento angular importarão lá dentro. Você pode estar se perguntando o que veria se caso caísse dentro de um buraco negro. A questão é que, muito provavelmente, a medida que cruzasse o horizonte de eventos, seu corpo seria transformado em um *espaguete,* pois, supondo que você ultrapasse o horizonte primeiro com os pés, haveria uma tremenda diferença entre a atração gravitacional entre seus pés e sua cabeça. Quanto mais próximo se chega para perto do núcleo, maior são as consequências gravitacionais. O fato é que, até onde se sabe, não sobreviveríamos a essa tremenda força da natureza[102].

E se pudéssemos observar alguém (ou coisa) caindo em um buraco negro, o que veríamos? Neste cenário, devemos nos lembrar que, de acordo com a Teoria Geral da Relatividade, a luz e o espaço-tempo são afetados pela gravidade, e quanto mais forte um campo gravitacional, maiores são os efeitos. Como buracos negros são corpos que produzem um grande campo gravitacional, perceberemos dois fatos bem *estranhos.* O primeiro é que, conforme o astronauta o qual observamos vai se aproximando do horizonte de eventos, enxergaríamos a sua figura com a cor deslocada para o vermelho, devido ao *desvio gravitacional da luz,* o qual vimos no tópico de Relatividade Geral. Não importa qual a cor original, a gravidade desviaria a luz para o vermelho.

---

[100] Para mais informações, acessar https://jornal.usp.br/atualidades/quasares-buracos-negros-que-brilham-mais-do-que-as-galaxias/

[101] Grandeza física relacionada ao movimento de rotação do corpo.

[102] Talvez, em um buraco negro muito grande, seria possível adentrar sem ser destruído completamente pela força gravitacional. Hawking comenta essa hipótese em "Buracos Negros – palestras da BBC Reith Lectures, p. 26-27."

O segundo fato, mais peculiar ainda, é que nunca veríamos, de fato, o astronauta ultrapassar o horizonte de eventos, pois, para nós, o tempo no referencial do astronauta teria desacelerado de tal forma que enxergaríamos sua figura desacelerar, até parecer parar. A imagem, cada vez mais avermelhada, iria se tornando mais fraca até sumir, pois, a luz não conseguiria chegar até nós.

Stephen Hawking, Malcon Perry, Sasha Haco e Andrew Stominger são cientistas que realizaram um estudo, o qual foi publicado um pouco depois da morte de Hawking, discutindo a hipótese de que buracos negros podem *ter cabelo*[103]. Essa história de *cabelos* em buracos negros é uma expressão (engraçada) referente à metáfora utilizada para descrever o conceito oposto, o da *calvície* em buracos negros. Se referem assim para designar que, como buracos negros não conservam muitas informações sobre a matéria que cai neles, somente massa, carga elétrica e momento angular, todos seriam *muito parecidos*, como pessoas carecas! O artigo mencionado desenvolve um trabalho, mostrando como, talvez, outras informações pudessem ser obtidas da matéria que cruza o horizonte de eventos, ou seja, os buracos negros não seriam tão iguais assim, possuindo *cabelo*! Basicamente, partem da ideia que a matéria, quando cai em um buraco negro, afetaria a sua temperatura, a qual seria possível medir ao redor do horizonte de eventos, utilizando a radiação Hawking. Assim, a informação do que caiu estaria, em certo grau, armazenada ao redor do buraco negro e possível de ser acessada por observadores externos. São estudos sem nenhuma comprovação observacional e ainda serão desenvolvidos em maior grau.

**A primeira imagem de um buraco negro**

Em uma quarta feira (10) do mês de abril, no ano de 2019, foi publicada a primeira imagem de um buraco negro. Até então, o que se sabia graficamente deles provinha de simulações computacionais realizadas a partir dos modelos matemáticos e físicos, mas a sua *face* real nunca havia sido conhecida. O projeto era coordenado pelo *Event Horizon Telescope* (Telescópio de Horizonte de Eventos), que consiste numa série de radiotelescópios espalhados por todo globo terrestre, os quais captaram os sinais vindos de um buraco negro e processaram a sua imagem. Esse buraco negro está localizado no centro da galáxia M87, e se trata de um corpo supermassivo. Seu diâmetro é de 40 bilhões de quilômetros e está afastado 50 milhões de anos-luz do nosso planeta.

Foram dois anos captando os sinais, filtrando as interferências e combinando os *seus pedaços*, para que a imagem fosse processada. Isso, porque cada radiotelescópio[104] espalhado pelo planeta recebia uma parte do todo que compunha o sinal vindo do buraco negro. O desafio se assemelha muito a como se você e um grupo de amigos fossem montar um quebra-cabeça de 100.000 peças, por exemplo (esse número não é o de sinais do projeto, apenas uma exemplificação de um número alto). Cada integrante do grupo possui um certo número de peças sortidas do quebra-cabeça e mais algumas que não correspondem a ele, apenas peças erradas para confundir ainda mais. Para montá-lo, vocês precisam se reunir, encontrar as peças que não fazem parte, removê-las e também começar a encaixar as peças corretas. Que desafio[105]!

---

[103] Acessar https://brasil.elpais.com/brasil/2018/10/11/ciencia/1539279267_381059.html
[104] Telescópio que opera recebendo luz na frequência de rádio, e não na frequência visível como estamos acostumados a ver na maioria dos telescópios que funcionam assim.
[105] Um pouco do processo de como os dados foram filtrados e organizados pode ser lido aqui https://www.bbc.com/portuguese/geral-47886045

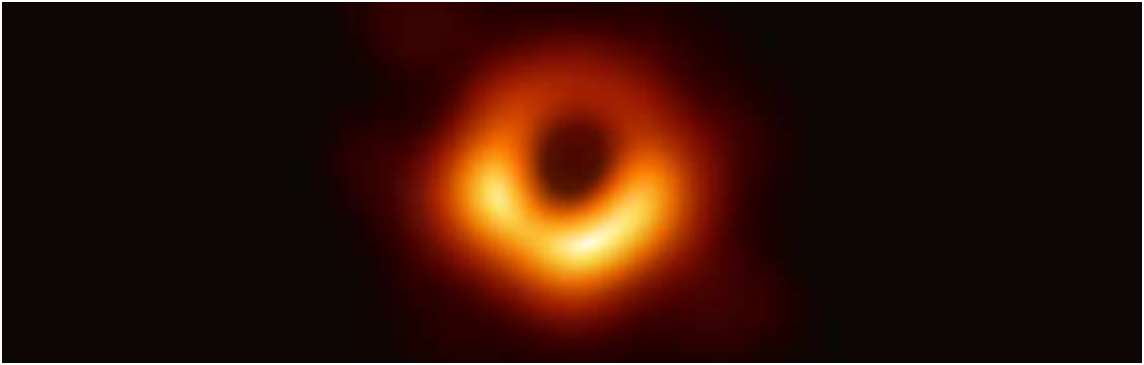

A primeira imagem de um buraco negro. Ela é compatível com as imagens obtidas através das simulações dos modelos matemáticos e físicos obtidos para eles. Pode-se notar a região escura, de onde nem a luz escapa, o disco de acreção, ao redor, em forma de halo, que está escorrendo para dentro do buraco negro. É possível ver o limite do horizonte de eventos e a luz desviada para o vermelho. (Fonte: Event Horizon Telescope: https://eventhorizontelescope.org/files/eht/files/20190410-78m4000x2330.png?m=1554877782)

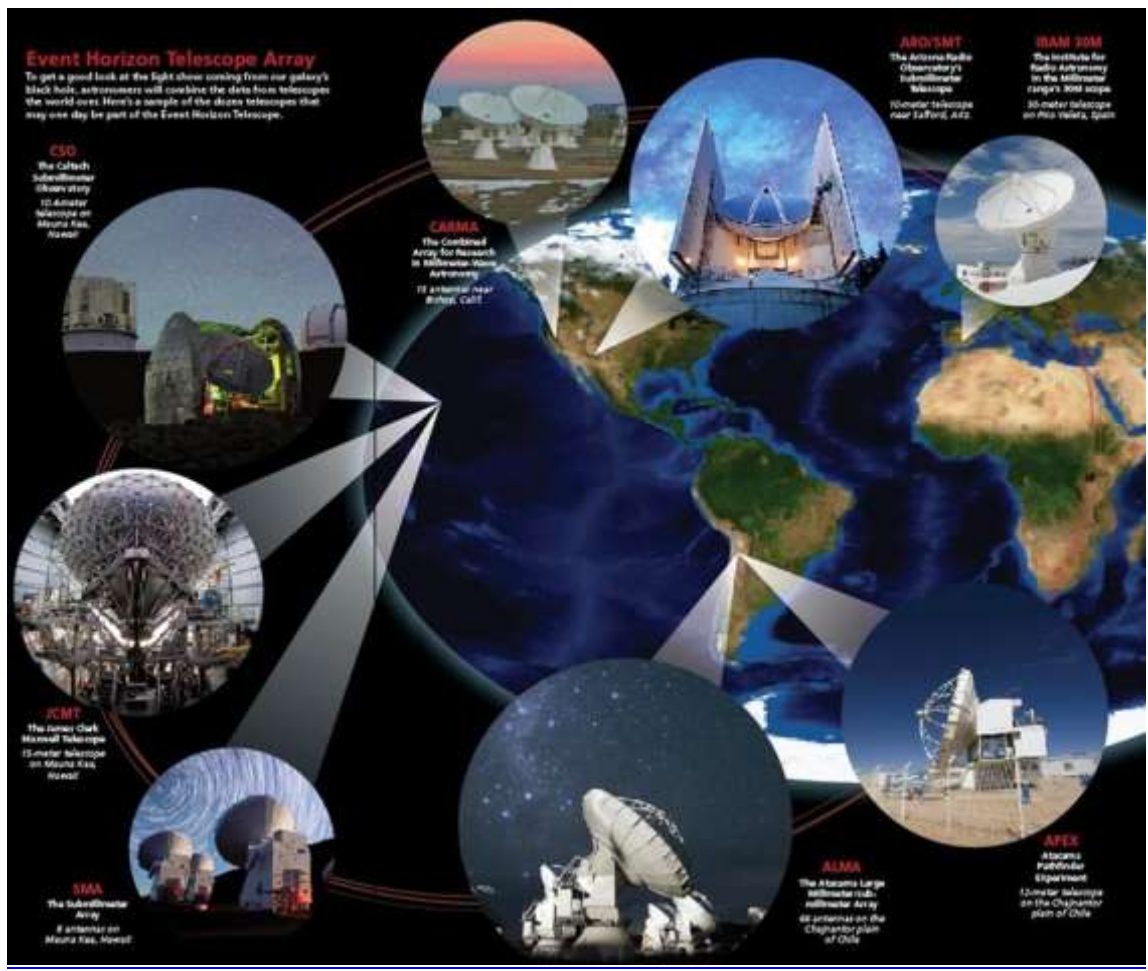

Montagem fotográfica publicada na página da revista de astronomia *Astronomy Magazine,* no dia 10 de abril de 2019. Ela mostra os radiotelescópios envolvidos no trabalho, seus nomes e localização no globo terrestre. (Fonte: Astronomy Magazine -https://astronomy.com/news/2019/04/event-horizon-telescope-releases-first-ever-black-hole-image)

Com os dados do telescópio espacial Hubble, lançado em 1990, foi possível avançar os conhecimentos sobre o universo em geral, incluindo os buracos negros. Com as observações feitas a partir do Hubble, indiretamente, foi possível aprender mais sobre eles[106]. Por exemplo, eles são estruturas muito comuns nas galáxias. É muito comum encontrá-los em seus centros. No centro da nossa própria Via Láctea existe um objeto supermassivo, o qual acredita-se ser um buraco negro (Sargitarius A*). A massa desses buracos negros é proporcional ao tamanho da galáxia em que está, assim, quanto maior a galáxia maior o buraco negro. Outro fenômeno que também acontece com eles é a fusão. Quando galáxias se fundem, os buracos negros também podem se fundir, formando um maior. Em 2018, alguns cientistas da Nasa, trabalhando com dados provenientes do Hubble, captaram os dados de duas galáxias com buracos negros em seus centros, se fundindo[107]. Os dados revelaram que, a medida que isso acontecia, a massa dos buracos negros ia aumentando rapidamente. Com dados prévios de simulações computacionais, eles concluíram que esse era o cenário exato de quando buracos negros estão prestes a se fundir.

Aprender sobre os buracos negros e sua dinâmica é uma atividade intrigante, até mesmo para quem se dedica profissionalmente a isso. Muito ainda precisa ser avançado para que nosso campo de conhecimento seja mais sólido. A detecção da primeira imagem direta de um buraco negro em 2019 nos mostra como esse assunto é relativamente novo na jornada dos seres humanos ruma ao cosmo.

**Prêmio Nobel de 2020**

Para se ter uma ideia do quão atual é o assunto sobre buracos negros, quando este livro já estava finalizado, saiu o resultado dos ganhadores do prêmio Nobel de Física de 2020 (6 de outubro de 2020). Assim, foi necessário atualizá-lo, através deste capítulo, pois, os ganhadores foram três cientistas que desenvolveram trabalhos com essas estruturas. Roger Penrose (colaborador direto de Stephen Hawking), Andrea Ghez e Reinhard Genzel foram escolhidos por suas colaborações significativas com o entendimento sobre os buracos negros. Roger Penrose demonstrou matematicamente como a Teoria Geral da Relatividade leva à formação de buracos negros. Já Andrea Ghez (a quarta mulher da história a ganhar um prêmio Nobel) e Reinhard Genzel descobriram a existência de um objeto supermassivo no centro de nossa galáxia, para o qual a explicação mais plausível é de que seja um buraco negro supermassivo.

---

[106]Página da Nasa com informações - https://nasa.tumblr.com/post/187921705234/hubbles-5-weirdest-black-hole-discoveries

[107] Reportagem original https://www.nasa.gov/feature/goddard/2018/astronomers-unveil-growing-black-holes-in-colliding-galaxies

> "Estou animada em receber o prêmio – levo muito a sério a responsabilidade de ser a quarta mulher a ganhar o prêmio Nobel [em física]. Espero poder inspirar outras jovens mulheres para uma área que tem tantos prazeres, se você tem paixão pela ciência. Há muito para ser feito."

*Declaração de Andrea Ghez, logo após o anúncio dos vencedores.*

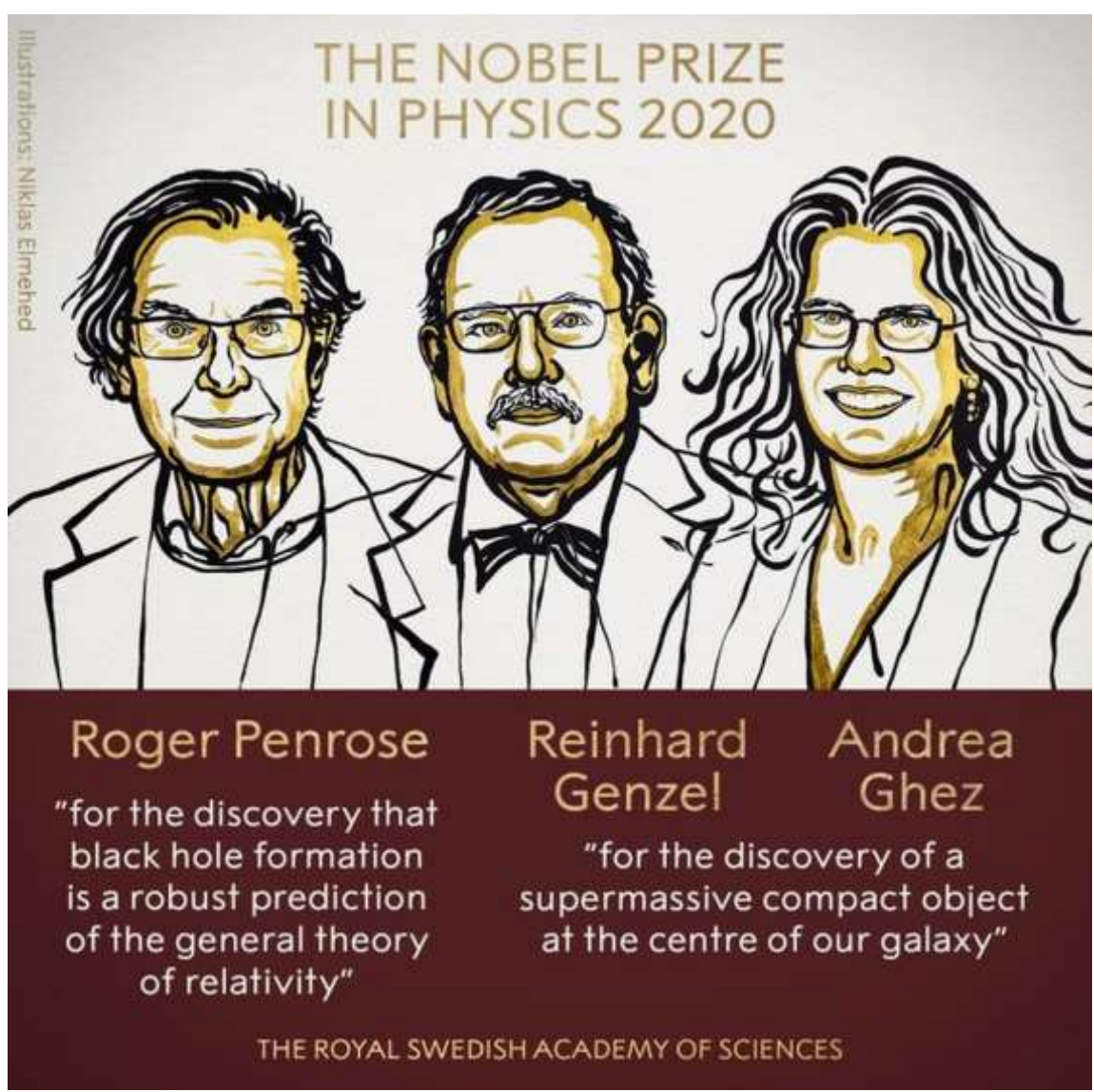

Fonte: Tweeter/Nobel

**Para conhecer mais:**

**Interestelar – filme de Christopher Nolan, de 2014.**

**Perguntas de fixação**

a) Calcule qual seria o diâmetro de um buraco negro gerado a partir de você, caso toda sua massa fosse comprimida em um único ponto.
*Aplicação direta da fórmula de Schwarzschild $r = 2GM/c^2$, utilizando a massa M como a massa do corpo em questão. O raio resultante deve ser multiplicado por 2 para se ter o valor do diâmetro do buraco negro originado pela massa de uma pessoa, caso ela fosse comprimida até aquele valor.*

b) Uma fórmula utilizada para calcular a velocidade de escape[108] v de corpos esféricos de massa M e raio r como, por exemplo, a Terra è:

$$v = \sqrt[2]{2GM/r}$$

Utilize essa relação para encontrar a fórmula do raio de Schawarzschild, tal como apresentamos nesse capítulo. Explique passo a passo seu raciocínio.

*O processo bem simples. Basta substituir v por c, pois, considerando o enorme campo gravitacional do buraco negro, supomos **maior velocidade** de matéria/informação que conhecemos (c – velocidade da luz. Daí, basta elevarmos c ao quadrado e colocar tudo em função de r, obtendo:*

$$r = 2GM/r^2$$

---

[108] Essa velocidade representa o valor mínimo que um corpo, sem propulsão, deve atingir para vencer o campo gravitacional de, por exemplo, um planeta. A velocidade de escape do planeta Terra é, aproximadamente, 11km/s.

<div align="center">

**Capítulo 9**
# Matéria escura

</div>

*"Neste capítulo final, os alunos serão apresentados ao assunto da matéria escura e ao modelo padrão cosmológico. É imprescindível que se atentem para como, nesse caso, a partir da observação, surgem as hipóteses e depois como elas são testadas. O tópico sobre Modelo Padrão é uma pequena amostra do processo de teoria sendo validada a partir dos testes"*

**Introdução**

Além dos buracos negros, a chamada "matéria escura" representa outro dos fenômenos mais intrigantes do Universo. Como conhecer algo que, simplesmente, não interage conosco? Pelo menos, até onde se sabe, hoje, a matéria escura não interage significativamente com a matéria *comum*, a não ser gravitacionalmente (e provém desse tipo de interação gravitacional os indícios mais fortes da sua existência). Até consigo mesma, a matéria escura não possui interações significativas. Se formos mais profundos em nossa abordagem, chega a ser estranho chamar tal *coisa* de matéria! Como brinca o astrofísico e divulgador científico, Neil deGrasse Tyson, em uma das suas palestras sobre o assunto "Nós poderíamos chamá-la de Fred!", se referindo ao fato de que nem mesmo sabemos o que ela é[109] e qual a sua natureza. Quando pensamos que mais de 80% da gravidade exibida em nosso Universo não pode ser explicada pela quantidade de matéria visível[110], o assunto "matéria escura" ganha contornos imprescindíveis na nossa busca de conhecimento em Cosmologia.

**Breve histórico**

Embora o trabalho do astrônomo alemão Fritz Zwicky, de 1933, muitas vezes, seja apontado como o primeiro a propor a existência de matéria escura, outros cientistas anteriores a ele já discutiam a ideia de "matéria que não podia ser observada"[111]. É de Lord Kelvin, um físico importante na história da ciência, um dos primeiros relatos referindo-se ao fato de que a massa do grupo de estrelas, as quais ele observava, orbitando o centro da galáxia, não condizia com o valor estimado com base nas suas velocidades de afastamento apresentadas[112]. Ele conclui o estudo, levantando a hipótese de que esse efeito se daria por conta de estrelas que não estavam sendo vistas, pois seriam "corpos escuros". Henri Poincarè, outro nome muito importante da ciência, cujo trabalho foi dedicado, principalmente, à física, matemática e filosofia da ciência, também falou sobre a hipótese de Kelvin, em 1906, se referindo a ela como "matéria negra" (no original, em francês, matière obscure). Nas décadas posteriores, os astrônomos Jacobus Kapteyn e Jaan Ortz também falaram sobre a hipótese de matéria escura, ao estudarem as discrepâncias observadas nos valores de massa e velocidade de estrelas na galáxia. Porém, o trabalho mais conhecido foi proposto, em 1933, pelo astrônomo alemão Fritz Zwicky. Estudando a dinâmica do movimento de algumas galáxias em aglomerados e utilizando a física newtoniana, ele observou que a velocidade apresentada por essas galáxias não correspondia a estimativa de massa concebida quando se analisava a matéria visível nesse grupo[113]. Em outras palavras, o resultado para a massa total do sistema, calculado a partir das velocidades observadas das galáxias, era muito superior ao cálculo de massa feito através da observação direta da matéria visível do mesmo. Zwicky, então, propõe que algum tipo de matéria escura (que ele chamou em alemão como dunkle Materie) estava presente ao redor desses aglomerados, e os estava afetando gravitacionalmente.

---

[109] O vídeo com legendas do trecho da palestra pode ser visto aqui:
https://www.youtube.com/watch?v=xaqO4gLEF28

[110] Basicamente, os astrônomos não conseguem explicar a intensidade das interações gravitacionais somente com a matéria detectável, ficando grande parte dessa gravidade sem uma causa aparente.

[111] Página do Instituto de Tecnologia da Califórnia (Caltech) com um pouco da história da matéria negra - https://ned.ipac.caltech.edu/level5/Sept16/Bertone/Bertone2.html

[112] Kelvin, Lord (1904). Baltimore Lectures on Molecular Dynamics and the Wave Theory of Light.    London, England: C.J. Clay and Sons. p. 274. – Disponível em :
https://babel.hathitrust.org/cgi/pt?id=ien.35556038198842&view=1up&seq=304

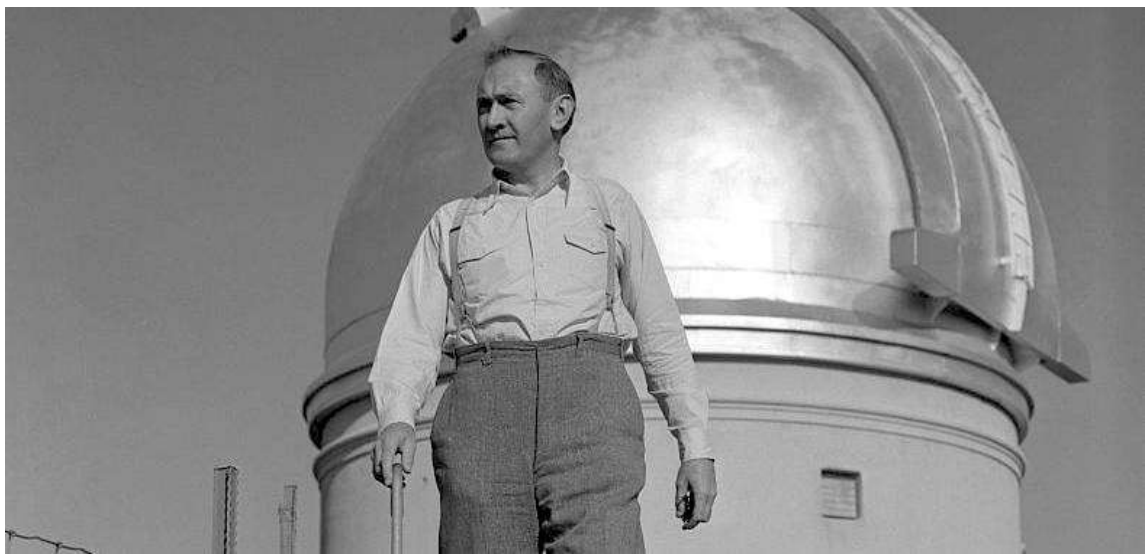

Fritz Zwicky em frente a um observatório. (Fonte: https://lithub.com/the-humble-origins-of-the-man-who-discovered-dark-matter/)

Os trabalhos de Kent Ford e Vera Rubin na década de 70 e início de 80 foram muito importantes para os estudos sobre a matéria escura. Basicamente, através da observação do movimento de galáxias espirais individuais, ficou comprovado que a velocidade de rotação delas também era maior do que se esperava, de acordo com as massas medidas. Hoje, esse estudo é considerado como um forte indício da existência da, ainda não detectada diretamente, matéria escura. Poucos anos depois da publicação de Rubin, uma teoria alternativa à matéria escura foi também publicada, tentando explicar essas discrepâncias entre velocidades e quantidade de matéria observada, em galáxias e aglomerados. Mordhai Milgrom, físico israelense, publicou em 1983 uma teoria onde, ao invés de se sugerir a existência de mais matéria não detectável (matéria escura), para que os dados correspondessem a teoria, propôs alterar a própria teoria gravitacional. A dinâmica modificada newtoniana (sigla em inglês *MOND – Modified Newtonian Dynamics*) previa que, para pequenas velocidades (tais como as das galáxias nos estudos), as equações deveriam ser modificadas[114]. Basicamente, a segunda lei de Newton passaria a ser um caso mais específico de outra lei mais abrangente[115]:

$$F = m.a.\mu(a/a1)$$

Onde, "m" e "a" são a massa e aceleração do corpo, respectivamente, e "a1" seria uma nova constante de aceleração, de valor igual a $10^{-10}$m/s$^2$. O termo adicionado $\mu(a/a1)$ é uma função que pode ser escrita de vária formas, entretanto, a maneira exata de escrevê-la não influi diretamente nas implicações da MOND[116]. O fato essencial é entender que quando a razão a/a1 for muito maior do que 1, ou seja, a >> a1, a equação original de Newton é válida, e $\mu(a/a1)$ = 1. Já quando a razão a/a1 for muito menor do que 1, ou seja, quando a << a1, o termo assumirá o valor do cálculo dessa razão, podendo ser escrito, de forma simplificada como:

$$\mu = \frac{a}{a1}$$

---

[113] Consultar - XIMENES, Samuel Jorge Carvalho. A Matéria Escura. 2016. Tese de Mestrado. Universidade Federal do Rio de Janeiro. Rio de Janeiro. – para ver os cálculos (p.34-35).

[114] Pode ser encontrado, com mais detalhes, aqui - XIMENES, Samuel Jorge Carvalho. A Matéria Escura. 2016. Tese de Mestrado. Universidade Federal do Rio de Janeiro. Rio de Janeiro. – para ver os cálculos (p.30-33)

[115] VELTEN, H. E. S. MOND: uma alternativa à mecânica newtoniana. Revista brasileira de ensino de Física, v. 30, n. 3, p. 3314.1-3314.5, 2008.

Resultando em

$$F = m\frac{a^2}{a1}$$

Note que a1 tem um valor muito reduzido, de forma que situações em nosso cotidiano em que a << a1 praticamente não existem. No entanto, em se tratando da aceleração de galáxias e aglomerados de galáxias, os valores são tão baixos quanto a1, sendo esse um caso onde os defensores dessa teoria consideram que ela funcione se a necessidade da hipótese da matéria escura. A MOND não explica muito bem alguns outros fatores relacionados ao comportamento de aglomerados de galáxias e também o efeito das lentes gravitacionais, que veremos mais adiante, ainda neste capítulo. Em 2018, um estudo com a participação de dois pesquisadores brasileiros, publicado na revista "*Nature Astrophysics*", disse ter descartado, com alta margem de confiabilidade, a hipótese da MOND, depois da análise de 193 galáxias em busca do fator $a_1$[117]. Porém, essa ainda é uma questão ainda estudada e requer mais esclarecimentos.

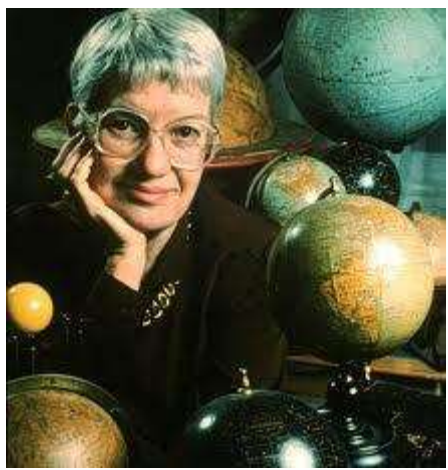

Vera Rubin, uma cientista que enfrentou muitas dificuldades pelos preconceitos da sua época. (Fonte: http://mulheresnaciencia-mc.blogspot.com/2013/02/vera-rubin.html)

**Um pouco de matemática para o problema das velocidades nos aglomerados de galáxias**

Consideramos um corpo de massa m, em órbita circular, ao redor de uma massa M (como uma galáxia em um aglomerado orbitando um centro massivo), de modo que M é significativamente maior do que m (M >> m) e r é a distância entre eles. O termo *v* é a velocidade de rotação de m ao redor de M. Aplicando a 2ª Lei de Newton:

---

[116] Idem nota anterior (7).

[117] Reportagem da Revista Brasileira do Ensino de Física sobre a pesquisa: http://www.sbfisica.org.br/v1/home/index.php/pt/destaque-em-fisica/726

$$m \cdot \frac{v^2}{r} = G \cdot m \cdot M / r^2$$

A energia cinética $E_c$ e potencial $E_p$ são:

$$E_C = \frac{1}{2} \cdot m \cdot a$$

e

$$E_p = -GmM/r$$

Fritz Zwicky fez uso do teorema virial para analisar a questão das velocidades das galáxias em aglomerados[118]. Iremos fazer uma dedução aproximada e mais simplificada (procure entender o conceito). O teorema virial nos diz que em um sistema de N partículas, interagindo gravitacionalmente, se não houver variações significativas na média da energia cinética $<E_c>$ e potencial $<E_p>$ do sistema, é válida a seguinte relação[119][120]:

$$2 <E_c> = -<E_p>$$

Na aplicação do aglomerado de galáxias, vamos utilizar como média o somatório das energias cinética e potencial das partículas:

$$<E_c> = \sum E_c \sim \frac{1}{2} M_t V$$

$$<E_p> = \sum E_p \sim - M_t^2 G/2R$$

O fator 2 no denominador da energia potencial é devido a existência de $N^2/2$ pares de partículas no aglomerado. $V$ é o valor resultante da soma das velocidades. Aplicando o teorema virial e deixando expresso em função da massa total do sistema $M_t$:

$$M_t = \frac{2RV^2}{G}$$

Aqui temos um resultado interessante para a aplicação. A massa total do aglomerado está expressa em função também da velocidade $V$. Ou seja, a partir dos valores medidos e somados das velocidades seria possível encontrar o valor da massa total do sistema. Não haveria discrepâncias se não existisse outra maneira confiável para também calcular a mesma massa. Porém, a partir do brilho de uma galáxia é possível calcular a sua massa total[121]. Uma galáxia comum emite um valor de brilho por unidade de massa proporcional a 0,3 o brilho por unidade de massa do Sol. Como sabemos a massa do Sol, conhecendo o brilho de uma galáxia conseguimos estipular a sua massa.

Os valores encontrados por Zwicky para a massa, aplicando o teorema virial, eram significativamente superior aos esperados, de acordo com o brilho emitido pelo aglomerado. A massa da "matéria luminosa"[122] era muito inferior ao valor encontrado para a massa total do aglomerado Os caminhos que levam a hipótese para matéria escura nasceram daí [123].

---

[118] Trabalho original pode ser acessado aqui - https://ui.adsabs.harvard.edu/abs/1937ApJ....86..217Z/abstract

[119] Pode ser visto aqui - http://www.if.ufrgs.br/fis02001/aulas/virial.htm

[120] Teorema com mais detalhes nessa referência - L. D. Landau, E. M. Lifshitz, Mechanics, cap. 10, 3a edição, Butterworth-Heinemann (1976)

[121] Consultar - B. Ryden, Introduction to Cosmology, p.127, Addison-Wesley (2003).

**Lentes Gravitacionais**

Outro fato indireto considerado como um forte indício da existência de matéria escura é o efeito das lentes gravitacionais. Esse efeito tem a ver com o mesmo fenômeno que estudamos no capítulo sobre a Teoria Geral da Relatividade em relação aos feixes de luz provenientes de estrelas serem curvados, ao passarem perto Sol (a curvatura do espaço-tempo, causada pela massa solar, desvia a trajetória dos raios luminosos vindo das estrelas, causando uma aparente diferença nas suas posições). Quando a fonte que está emitindo a luz e o observador estão bem alinhados, e os raios de luz que chegam para o observador sofrem algum desvio gravitacional no caminho, causado por um corpo massivo, o observador terá a impressão de que está enxergando *várias fontes luminosas*, quando, na verdade, só existe uma. Esse efeito é o das *lentes gravitacionais* (o nome lente é devido ao fato de que elas amplificam as imagens).

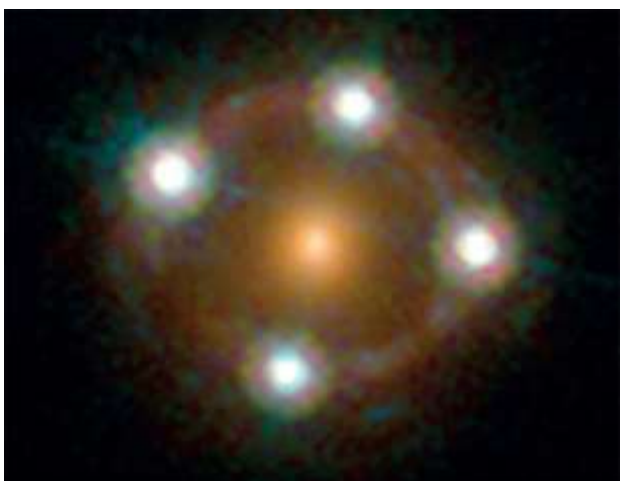

No cento da imagem, uma galáxia e, ao seu redor, quatro corpos luminosos. Na verdade, esses quatro corpos são somente um, e ele está num plano *atrás* ao da galáxia. A lente gravitacional causada pela massa da galáxia (centro) desvia a luz do quasar (disco de acreção energizado e emitindo luz) e cria essa ilusão de ótica. (Fonte: CERN Courier – disponível em https://cerncourier.com/a/gravitational-lens-challenges-cosmic-expansion/)

E, o que as lentes gravitacionais têm a ver com a matéria escura? Bem, através desse efeito também é possível modelar como é a distribuição de massa do corpo que o produz e, no caso das lentes gravitacionais causadas por galáxias, essa distribuição de matéria mostrou que maior parte dela não era composta por matéria luminosa (estrelas, gases entre outros) mas de matéria escura, a qual não emite qualquer tipo de radiação eletromagnética[124].

A figura abaixo mostra um mapeamento de matéria escura (azul) e de matéria não-escura (rosa), em seis colisões diferentes entre galáxias. Os pesquisadores utilizaram essas imagens captadas pelos telescópios Hubble e Chandra para analisar o comportamento da matéria escura[125].

---

[122] Não significa matéria que emite luz própria, mas é um termo usado para designar qualquer matéria que emita ou reflita luz. Matéria detectável.

[123] Utilizamos uma maneira bem aproximada para demonstrar como, teoricamente, se dão as discrepâncias entre os valores de massa total de um aglomerado calculados por dois métodos diferentes. Para detalhes, consultar o trabalho original de Fritz Zwicky já mencionado em nossas notas de rodapé.

[124] Acessar https://hubblesite.org/contents/articles/gravitational-lensing

[125] Acessar https://chandra.harvard.edu/press/15_releases/press_032615.html

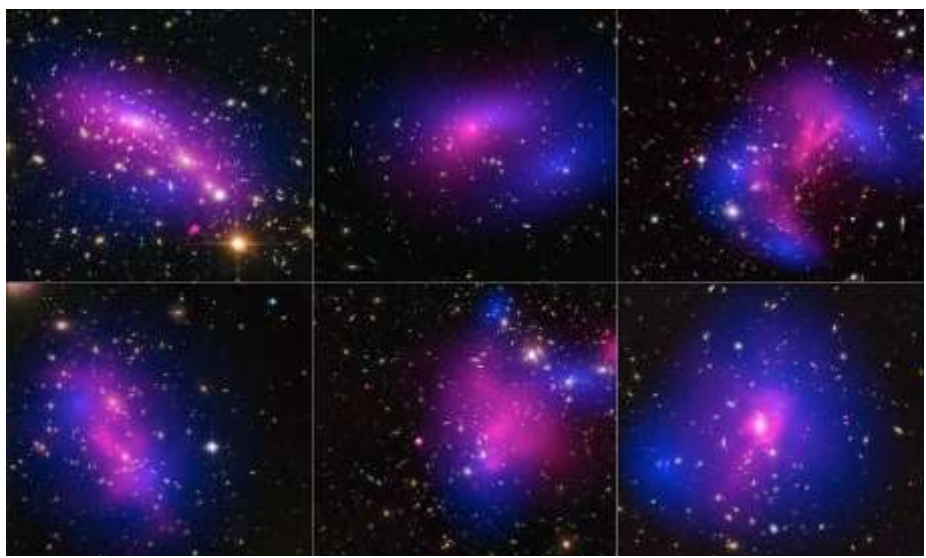

Imagens das seis colisões diferentes e os mapeamentos de matérias. (Fonte: Chandra Observatory

Eles concluíram que ela, além de não interagir com a matéria luminosa, também quase não interagiu consigo mesma. Por exemplo, observaram que uma porção de matéria escura, quando estava em colisão com outra porção também de matéria escura, não sofreu nenhuma desaceleração, revelando que, a princípio, elas não interagiram fisicamente como seria esperado no choque entre matéria. Esses fatos conferem ainda mais perplexidade e mistério ao assunto da matéria escura.

**O que se sabe sobre a matéria escura?**

Não existem evidências diretas da sua existência. O fenômeno das lentes gravitacionais e das velocidades em aglomerados de galáxias e galáxias individuais são os indícios mais fortes que indicam que *existe algo a mais* do que só a matéria que podemos detectar, interagindo apenas gravitacionalmente com ela. Lembrando Tyson, não sabemos nem se esse *algo a mais* pode ser considerado matéria. Pelas observações das colisões entre galáxias, existem também fortes indícios de que a matéria escura quase não interage consigo mesma. Sobre a sua natureza e do que é constituída existem apenas teorias que tentam explicá-la, mas ainda sem evidências observacionais (teorias modificadas da gravidade também podem ser hipóteses viáveis como alternativa à matéria escura). Enfim, esses estudos são um campo recente da cosmologia, os quais necessitam de mais desenvolvimento e pesquisas. A cada dia, surgem novas informações e trabalhos e, dessa forma, o campo do conhecimento vai avançando.

**Modelo Cosmológico Padrão ou Modelo $\Lambda CDM$**

Esse modelo é o que melhor explica o Universo, de acordo com as evidências observacionais da Cosmologia[126][127]. Um dos últimos resultados sobre as observações do espaço, para os estudos cosmológicos que temos, foram coletados pelo satélite Planck e publicados no ano de 2015[128]. A pesquisa conclui que o Modelo Cosmológico Padrão fornece uma excelente descrição dos dados coletados pelo satélite e, se existisse alguma física que esteja além desse modelo, o satélite não pôde detectá-la (ver página 59 do artigo original).

---

[126] Recomenda-se a leitura de - SILVA NETO, Gival Pordeus da. Estimando parâmetros cosmológicos a partir de dados observacionais. **Revista Brasileira de Ensino de Física**, v. 40, n. 2, 2018.

[127] Recomenda-se a leitura de - ROSENFELD, Rogério. A cosmologia. **Física na Escola**, v. 6, n. 1, p. 31-37, 2005.

[128] Artigo original - ADE, Peter AR et al. Planck 2015 results-xiii. cosmological parameters. Astronomy & Astrophysics, v. 594, p. A13, 2016.

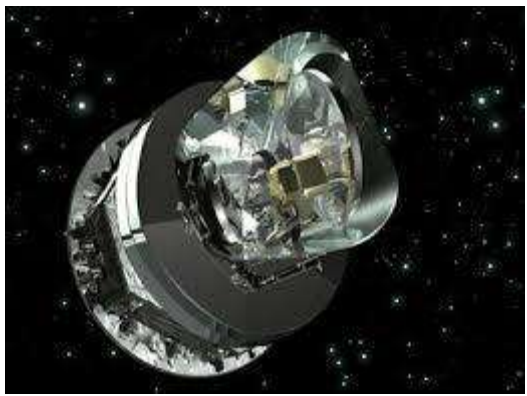

Satélite Planck. (Fonte: Agência Espacial Europeia - https://www.esa.int/ESA_Multimedia/Images/2007/01/Planck_spinning_in_space)

Em relação ao nome Modelo $\Lambda CDM$, o termo $\Lambda$ (lambda) se refere ao que chamamos de energia escura. Essa energia é considerada como sendo a maior componente do Universo, e seria ela a responsável pela sua expansão estar acelerando, conforme o tempo passa. Conforme vimos no capítulo sobre a expansão do Universo, atualmente, ela está ocorrendo de maneira acelerada e, para que isso aconteça, é necessário que algum tipo de energia com pressão negativa vença a atração gravitacional, gerada pelas massas distribuídas no universo. Do contrário, seria de se esperar que a expansão fosse desacelerando, porém, não é isso que as evidências observacionais mostram. Os dados revelam que o Universo é composto por, aproximadamente, 70% dessa energia escura, mas não sabemos ao certo nem sua origem e nem sua natureza. Já o termo **CDM** é uma sigla para, em inglês, "Cold Dark Matter" ou "Matéria Escura Fria". O nome é devido que existem outras duas teorias alternativas, a matéria escura morna e matéria escura quente, sendo que a teoria da matéria escura fria é a que melhor estima as quase inexistentes interações desse tipo de matéria e as possíveis partículas que a formaria. De acordo com as observações, aproximadamente, 25% do Universo é composto por ela. Apenas 5% é composto por matéria comum, a mesma que forma você, seu livro, o Sol, os oceanos, animais etc[129].

O Modelo $\Lambda CDM$ também sustenta a Teoria Inflacionária do Big Bang tal como vimos no capítulo sobre a origem do Universo. A detecção da abundância de hidrogênio (deutério), hélio e as quantidades menores dos demais elementos reforçam a teoria da nucleossíntese primordial. A radiação cósmica de fundo é outra evidência muito forte em favor do modelo. A partir dela, a idade do Universo é calculada como sendo, aproximadamente, 13,7 bilhões de anos. A expansão acelerada também é sustentada observacionalmente, sendo a constante de Hubble (taxa de expansão) de, aproximadamente, 71 Km/s.

Em relação à forma do Universo, a observação da radiação cósmica de fundo leva à predição de que, em larga escala, o Universo é plano. Isso quer dizer que, dois raios de luz viajando paralelamente um ao outro nunca irão se cruzar. Você pode estar se perguntando sobre o que aprendeu em Teoria Geral da Relatividade, sobre o espaço-tempo curvado pela presença de matéria. Está correto, e, em análises de escalas menores, essas curvaturas prevalecem. Quando aumentamos a escala, aí sim observamos que o Universo é plano.

Dessa forma, termina aqui nossa "microscópica" introdução à Cosmologia Moderna. Existem muito mais conhecimento aprofundado esperando por você! No entanto, o que apresentamos aqui pode ser usado como base para que os estudos prossigam! Esperamos que depois da experiência com esse material, você tenha adquirido alguns conhecimentos e ideias que o auxiliem a questionar, pesquisar e argumentar sobre o Universo em que vive. Existe muito, mas muito ainda a ser estudado e explorado, de forma que continuar nessa jornada é imprescindível, seja profissionalmente, aos que quiserem seguir por esse caminho, ou mesmo nas horas vagas e recreativas – querer desvendar o Universo é essencial à nossa vida! Boa viagem!

---

[129] Sugestão de leitura: https://astropontos.org/2019/10/28/o-premio-nobel-pelos-fundamentos-da-cosmologia/

*"O Cosmos é tudo o que existe, existiu ou existirá. Nossa contemplação do Cosmos nos comove – provoca calafrios, nos deixa sem voz, causa uma sensação de vertigem, como uma memória remota de estarmos caindo de uma grande altura. Sabemos que estamos nos aproximando do maior de todos os mistérios"*

Carl Sagan — No episódio "As Margens do Oceano Cósmico", primeiro da série Cosmos: A Personal Voyage, em 1980.

**Para conhecer mais**

- *O que é Cosmologia?* – **Livro de Mario Novello, cosmólogo brasileiro.**

- *O Universo numa casca de noz* – **Livro de Stephen Hawking.**

**Perguntas de fixação**

    **a) Descreva o que é a matéria escura? Podemos afirmar a sua existência? Justifique.**

"*A matéria escura é uma hipótese que explica a discrepância das velocidades, nas periferias de aglomerados de galáxias e galáxias unitárias. Como a teoria newtoniana prevê que, nesses casos, a velocidade dos corpos, ao redor da galáxia ou do aglomerado, é proporcional à massa dos mesmos, era de se esperar que isso se comprovasse na prática, quando se observasse essas estruturas. No entanto, como vimos no capítulo, existe uma grande discrepância entre a quantidade de matéria visível (massa) e a velocidade apresentada pelos corpos nas periferias dessas estruturas. Como a velocidade deles é muito maior do que o esperado, uma hipótese é a de que exista mais matéria naquela estrutura do que podemos ver – seria ela a matéria escura. Diretamente, sua exustência não foi comprovada com observações. Existem alguns indícios, como as lentes gravitacionais que sustentam a hipotese, porém não a evidenciam.*